\documentclass[11pt]{book} % Default font size 

\usepackage[top=3cm,bottom=3cm,left=3cm,right=3cm,headsep=10pt,a4paper]{geometry} % Page margins

\usepackage{graphicx} % Required for including pictures
\graphicspath{{./Pictures/}} % Specifies the directory where pictures are stored

\usepackage{tikz} % Required for drawing custom shapes

\usepackage[english]{babel} % English language/hyphenation
\usepackage{hyphenat}

\usepackage{enumitem} % Customize lists
\setlist{nolistsep} % Reduce spacing between bullet points and numbered lists

\usepackage{booktabs} % Required for nicer horizontal rules in tables

\usepackage{xcolor} % Required for specifying colors by name
\definecolor{bluinfn}{RGB}{0,42,72}  % Define the color used for highlighting throughout the book
\definecolor{azinfn}{RGB}{65,150,180}  % Define the color used for highlighting throughout the book
\definecolor{ocre}{RGB}{0,0,255}  % Define the color used for highlighting throughout the book
\definecolor{purple}{RGB}{120,0,240}  % Define the color used for highlighting throughout the book
\definecolor{purple-bckg}{RGB}{180,160,240}  % Define the color used for highlighting throughout the book

\usepackage{import} %import files
\usepackage{tabulary} %formatted tables
\usepackage{tabularx} %formatted tables
\usepackage[squaren,mediumqspace]{SIunits} %SI units formatting
\usepackage{subcaption} %multiple figures
\usepackage{multirow} %multirow in a single table cell
\usepackage{makecell} %another package for cell formatting in tables
\usepackage[gen]{eurosym} %€ symbol, use: \euro{}
\usepackage{pdfpages} %include pdf documents

\usepackage{textcomp}
\usepackage{adjustbox}
\usepackage[normalem]{ulem}
\usepackage{avant} % Use the Avantgarde font for headings
\usepackage{mathptmx} % Use the Adobe Times Roman as the default text font together with math symbols from the Sym­bol, Chancery and Com­puter Modern fonts

\usepackage{microtype} % Slightly tweak font spacing for aesthetics
\usepackage[utf8]{inputenc} % Required for including letters with accents
\usepackage[T1]{fontenc} % Use 8-bit encoding that has 256 glyphs

\usepackage[square,numbers,sort&compress]{natbib}

\usepackage{titletoc} % Required for manipulating the table of contents

\contentsmargin{0cm} % Removes the default margin

\titlecontents{part}[0cm]
{\addvspace{20pt}\centering\large\bfseries\color{white}}
{}
{}
{}

\titlecontents{chapter}[1.25cm] % Indentation
{\addvspace{12pt}\large\sffamily\bfseries} % Spacing and font options for chapters
{\color{ocre!80}\contentslabel[\Large\thecontentslabel]{1.25cm}\color{ocre}} % Chapter number
{\color{ocre}}  
{\color{ocre!80}\normalsize\;\titlerule*[0.5pc]{.}\;\thecontentspage} % Page number

\titlecontents{section}[1.25cm] % Indentation
{\addvspace{3pt}\sffamily\bfseries} % Spacing and font options for sections
{\contentslabel[\thecontentslabel]{1.25cm}} % Section number
{}
{\hfill\color{black}\thecontentspage} % Page number
[]

\titlecontents{subsection}[1.25cm] % Indentation
{\addvspace{1pt}\sffamily\small} % Spacing and font options for subsections
{\contentslabel[\thecontentslabel]{1.25cm}} % Subsection number
{}
{\ \titlerule*[.5pc]{.}\;\thecontentspage} % Page number
[]

\titlecontents{figure}[1em]
{\addvspace{-1pt}\sffamily}
{\thecontentslabel\hspace*{1em}}
{}
{\ \titlerule*[0.5pc]{.}\;\thecontentspage}
[]

\titlecontents{table}[1em]
{\addvspace{-1pt}\sffamily}
{\thecontentslabel\hspace*{1em}}
{}
{\ \titlerule*[0.5pc]{.}\;\thecontentspage}
[]

\titlecontents{lchapter}[0em] % Indenting
{\addvspace{15pt}\large\sffamily\bfseries} % Spacing and font options for chapters
{\color{ocre}\contentslabel[\Large\thecontentslabel]{1.25cm}\color{ocre}} % Chapter number
{}  
{\color{ocre}\normalsize\sffamily\bfseries\;\titlerule*[.5pc]{.}\;\thecontentspage} % Page number

\titlecontents{lsection}[0em] % Indenting
{\sffamily\small} % Spacing and font options for sections
{\contentslabel[\thecontentslabel]{1.25cm}} % Section number
{}
{}

\titlecontents{lsubsection}[.5em] % Indentation
{\normalfont\footnotesize\sffamily} % Font settings
{}
{}
{}

\usepackage{fancyhdr} % Required for header and footer configuration

\fancypagestyle{plain}{\fancyhead{}} % Style for when a plain pagestyle is specified

\makeatletter
\renewcommand{\cleardoublepage}{
\clearpage\ifodd\c@page\else
\hbox{}
\vspace*{\fill}
\thispagestyle{empty}
\newpage
\fi}

\usepackage{amsmath,amsfonts,amssymb,amsthm} % For math equations, theorems, symbols, etc

\newtheoremstyle{ocrenumbox}% % Theorem style name
{0pt}% Space above
{0pt}% Space below
{\normalfont}% % Body font
{}% Indent amount
{\small\bf\sffamily\color{ocre}}% % Theorem head font
{\;}% Punctuation after theorem head
{0.25em}% Space after theorem head
{\small\sffamily\color{ocre}\thmname{#1}\nobreakspace\thmnumber{\@ifnotempty{#1}{}\@upn{#2}}% Theorem text (e.g. Theorem 2.1)
\thmnote{\nobreakspace\the\thm@notefont\sffamily\bfseries\color{black}---\nobreakspace#3.}} % Optional theorem note
\newtheoremstyle{blacknumex}% Theorem style name
{5pt}% Space above
{5pt}% Space below
{\normalfont}% Body font
{} % Indent amount
{\small\bf\sffamily}% Theorem head font
{\;}% Punctuation after theorem head
{0.25em}% Space after theorem head
{\small\sffamily{\tiny\ensuremath{\blacksquare}}\nobreakspace\thmname{#1}\nobreakspace\thmnumber{\@ifnotempty{#1}{}\@upn{#2}}% Theorem text (e.g. Theorem 2.1)
\thmnote{\nobreakspace\the\thm@notefont\sffamily\bfseries---\nobreakspace#3.}}% Optional theorem note

\newtheoremstyle{blacknumbox} % Theorem style name
{0pt}% Space above
{0pt}% Space below
{\normalfont}% Body font
{}% Indent amount
{\small\bf\sffamily}% Theorem head font
{\;}% Punctuation after theorem head
{0.25em}% Space after theorem head
{\small\sffamily\thmname{#1}\nobreakspace\thmnumber{\@ifnotempty{#1}{}\@upn{#2}}% Theorem text (e.g. Theorem 2.1)
\thmnote{\nobreakspace\the\thm@notefont\sffamily\bfseries---\nobreakspace#3.}}% Optional theorem note

\newtheoremstyle{ocrenum}% % Theorem style name
{5pt}% Space above
{5pt}% Space below
{\normalfont}% % Body font
{}% Indent amount
{\small\bf\sffamily\color{ocre}}% % Theorem head font
{\;}% Punctuation after theorem head
{0.25em}% Space after theorem head
{\small\sffamily\color{ocre}\thmname{#1}\nobreakspace\thmnumber{\@ifnotempty{#1}{}\@upn{#2}}% Theorem text (e.g. Theorem 2.1)
\thmnote{\nobreakspace\the\thm@notefont\sffamily\bfseries\color{black}---\nobreakspace#3.}} % Optional theorem note
\makeatother

\newcounter{dummy} 
\numberwithin{dummy}{section}
\theoremstyle{ocrenumbox}
\newtheorem{theoremeT}[dummy]{Theorem}

\newtheorem{exerciseT}{Exercise}[chapter]
\theoremstyle{blacknumex}
\newtheorem{exampleT}{Example}[chapter]
\theoremstyle{blacknumbox}

\newtheorem{definitionT}{Definition}[section]
\newtheorem{corollaryT}[dummy]{Corollary}
\theoremstyle{ocrenum}

\RequirePackage[framemethod=default]{mdframed} % Required for creating the theorem, definition, exercise and corollary boxes

\newmdenv[skipabove=7pt,
skipbelow=7pt,
backgroundcolor=black!5,
linecolor=ocre,
innerleftmargin=5pt,
innerrightmargin=5pt,
innertopmargin=5pt,
leftmargin=0cm,
rightmargin=0cm,
innerbottommargin=5pt]{tBox}

\newmdenv[skipabove=7pt,
skipbelow=7pt,
rightline=false,
leftline=true,
topline=false,
bottomline=false,
backgroundcolor=ocre!10,
linecolor=ocre,
innerleftmargin=5pt,
innerrightmargin=5pt,
innertopmargin=5pt,
innerbottommargin=5pt,
leftmargin=0cm,
rightmargin=0cm,
linewidth=4pt]{eBox}	

\newmdenv[skipabove=7pt,
skipbelow=7pt,
rightline=false,
leftline=true,
topline=false,
bottomline=false,
linecolor=ocre,
innerleftmargin=5pt,
innerrightmargin=5pt,
innertopmargin=0pt,
leftmargin=0cm,
rightmargin=0cm,
linewidth=4pt,
innerbottommargin=0pt]{dBox}	

\newmdenv[skipabove=7pt,
skipbelow=7pt,
rightline=false,
leftline=true,
topline=false,
bottomline=false,
linecolor=gray,
backgroundcolor=black!5,
innerleftmargin=5pt,
innerrightmargin=5pt,
innertopmargin=5pt,
leftmargin=0cm,
rightmargin=0cm,
linewidth=4pt,
innerbottommargin=5pt]{cBox}

\makeatletter
\renewcommand{\@seccntformat}[1]{\llap{\textcolor{purple}{\csname the#1\endcsname}\hspace{1em}}}                    
\renewcommand{\section}{\@startsection{section}{1}{\z@}
{-4ex \@plus -1ex \@minus -.4ex}
{1ex \@plus.2ex }
{\normalfont\large\sffamily\bfseries}}
\renewcommand{\subsection}{\@startsection {subsection}{2}{\z@}
{-3ex \@plus -0.1ex \@minus -.4ex}
{0.5ex \@plus.2ex }
{\normalfont\sffamily\bfseries}}
\renewcommand{\subsubsection}{\@startsection {subsubsection}{3}{\z@}
{-2ex \@plus -0.1ex \@minus -.2ex}
{.2ex \@plus.2ex }
{\normalfont\small\sffamily\bfseries}}                        
\renewcommand\paragraph{\@startsection{paragraph}{4}{\z@}
{-2ex \@plus-.2ex \@minus .2ex}
{.1ex}
{\normalfont\small\sffamily\bfseries}}

\newcommand{\@mypartnumtocformat}[2]{%
\setlength\fboxsep{0pt}%
\noindent\colorbox{purple-bckg}{\strut\parbox[c][.7cm]{\ecart}{\color{purple}\Large\sffamily\bfseries\centering#1}}\hskip\esp\colorbox{purple!85}{\strut\parbox[c][.7cm]{\linewidth-\ecart-\esp}{\Large\sffamily\centering#2}}}% boxes main ToC color (number box, number, box title)
\newcommand{\@myparttocformat}[1]{%
\setlength\fboxsep{0pt}%
\noindent\colorbox{purple-bckg}{\strut\parbox[c][.7cm]{\linewidth}{\Large\sffamily\centering#1}}}%
\newlength\esp
\newlength\ecart
\def\@part[#1]#2{%
\ifnum \c@secnumdepth >-2\relax%
\refstepcounter{part}%
\addcontentsline{toc}{part}{\texorpdfstring{\protect\@mypartnumtocformat{\thepart}{#1}}{\partname~\thepart\ ---\ #1}}
\else%
\addcontentsline{toc}{part}{\texorpdfstring{\protect\@myparttocformat{#1}}{#1}}%
\fi%
\startcontents%
\markboth{}{}%
{\thispagestyle{empty}%
\begin{tikzpicture}[remember picture,overlay]%
\node at (current page.north west){\begin{tikzpicture}[remember picture,overlay]%	
\fill[purple-bckg!50](0cm,0cm) rectangle (\paperwidth,-\paperheight);%page color
\node[anchor=north] at (5cm,-3.25cm){\color{purple}\fontsize{220}{100}\sffamily\bfseries\thepart};%roman numbering
\node[anchor=south east] at (\paperwidth-1cm,-\paperheight+1cm){\parbox[t][][t]{8.5cm}{
\printcontents{l}{0}{\setcounter{tocdepth}{0}}% toc part print
}};
\node[anchor=north] at (15cm,-3.2cm){\parbox[t][][t]{11.5cm}{\strut\raggedright\color{white}\fontsize{50}{50}\sffamily\bfseries\nohyphens{#2}}};%title part print
\end{tikzpicture}};
\end{tikzpicture}}%
\@endpart}
\def\@spart#1{%
\startcontents%
\phantomsection
{\thispagestyle{empty}%
\begin{tikzpicture}[remember picture,overlay]%
\node at (current page.north west){\begin{tikzpicture}[remember picture,overlay]%	
\fill[purple-bckg](0cm,0cm) rectangle (\paperwidth,-\paperheight);
\node[anchor=north east] at (\paperwidth-1.5cm,-3.25cm){\parbox[t][][t]{15cm}{\strut\raggedleft\color{white}\fontsize{30}{30}\sffamily\bfseries#1}};
\end{tikzpicture}};
\end{tikzpicture}}
\addcontentsline{toc}{part}{\texorpdfstring{%
\setlength\fboxsep{0pt}%
\noindent\protect\colorbox{ocre!40}{\strut\protect\parbox[c][.7cm]{\linewidth}{\Large\sffamily\protect\centering #1\quad\mbox{}}}}{#1}}%
\@endpart}
\def\@endpart{\vfil\newpage
\if@twoside
\if@openright
\null
\thispagestyle{empty}%
\newpage
\fi
\fi
\if@tempswa
\twocolumn
\fi}

\newif\ifusechapterimage
\usechapterimagetrue
\newcommand{\thechapterimage}{}%
\newcommand{\chapterimage}[1]{\ifusechapterimage\renewcommand{\thechapterimage}{#1}\fi}%
\newcommand{\autodot}{.}
\def\@makechapterhead#1{%
{\parindent \z@ \raggedright \normalfont
\ifnum \c@secnumdepth >\m@ne
\if@mainmatter
\begin{tikzpicture}[remember picture,overlay]
\node at (current page.north west)
{\begin{tikzpicture}[remember picture,overlay]
\node[anchor=north west,inner sep=0pt] at (0,0) {\ifusechapterimage\includegraphics[width=\paperwidth]{\thechapterimage}\fi};
\draw[anchor=west] (\Gm@lmargin,-9cm) node [line width=2pt,rounded corners=15pt,draw=purple,fill=white,fill opacity=0.75,%inner sep=15pt
]{\strut\parbox[l]{22cm}{\parbox[l]{17cm}{\vspace{5pt}\hspace{5pt}\huge\sffamily\bfseries\color{purple}\thechapter\autodot~\nohyphens{#1}\strut\vspace{2pt}}}};
\end{tikzpicture}};
\end{tikzpicture}
\else
\begin{tikzpicture}[remember picture,overlay]
\node at (current page.north west)
{\begin{tikzpicture}[remember picture,overlay]
\node[anchor=north west,inner sep=0pt] at (0,0) {\ifusechapterimage\includegraphics[width=\paperwidth]{\thechapterimage}\fi};
\draw[anchor=west] (\Gm@lmargin,-9cm) node [line width=2pt,rounded corners=15pt,draw=purple,fill=white,fill opacity=0.75,inner sep=15pt]{\strut\makebox[22cm]{}};
\draw[anchor=west] (\Gm@lmargin+.3cm,-9cm) node {\huge\sffamily\bfseries\color{purple}#1\strut};
\end{tikzpicture}};
\end{tikzpicture}
\fi\fi\par\vspace*{270\p@}}}

\def\@makeschapterhead#1{%
\begin{tikzpicture}[remember picture,overlay]
\node at (current page.north west)
{\begin{tikzpicture}[remember picture,overlay]
\node[anchor=north west,inner sep=0pt] at (0,0) {\ifusechapterimage\includegraphics[width=\paperwidth]{\thechapterimage}\fi};
\draw[anchor=west] (\Gm@lmargin,-9cm) node [line width=2pt,rounded corners=15pt,draw=purple,fill=white,fill opacity=0.75,inner sep=15pt]{\strut\makebox[22cm]{}};
\draw[anchor=west] (\Gm@lmargin+.3cm,-9cm) node {\huge\sffamily\bfseries\color{purple}#1\strut};
\end{tikzpicture}};
\end{tikzpicture}
\par\vspace*{270\p@}}
\makeatother

\usepackage{hyperref}
\hypersetup{hidelinks,colorlinks=false,breaklinks=true,urlcolor= ocre,bookmarksopen=false,pdftitle={Title},pdfauthor={Author}}
\usepackage{bookmark}
\bookmarksetup{
open,
numbered,
addtohook={%
\ifnum\bookmarkget{level}=0 % chapter
\bookmarksetup{bold}%
\fi
\ifnum\bookmarkget{level}=-1 % part
\bookmarksetup{color=ocre,bold}%
\fi
}
}

\begin{document}
%----------------------------------------------------------------------------------------
%	TITLE PAGE
%----------------------------------------------------------------------------------------

\begingroup
\thispagestyle{empty}
\begin{tikzpicture}[remember picture,overlay]
%\node[inner sep=0pt] (background) at (current page.center) {\includegraphics[width=\paperwidth]{copertina}};     %copertina.pdf background image for the cover page (not used)
\node[anchor=south] at (7.4cm,-\paperheight+4.7cm){\includegraphics[width=0.5\paperwidth]{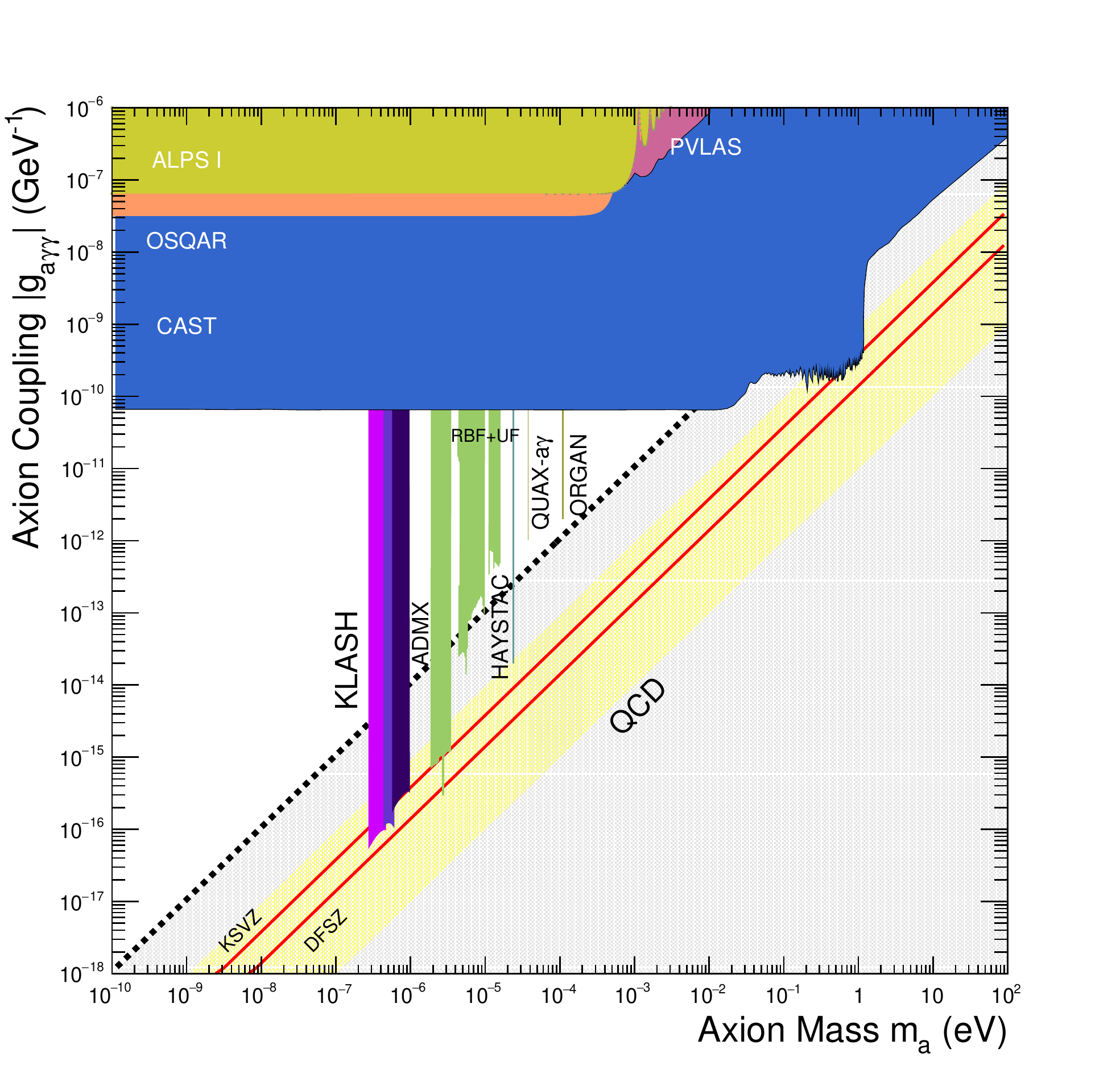}}; %add image below the centar title
\node[anchor=north west] at (-2.5cm,2.5cm){\includegraphics[width=0.2\paperwidth]{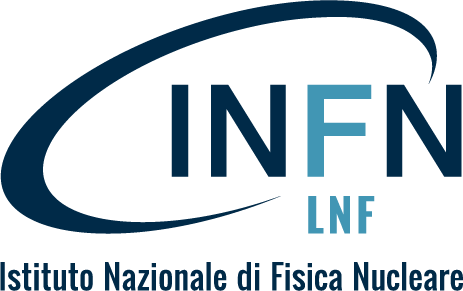}};% infn logo in the top left angle
\draw[anchor=north] (14cm,.5cm) node {\parbox{5cm}{\Large{INFN-19-18/LNF\\ \today}}};% CDR number in top right position
\draw (current page.center) node [fill=purple-bckg,fill opacity=0.6,text opacity=1,inner sep=1cm]{\Huge\centering\bfseries\sffamily\parbox[c][][t]{\paperwidth}{\centering\color{purple} {\bf KLASH}\\[15pt] % Book title
{\Large Conceptual Design Report}\\[20pt] % Subtitle
}}; 
\end{tikzpicture}
\vfill
\endgroup

%----------------------------------------------------------------------------------------
%	Authors
%----------------------------------------------------------------------------------------

\newpage

\thispagestyle{empty}

{\Huge Authors}\\
\\
\\
\\
{\large 
	 D. Alesini$^1$,
	 D. Babusci$^1$,
	 P. Beltrame S.J.$^2$,
	 F. Bj\"orkeroth$^1$,
	 F. Bossi$^1$, 
	 P. Ciambrone$^1$,
	 G. Delle Monache$^1$,
	 D. Di Gioacchino$^1$, 
	 P. Falferi$^3$,
	 A. Gallo$^1$,
	 C. Gatti$^1$,
	 A. Ghigo$^1$,
	 M. Giannotti$^4$
         G. Lamanna$^5$, 
         C. Ligi$^1$,
         G. Maccarrone$^1$,
         A. Mirizzi$^6$,
         D. Montanino$^7$,
         D. Moricciani$^1$,
         A. Mostacci$^7$,
         M. M\"uck$^8$, 
         E. Nardi$^1$,
         F. Nguyen$^9$,
         L. Pellegrino$^1$,
         A. Rettaroli$^{1,10}$,
         R. Ricci$^1$,
         L. Sabbatini$^1$,
         S. Tocci$^1$,
         L. Visinelli$^{11}$
}
\\
\\
\\
$^1$ \emph{INFN - Laboratori Nazionali di Frascati, via E. Fermi 40, 00044 Frascati, Italy.}\\%lnf
$^2$ \emph{Specola Vaticana, V-00120 Vatican City, Vatican City State and Vatican Observatory Research Group Steward Observatory, The University Of Arizona, 
933 North Cherry Avenue, Tucson, Arizona 85721, USA}\\ %specola
$^3$ \emph{Istituto di Fotonica e Nanotecnologie, CNR—Fondazione Bruno Kessler, and INFN-TIFPA, \\I-38123 Povo (Trento), Italy.}\\%tifpa-fbk
$^4$ \emph{Physical Sciences, Barry University, 11300 NE 2nd Ave., Miami Shores, FL 33161, USA.}\\
$^5$ \emph{Universit\`a di Pisa and INFN Pisa, Italy.}\\%pisa
$^6$ \emph{Dipartimento Interateneo di Fisica ``Michelangelo Merlin'', Universit\`a degli Studi di Bari, Bari, Italy.}\\%bari
$^7$ \emph{Universit\`a del Salento and INFN Lecce, Italy.}\\%lecce
$^8$ \emph{ez SQUID Mess- und Analyseger\"ate, D-35764 Sinn, Germany.}\\
$^9$ \emph{ENEA - Centro Ricerche Frascati, via E. Fermi 45, 00044 Frascati, Italy}\\%ENEA-Frascati
$^{10}$ \emph{Universit\`a ``Roma3" e INFN Roma3, Italy.}\\%roma3
$^{11}$ \emph{Gravitation Astroparticle Physics Amsterdam (GRAPPA), Institute for Theoretical Physics Amsterdam and Delta Institute 
for Theoretical Physics, University of Amsterdam, Science Park 904, 1098 XH Amsterdam, The Netherlands.}

%\newpage
%\vspace*{2.0cm}
%\section*{Introduction}
The axion, a pseudoscalar particle originally introduced by  Peccei, Quinn \cite{Peccei:1977ur,Peccei:1977hh}, Weinberg \cite{Weinberg:1977ma}, and 
Wilczek \cite{Wilczek:1977pj} to solve the ``strong CP problem'', is a well motivated dark-matter (DM) candidate with a mass lying in a broad 
range from peV to few meV \cite{PDG2018}. The last decade witnessed an increasing interest in axions and axion-like particles with many theoretical works 
published and many new experimental proposals \cite{Irastorza:2018dyq} that started a real \emph{race} towards their discovery.  Driven by this new challenge 
and stimulated by the availability, at the Laboratori Nazionali di Frascati (LNF), of large superconducting magnets previously used for particle detectors 
\cite{KLOETDR,FINUDA} at the DAFNE collider, we proposed to build a large \emph{haloscope} \cite{Sikivie} to observe galactic axions in the mass 
window between 0.2 and 1 $\mu$eV \cite{KLASHProposal}.

This paper is the Conceptual Design Report (CDR) of the KLASH (KLoe magnet for Axion SearcH) experiment, designed having in mind the performance and dimensions 
of the KLOE magnet, a large volume superconducting magnet with a moderate magnetic field of 0.6 T. In the first part of this Report we discuss the physics case of KLASH, the 
theoretical motivation for an axion in the mass window $0.1 \div1\,\mu$eV based on a review of \emph{standard} and \emph{non-standard} axion-cosmology (Sec. \ref{ch1:sec_theory}), 
and the physics reach of the KLASH experiment (Sec. \ref{ch2:sec_phys-reach}), including both the sensitivity to QCD axions and to Dark-Photon DM. The sensitivity plots are 
based on the detector performance discussed in the second part of the CDR. Here, we summarize the results obtained with calculations and simulations of several 
aspects of the experiment: the mechanical construction of cryostat and cavity based on the study commissioned to the mechanical engineers of the \texttt{Fantini-Sud} 
company \cite{Fantini} (Sec. \ref{cha:mech}); the cryogenics plant (Sec. \ref{cha:cryo}); the RF cavity design and tuning based on detailed simulations with code Ansys-HFSS 
(Sec. \ref{cha:rf}); the signal amplification, in particular the first stage based on a Microstrip SQUID Amplifier (Sec. \ref{cha:sig_amp}). 

Finally, in Sec. \ref{cha:ana}, mainly based on the experience of existing experiments \cite{ADMX,HAYSTAC,QUAX}, we discuss the data taking, analysis procedure and 
computing requirements.

The main conclusion we draw from this report is the possibility to build and put in operation at LNF in 2-3 years a large haloscope with the sensitivity to KSVZ axions in the low mass 
range between 0.2 and 1$\mu$eV in a region complementary to that of other experiments with a cost of about 3 M\euro. Timeline and cost are competitive with respect to other 
proposals in the same mass region \cite{ABRA,DMradio} thanks to the availability of most of the infrastructure, in particular the superconducting magnet and the cryogenics plant.

During the writing of this CDR, in July 2019, we were informed about the decision of INFN management to devote the KLOE magnet to the DUNE experiment at Fermilab. 
The KLOE magnet has always been the preferred choice for several reasons: it was in operation until 2018; its mechanical structure is able to support the several-tons weight of the 
cryostat and cavity; it is placed in the KLOE assembly-hall that can be used as the experimental area of KLASH. However, another option is given by the FINUDA magnet. The are few 
aspects to be explored (mechanical strength, move to experimental area,  put in operation after more than 10 years), but it has a higher nominal field of 1.1 T in a large volume with an 
inner radius 1385 mm and length 3800 mm. A preliminary estimate of sensitivity to axions of FLASH, the haloscope built with the FINUDA magnet, gives results similar to those obtained 
for KLASH. This option will be eventually investigated in another document.
\vspace*{0.5cm}

\section*{Acknowledgements}
L.V. thanks for the hospitality at Istituto Nazionale di Fisica Nucleare (INFN), Laboratori Nazionali di Frascati (LNF). L.V. acknowledges support by the Vetenskapsr\r{a}det 
(Swedish Research Council) through contract No. 638-2013-8993 and the Oskar Klein Centre for Cosmoparticle Physics. This work is part of the research program 
"The Hidden Universe of Weakly Interacting Particles" with project number 680.92.18.03 (NWO Vrije Programma), which is (partly) financed by the Dutch Research 
Council (NWO). 
%----------------------------------------------------------------------------------------
%	TABLE OF CONTENTS
%----------------------------------------------------------------------------------------

\pagestyle{empty} % No headers
\chapterimage{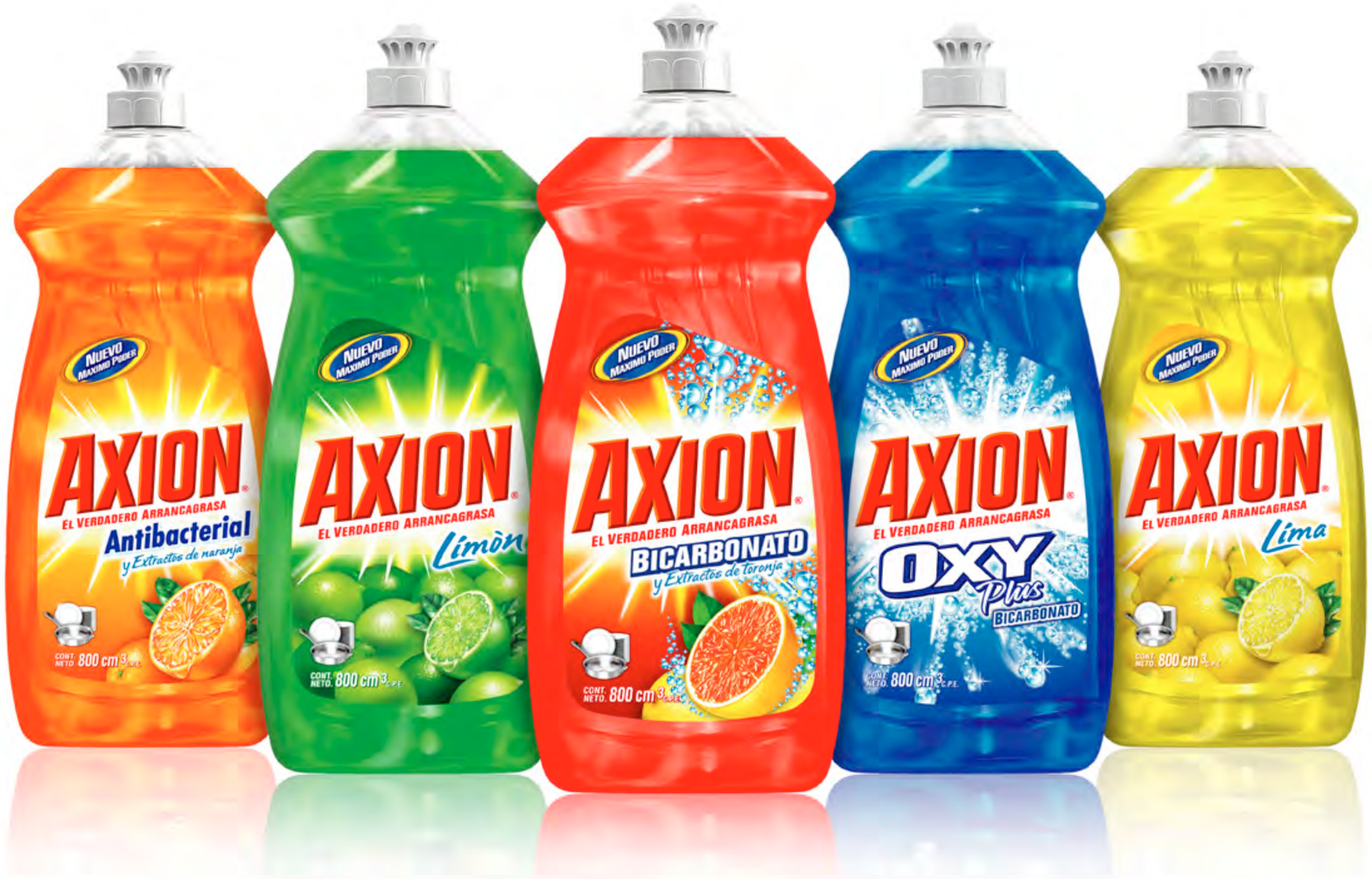} % Table of contents heading image
\tableofcontents

\cleardoublepage % Forces the first chapter to start on an odd page so it's on the right

\pagestyle{fancy} % Print headers again

%----------------------------------------------------------------------------------------
%	PARTS
%----------------------------------------------------------------------------------------
%
%
\chapterimage{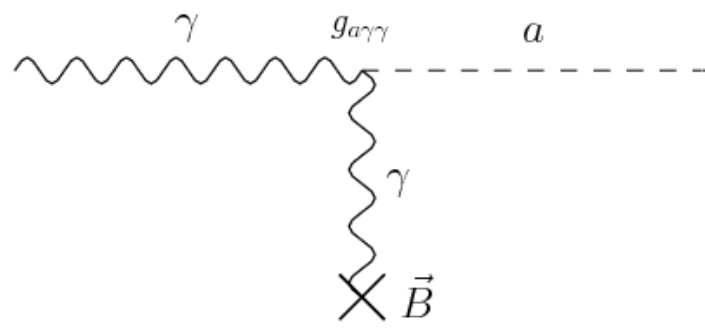}
 \part{Physics Case}
 
\everymath{\displaystyle} 
\newcommand{\pr}[1]{\ensuremath{\left[#1\right]}} 
\newcommand{\pc}[1]{\ensuremath{\left(#1\right)}} 
\newcommand{\px}[1]{\ensuremath{\left\lbrace#1\right\rbrace}} 
\newcommand{\bra}[1]{\ensuremath{\left\langle#1\right\vert}} 
\newcommand{\ket}[1]{\ensuremath{\left\vert#1\right\rangle}} 
\newcommand{\md}[1]{\ensuremath{\left\vert#1\right\vert}} 
\newcommand{\av}[1]{\ensuremath{\left\langle#1\right\rangle}} 
\newcommand{\dev}[1]{\ensuremath{\left.#1\right|}}
\newcommand{\vp}{\varphi}
\newcommand{\gae}{g_{ae}}
\newcommand{\ga}{g_{a\gamma}}
\newcommand{\RR}{\mathcal{R}}
\newcommand{\dH}{\dot{H}}
\newcommand{\du}{\dot{u}}
\newcommand{\bB}{{\bf B}}
\newcommand{\bE}{{\bf E}}
\newcommand{\MP}{M_{\rm P}}
\newcommand{\bs}{{\bf \sigma}}
\newcommand{\bn}{{\bf \nabla}}
\newcommand{\bv}{{\bf v}}
\newcommand{\bz}{\bar{\zeta}_1}
\newcommand{\tU}{\tilde{U}}
\newcommand{\mA}{\mathcal{A}}
\newcommand{\mD}{\mathcal{D}}
\newcommand{\mL}{\mathcal{L}}
\newcommand{\Curl}{{\rm Curl}}
\newcommand{\tF}{\tilde{F}}
\newcommand{\Rc}{R_{\rm mc}}
\newcommand{\Mc}{M_{\rm mc}}
\newcommand{\TL}{T_{C}}
\newcommand{\TRH}{T_{\rm RH}}
\newcommand{\n}{{\bf \nabla}}
\renewcommand\({\left(}
\renewcommand\){\right)}
\renewcommand\[{\left[}
\renewcommand\]{\right]}
\newcommand{\be}{\begin{equation}}
\newcommand{\ee}{\end{equation}}
\newcommand{\bea}{\begin{eqnarray}}
\newcommand{\eea}{\end{eqnarray}}

\chapter{The KLASH Physics Case}\label{ch1:sec_theory}

\section{Brief review on axion cosmology}

The ``invisible'' QCD axion~\cite{Weinberg:1977ma, Wilczek:1977pj} is
a light Goldstone boson arising within the solution to the strong CP
problem proposed by Peccei and Quinn~\cite{Peccei:1977ur,
  Peccei:1977hh}, and a possible dark matter
candidate~\cite{Abbott:1982af, Dine:1982ah, Preskill:1982cy}. In the
standard cosmological scenario, refined cosmological simulations yield
a narrow range in which the QCD axion would be the CDM particle, with
an axion mass in the range $m_A \approx (10-100)\,\mu {\rm eV}$,
as recently proven by refined cosmological
simulations~\cite{Klaer:2017qhr, Klaer:2017ond, Gorghetto:2018myk,
  Vaquero:2018tib}. Along with the QCD axion, a set of particles which
share the same phenomenology would arise from string compactification
and form the so-called ``axiverse''~\cite{Svrcek:2006hf,
  Svrcek:2006yi, Arvanitaki:2009hb, Arvanitaki:2009fg, Higaki:2012ar,
  Cicoli:2012aq, Cicoli:2012sz, Higaki:2013lra, Stott:2017hvl,
  Visinelli:2018utg, Kinney:2018nny}. These possibilities have been
proven interesting for a number of experiments that are planned to
explored the parameter space of the QCD axion and other axion-like
particles away from the preferred region~\cite{DiLuzio:2016sbl,DiLuzio:2017pfr,DiLuzio:2017ogq}, as discussed in depth in
various reviews on the subject~\cite{Raffelt:1995ym, Raffelt:2006rj,
  Sikivie:2006ni, Kim:2008hd, Wantz:2009it, Kawasaki:2013ae,
  Marsh:2015xka, Kim:2017yqo, Irastorza:2018dyq}.

\subsection{The temperature dependence of the QCD axion mass}

\begin{figure}[ht!]
\begin{center}
	\includegraphics[width=0.75\linewidth]{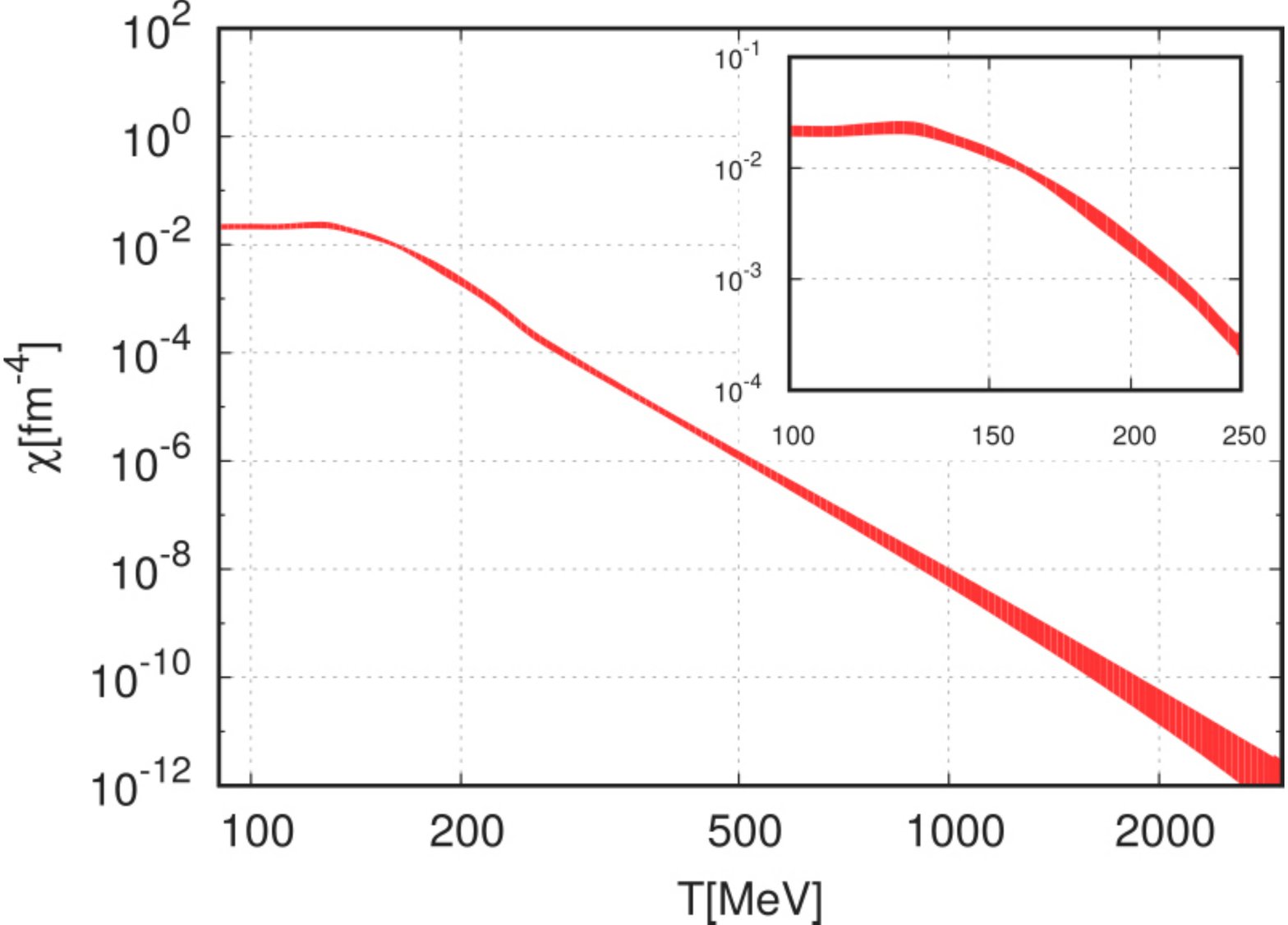}
	\caption{The topological susceptibility $\chi(T)$ as a
          function of temperature. The width of the line represents
          the combined statistical and systematic errors. The insert
          shows the behavior around the transition temperature. For additional details see Ref.~\cite{Borsanyi:2016ksw}.}
	\label{fig:susceptibility}
\end{center}
\end{figure}

The QCD axion mass originates from non-perturbative effects during the
QCD phase transition. At zero temperature, the axion gets a mass $m_0$
from mixing with the neutral pion~\cite{Weinberg:1977ma},
\begin{equation}
	m_0 = \frac{\Lambda^2}{f_A} = \frac{\sqrt{z}}{1+z}\,\frac{m_\pi f_\pi}{f_A},
	\label{eq:m0}
\end{equation}
where $z = m_u/m_d$ is the ratio of the masses of the up and down
quarks, $m_\pi$ and $f_\pi$ are respectively the mass and the decay
constant of the pion, and $f_A$ is the QCD axion energy scale. Since
the mass of the axion is tied to the underlying QCD theory, the energy
scale $\Lambda$ is related to the QCD scale $\Lambda_{\rm QCD}$. Using
$z = 0.48(5)$, $m_\pi = 132\,$MeV, and $f_\pi = 92.3\,$MeV gives
$\Lambda = 75.5\,$MeV~\cite{Weinberg:1977ma}. This value has been
confirmed by dedicated lattice simulations~\cite{Bonati:2015vqz,
  Borsanyi:2015cka, Borsanyi:2016ksw, Petreczky:2016vrs}, in which the
goal is to compute the QCD topological susceptibility $\chi(T)$
normalised so that at zero temperature $\chi(0) = \Lambda^4$. We show
the topological susceptibility around the QCD phase transition in
Fig.~\ref{fig:susceptibility}, which contains the results obtained by
the lattice simulation in Ref.~\cite{Borsanyi:2016ksw}.

At temperatures higher than that of the QCD phase transition, the
effects of instantons become severely suppressed. This reflects onto
the value of the axion mass, whose dependence on the temperature is
fixed in terms of the QCD topological susceptibility as
$\chi(T) \equiv f_A^2m_A^2(T)$. The axion mass decreases quickly with
an increasing value of the temperature of the plasma above the
confinement temperature $T_C$~\cite{Gross:53.43, Fox:2004kb}. Here, we
parametrise this dependence as~\cite{Gross:53.43}
\begin{equation}
m_A(T) =
m_0\left(\frac{T_C}{T}\right)^{\gamma},
	 \label{eq:QCDaxion_mass}
\end{equation}
where $m_0$ is the mass of the axion at zero temperature, $T_C$ is a
confinement temperature, and the exponent depends on temperature as
\begin{equation}
\gamma = \begin{cases}
  0 & \hbox{for $T \leq T_C$},\\
  \gamma_{\infty}, & \hbox{for $T \geq T_C$}.
	\end{cases}
	\label{eq:gamma_exponent}        
\end{equation} 
The exponent $\gamma_{\infty}$ has been obtained with various
techniques, as for example from lattice computations or by using the
dilute instanton gas approximation~\cite{Gross:53.43, Turner:1986,
  Bae:2008ue}, or relying on the interacting instanton liquid
model~\cite{Wantz:2009mi}. In the QCD axion theory, the mass $m_0$
depends on the axion energy scale $f_A$, which represents the energy
at which the Peccei-Quinn U(1) symmetry breaks, through
Eq.~\eqref{eq:m0}, $m_0 f_A = \Lambda^2$. Here, we choose to present
results in terms of the axion mass $m_0$, in place of the energy scale
$f_A$, and we assume the values $T_C = 140{\rm ~MeV}$ and
$\gamma_{\infty} = 4$, in line with the numerical results in
Ref.~\cite{Borsanyi:2016ksw}.

\subsection{Production of cold axions in the early universe}

Cold axions are produced in the early universe through various
mechanisms, which include the vacuum realignment mechanism (VRM), the
decay of topological defects, or the decay from a parent particle. In
the VRM, the present energy density stored in the axion field
$\phi(x)$ from the coherent oscillations of the field can be obtained
by solving the equation of motion in an expanding universe
\begin{equation}
	\ddot \phi + 3H\dot \phi + \frac{dV(\phi)}{d\phi} = 0,
	\label{eq:equation_motion}
\end{equation} 
where a dot means a derivative with respect to cosmic time $t$. The
axion potential in Eq.~\eqref{eq:equation_motion} has been computed in
the next-to-leading order as~\cite{DiVecchia:1980yfw,
  diCortona:2015ldu}
\begin{equation}
	V(\phi) = \frac{\Lambda^4}{c_z} \left(1-\sqrt{1-4c_z\sin^2(\phi/2f_A)}\right),
	\label{Vqcd}
\end{equation}
where we introduced $c_z \approx z/(1+z)^2 \approx 0.22$.

The axion potential becomes a relevant term in the equation of motion
at the temperature $T_{\rm osc}$ when the universe has sufficiently
cooled so that the axion mass is of the same order as the Hubble rate
\begin{equation}
	m_A(T_{\rm osc}) \approx 3H(T_{\rm osc}),
	\label{eq:condition_oscillations}
\end{equation}
where $H(T)$ is the Hubble expansion rate for the temperature of the
plasma $T$. We assume that the coherent oscillations in the axion
field take place in a radiation-dominated universe, for which the
expansion rate is given by
\begin{equation}
	H(T) = \sqrt{\frac{4\pi^3}{45}\,g_*(T)}\,\frac{T^2}{\MP},
	\label{eq:HubbleRateStandard}
\end{equation}
where $\MP \approx 1.221 \times 10^{19}\,$GeV is the Planck mass and
$g_*(T)$ is the number of relativistic degrees of freedom at
temperature $T$, corresponding to the number of relativistic species
in thermal equilibrium at $T$ weighted with a statistical factor (1
for bosons, 7/8 for fermions). The initial conditions for solving the
equation of motion are the initial value $\theta_i$ of the
misalignment of the axion angular variable $\theta = \phi(x)/f_A$,
which is drawn randomly from a uniform distribution in the interval
$\[-\pi, \pi\]$,
and a vanishing initial value for the velocity $\dot \phi(x)$.

Inserting the expressions in Eqs.~\eqref{eq:QCDaxion_mass}
and~\eqref{eq:HubbleRateStandard} into
Eq.~\eqref{eq:condition_oscillations}, we obtain the value
\begin{equation}
T_{\rm osc} =
T_C\,\[\sqrt{\frac{5}{4\pi^3\,g_*(T_{\rm osc})}}
\frac{m_0 \MP}{T_C^2}\]^{\frac{1}{2+\gamma}}
\approx 700{\rm \, MeV}\,\(\frac{m_0}{1{\rm \, \mu eV}}\)^{1/6},
	\label{eq:oscillation_onset}
\end{equation}
where in the numerical estimate we have set $\gamma = 4$ and
$g_*(T_{\rm osc}) = 61.75$, which is the number of relativistic
degrees of freedom in the Standard Model at a temperature of order
GeV. The number density of axions at the onset of oscillations is then
\begin{equation}
	n_A(T_{\rm osc}) = \frac{b}{2}m_A(T_{\rm osc})\,f_A^2\,\langle\theta_i^2\rangle,
\end{equation}
where $b$ is a factor of order one~\cite{Turner:1986} that captures
the uncertainties derived from using the approximation in
Eq.~\eqref{eq:condition_oscillations} instead of numerically solving
Eq.~\eqref{eq:equation_motion}, and $\langle\theta_i^2\rangle$ is the
average of the square of the initial value of the initial misalignment
angle $\theta_i$ over the observable universe (see the discussion in
Sec.~\ref{sec:cosmology} below).

The present number density of axions is obtained by assuming that
their number in a comoving volume is constant, 
\begin{equation}
	n_A(T_0) = n_A(T_{\rm osc})\,\(\frac{a(T_{\rm osc})}{a(T_0)}\)^3,
	\label{eq:numberdensity0}
\end{equation}
where $T_0$ is the present CMB temperature. The ratio of the two
comoving volumes at temperatures $T_{\rm osc}$ and $T_0$ is given in the
standard cosmological model by assuming entropy conservation 
in the same comoving volume, that is
\begin{equation}
	g_S(T_{\rm osc})\,a^3(T_{\rm osc})\,T_{\rm osc}^3 = g_S(T_0)\,a^3(T_0)\,T_0^3,
	\label{eq:entropy_conservation}
\end{equation}
where 
$g_S(T) T^3$ is the entropy density, with  
$g_S(T)$  the number of entropy degrees of freedom at
temperature $T$~\cite{Kolb:1990vq}. The present energy density of
axions $\rho_A^{\rm mis} = m_0 n_A(T_0)$ obtained from the VRM results in 
\begin{eqnarray} 
\rho_A^{\rm mis} &=&
\frac{b}{2}\,m_0^2\,f_A^2\,\left(\frac{T_{\rm
      osc}}{T_C}\right)^{-\gamma}\,\(\frac{a(T_{\rm
    osc})}{a(T_0)}\)^3\,\langle\theta_i^2\rangle
= \frac{b}{2}\,\Lambda^4\,\frac{g_S(T_0)}{g_S(T_{\rm
    osc})}\,\(\frac{T_0}{T_C}\)^3\,
\left(\frac{T_C}{T_{\rm osc}}\right)^{3+\gamma}\,\langle\theta_i^2\rangle  \nonumber\\
&=& b
\,\bar{\rho}\,\(\frac{m_0}{\MP}\)^{-\frac{3+\gamma}{2+\gamma}}\,\langle\theta_i^2\rangle
= 6.1\times 10^{-47}{\rm \, GeV^4}\,b\,
\(\frac{m_0}{1{\rm \, \mu eV}}\)^{-7/6} \,\langle\theta_i^2\rangle.
	\label{eq:misalignment}
\end{eqnarray}
In the last expression, we have introduced the quantity
\begin{equation}
\bar{\rho} =
\frac{\Lambda^4}{2}\,\frac{g_S(T_0)}{g_S(T_{\rm osc})}\,
\(\frac{T_0}{T_C}\)^3\,\[\sqrt{\frac{5}{4\pi^3\,g_*(T_{\rm osc})}}\frac{\MP^2}{T_C^2}\]^{-\frac{3+\gamma}{2+\gamma}}
= 1.05 \times 10^{-50}{\rm \,eV^4}, 
\end{equation}
where we have set $g_S(T_0) = 3.91$,
$g_S(T_{\rm osc}) = g_*(T_{\rm osc}) = 61.75$, and
$T_0 = 236 {\rm \,\mu eV}$. This computation only relies on the
contribution to the energy density in axions from the energy density
stored in the coherent oscillations of the field. Additional
contributions come from the low-energy spectra of axions radiated from
the decay of various topological defects, like axionic strings and
domain walls~\cite{Vilenkin:1982ks, Harari:1987ht, Vilenkin:2000jqa},
as described in Sec.~\ref{sec:cosmology} below.  We account for this
contribution by introducing a factor $\alpha_{\rm td}$ multiplying the
energy density from the VRM, so that the total axion energy density
today is written as
$\rho_A^{\rm tot} = \alpha_{\rm td}\rho_A^{\rm mis}$. The requirement
that axions account for the full amount of dark matter then reads
\begin{equation}
	\rho_A^{\rm tot} = \rho_{\rm CDM} = \Omega_{\rm CDM}\frac{3H_0^2 \MP^2}{8\pi} \approx 9.71 \times 10^{-48}{\rm \,GeV^4}.
	\label{eq:CDM_axion_mass}
\end{equation}
The numerical estimate corresponds to the value of the dark matter
energy density in units of the present critical energy density
$\Omega_{\rm CDM}h^2 \approx 0.12$. The contribution from both
topological defects and from the VRM to compute the energy density in
cold axions $\rho_A^{\rm tot}$ has been the subject of various
numerical simulations~\cite{Battye:1993jv, Battye:1994au,
  Yamaguchi:1998gx, Yamaguchi:1999dy, Hiramatsu:2010yn,
  Hiramatsu:2012sc, Hiramatsu:2012gg, Klaer:2017qhr, Klaer:2017ond,
  Gorghetto:2018myk, Vaquero:2018tib}.

\subsection{Axion cosmology and the inferred dark matter axion mass} \label{sec:cosmology}

The actual value of $\langle\theta_i^2\rangle$ strongly depends on the
cosmological history of the axion. We first define two conditions:
\vspace{4pt}
\begin{itemize} 
\item [{\bf a})] The Peccei-Quinn symmetry is spontaneously broken during inflation;
\item[{\bf b})] The Peccei-Quinn symmetry is never restored after its spontaneous breaking occurs.
\end{itemize}
\vspace{4pt} 
Condition {\bf a}) is realised whenever the Peccei-Quinn
scale is larger than the Hubble rate at the end of inflation,
$f_A > H_I$, while condition {\bf b}) is realised whenever the
Peccei-Quinn scale is larger than the maximum temperature of the
post-inflationary universe $T_{\rm MAX}$, which is generally higher
than the reheating temperature $\TRH$~\cite{Giudice:2000ex}. Broadly
speaking, one of these two possible scenarios occurs:
\begin{description}

\item[Scenario I] If both {\bf a}) and {\bf b}) are satisfied,
  inflation selects one patch of the universe within which the
  spontaneous breaking of the Peccei-Quinn symmetry leads to a
  homogeneous value of the initial misalignment angle $\theta_i$. This
  is the so-called {\em pre-inflationary scenario}, in which the
  distribution of $\theta_i$ has the second moment
  $\langle\theta_i^2\rangle = \theta_i^2 + \sigma_\theta^2$, where
  $\sigma_\theta$ is the standard deviation of the distribution of the
  angle centered around $\theta_i$. In this scenario, topological
  defects are inflated away and do not contribute to the axion energy
  density. Namely we can set $\alpha_{\rm td} =1$ and
  $\rho_A^{\rm tot} = \rho_A^{\rm mis}$. Inserting these conditions
  into Eqs.~\eqref{eq:misalignment} and~\eqref{eq:CDM_axion_mass}
  gives
  \begin{equation}
  m_0 =
  \MP\,\[b\frac{\bar{\rho}}{\rho_{\rm
      CDM}}\,\(\theta_i^2 +
  \sigma_\theta^2\)\]^{\frac{2+\gamma}{3+\gamma}}
  \approx 5{\rm \,\mu eV}\,\theta_i^{12/7}.
	\label{eq:CDM_axion_mass_ScenarioI}
\end{equation}
For the numerical estimate, we have set $b = 1$ and $\gamma = 4$, and
we have assumed that the standard deviation of the distribution in the
initial misalignment angle is much smaller than the value realised,
which is justified whenever $H_I \ll \theta_i\,f_A$.  As we will
detail in the next section, within this scenario, KLASH would be able
to test the QCD axion for values of the initial misalignment angle
within the range $|\theta_i| \in \(0.2 - 0.4\)$
which, although it covers only a few percent of the full range, it is
outside the reach of all other axion search experiments proposed so
far.
\begin{figure}[!ht]
\begin{center}
	\includegraphics[width=0.85\linewidth]{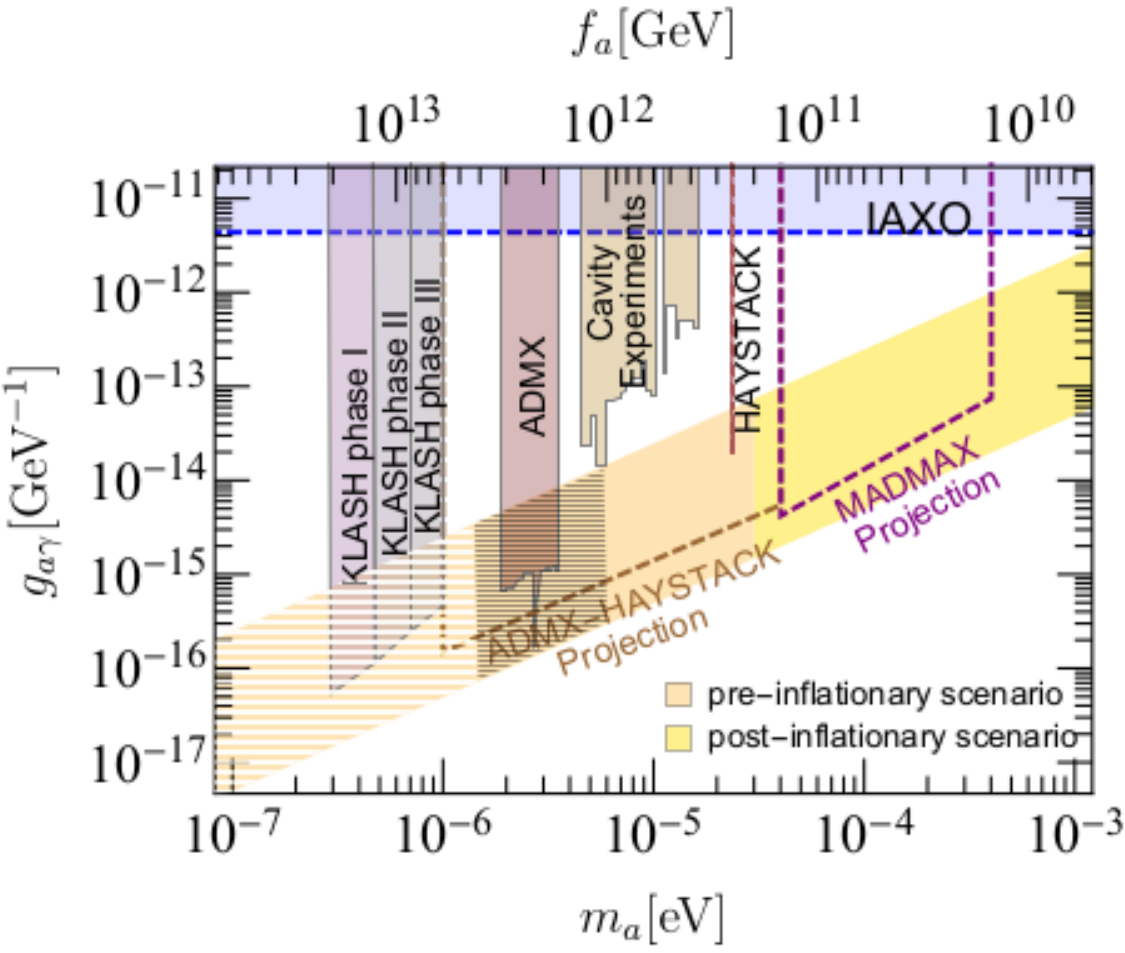}
	\caption{Sensitivities of various haloscope experiments. The
          region hatched in black represents the canonical dark matter
          axion window in post-inflationary scenarios. The region
          hatched in white represents the additional mass range for
          axion dark matter which becomes viable in the various
          theoretical scenarios discussed in the text.  }
	\label{fig:Haloscopes}
\end{center}
\end{figure}
\item[Scenario II] If at least one of the conditions {\bf a}) or {\bf
    b}) is violated, the Peccei-Quinn symmetry breaks with different
  values of $\theta_i$ taking different values in patches that are
  initially out of causal contact, but that today populate the volume
  enclosed by our Hubble horizon. In this scenario, which is
  generally referred to as {\em post-inflationary scenario},
  $\langle\theta_i^2\rangle$ is the average of the initial misalignment angle squared
  over the circle, assuming that $\theta_i$ is drawn from a uniform
  distribution. For a quadratic potential one has
  $\langle\theta_i^2\rangle= \pi^2/3$, however, in the periodic
  potential that defines the QCD axion, this result is modified due to
  the presence of non-harmonic terms~\cite{Turner:1986}. In
  particular, for a cosine potential we have
  $\langle\theta_i^2\rangle \approx (2.15)^2$~\cite{diCortona:2015ldu}.  We
  will use this value in our computations.

  In this scenario, topological defects form after inflation and
  remain within the horizon. Hence they contribute to the present
  energy density ($\alpha_{\rm td}\neq 1$). The detailed mechanism of
  cold axion production from topological defects has yet to be
  assessed with sufficient confidence and precision, and an agreement
  is yet to be achieved in the community. Available estimates differ
  by as much as one order of magnitude, spanning a range
  $\alpha_{\rm td} = \(0.5 - 10\)$
  (see Eq.~\eqref{eq:misalignment}). Computing the mass of the dark
  matter axion from Eq.~\eqref{eq:CDM_axion_mass} in this scenario
  gives
\begin{equation}
	m_0 = \MP \,\(b \,\frac{\alpha_{\rm td}\,\bar{\rho}}{\rho_{\rm
            CDM}}\,\langle\theta_i^2\rangle\)^{\frac{2+\gamma}{3+\gamma}}
        \approx \(5 - 50\){\rm \, \mu eV}, 
	\label{eq:CDM_axion_mass_ScenarioII}
\end{equation}
where, as in the previous scenario, we have taken $b=1$ and
$\gamma = 4$, and we have set
$\langle\theta_i^2\rangle \approx (2.15)^2$ and $\alpha_{\rm td}$
spanning the range given above. Therefore, as is depicted in
Fig.~\ref{fig:Haloscopes}, in this scenario the possible values of the
axion mass hinted by cosmological considerations remain well above the
mass region accessible to KLASH.

\end{description}

\section{Dark matter axion mass in the range of KLASH}
\label{section:III}

We have seen in the previous section that assuming a standard
cosmological evolution, only in the case of pre-inflationary scenarios
are canonical axion models within the reach of KLASH sensitivity, and
the region of the initial misalignment angle for which this happens is
rather restricted.  At the root of this conclusion lies the axion
equation of motion Eq.~\eqref{eq:equation_motion} which depends on
three crucial ingredients: (I)~the initial conditions for the axion
amplitude $a_i = \theta_i f_A$, see Sec.~\ref{sec:Initial misalignment
  angle}; (II)~the cosmological evolution of the Hubble parameter
$H(T)$ that, for a standard cosmological model, is given in
Eq.~\eqref{eq:HubbleRateStandard}, see Sec.~\ref{sec:Modified
  cosmological evolution}; (III)~the relation between the
temperature-dependent axion mass in Eq.~\eqref{eq:QCDaxion_mass} and
the scale of the initial amplitude, $m_A(T) \sim 1/f_A$, as described
in Sec.~\ref{sec:Extended QCD axion models}.  Namely, besides
intervening on the initial conditions as in Sec.~\ref{sec:Initial
  misalignment angle}, predictions would be different for a universe
with a non-standard cosmological evolution, Sec.~\ref{sec:Modified
  cosmological evolution}.  This would happen, for example, if the
particle content around the era of the onset of oscillations is
different from what it is generally assumed, see
Sec.~\ref{sec:entropygeneration}, or more drastically if during the
same epoch the cosmological evolution is controlled by some theory
which departs from general relativity (e.g. a scalar-tensor theory),
see Sec.~\ref{sec:non-standard thermal history}.  Finally, we envisage
a third possibility~(III) of particle theory models in which the
relation between the axion mass and the initial amplitude of the
oscillations $a_i \sim f_A$ is modified, see Sec.~\ref{sec:Extended
  QCD axion models}.  This can follow from a different dependence of
the axion mass on the temperature, i.e. from modifications of
Eq.~\eqref{eq:QCDaxion_mass}, without substantially altering the zero
temperature expression for $m_0$ in Eq.~\eqref{eq:m0}, or {\it ab
  initio} if the zero temperature $m_0$-$f_A$ relation differs from
the one given in Eq.~\eqref{eq:m0}. Any difference of these kinds
would be eventually reflected in a modification of the mass-energy
density relation.  All these possibilities represent motivations to
search for the QCD axion in non-canonical mass ranges as the one in
the reach of KLASH. We will describe them in more detail in the
following.

\subsection{I. \   Initial misalignment angle} \label{sec:Initial misalignment angle}

When the relation Eq.~\eqref{eq:m0} between the axion mass and the
axion decay constant $f_A$ holds and around the time of the onset
of axion oscillations the universe expansion rate is standard, as
described by Eq.~\eqref{eq:HubbleRateStandard} with
$g_* (T_{\rm osc}) = 61.75$, then the relation between the axion mass, the
initial misalignment angle, and the present axion energy density is
given by Eq.~\eqref{eq:CDM_axion_mass_ScenarioI}.
We have seen that in this case the KLASH range of sensitivity 
corresponds to the range $|\theta_i| \in \(0.2 - 0.4\)$. For a uniform
probability of $\theta_i$ within $\[-\pi,\pi\]$  
the probability that $\theta_i$ is drawn with the desired value is
then about 6\%. 

The actual computation of the probability derived
from the Fokker-Planck equation has been given in
Refs.~\cite{Graham:2018jyp, Guth:2018hsa, Tenkanen} and gives a probability 
somewhat lower than this. 
% of the order of 5\%.
Recall that this scenario is realised when both $H_I < f_A$ and
$\TRH < f_A$~\cite{Dine:1982ah}. As an example, for an axion mass 
$m_0 = 0.5{\rm \,\mu eV}$, the value of the axion decay constant is
$f_A = 1.1\times 10^{13}{\rm GeV}$.

Recently, the work in Ref.~\cite{Hoof:2018ieb} presented a global
fit that uses a Bayesian analysis technique to explore the parameter
space of the QCD axion (both the KSVZ and the DFSZ models), based on
the code GAMBIT~\cite{Athron:2017ard} and its module
DarkBit~\cite{Workgroup:2017lvb}. In particular,
Ref.~\cite{Hoof:2018ieb} considers the scenario in which the
Peccei-Quinn symmetry breaks during a period of inflation, while
taking into account results from various observations and experiments
in the likelihood including the light-shining-through-wall
experiments, helioscopes, cavity searches, distortions of gamma-ray
spectra, supernovae, horizontal branch stars and the hint from the
cooling of white dwarfs. The marginalised posterior distribution
obtained in Ref.~\cite{Athron:2017ard} when demanding that the
totality of dark matter is in axions gives the range
$0.12{\rm \,\mu eV} \leq m_0 \leq 0.15{\rm \,meV}$ at the 95\%
equal-tailed confidence interval. We stress that a portion of the
range inferred by this analysis is well within the reach of of the
KLASH experiment.

Small initial values of $\theta_i$ might also occur naturally,
i.e. without any fine tuning, in low-scale inflation models in which
inflation lasts sufficiently long~\cite{Graham:2018jyp,
  Guth:2018hsa}. If $H_I \lesssim \Lambda_{\rm QCD}$ the axion
acquires a mass already during inflation, the $\theta_i$-distribution
flows towards the CP conserving minimum and, for long durations of
inflation, stabilises around sufficiently small $\theta_i$ values.  As
a result the QCD axion can naturally give the dark matter abundance
for axion masses well below the classical window, down to
$10^{-12}\,$eV~\cite{Graham:2018jyp}.

\subsection{II.\  Modified  cosmological evolution} \label{sec:Modified  cosmological evolution}

\subsubsection{A.\ Entropy generation} \label{sec:entropygeneration}

If a new species is present in the early universe and if it decays
into thermalised products prior to Big Bang Nucleosynthesis (BBN), the
relation in Eq.~\eqref{eq:entropy_conservation} expressing the
conservation of the entropy density in a comoving volume is
violated. If a relevant amount of entropy is generated after the
axions are produced, for example by the decay of a massive scalar
field, the axion density in Eq.~\eqref{eq:misalignment} would be
diluted by a factor $\Delta$~\cite{Dine:1982ah, Steinhardt:1983ia,
  Lazarides:1987zf, Lazarides:1990xp,Kawasaki:1995vt,
  Visinelli:2009zm, Visinelli:2009kt, Visinelli:2017imh}, which would
in turn lower the value for the dark matter axion mass. Repeating the
computation that leads to Eq.~\eqref{eq:CDM_axion_mass_ScenarioII}
when $\rho_A^{\rm tot} \to \rho_A^{\rm tot} / \Delta$, we find
\begin{equation}
	m_A \approx \frac{\(5 - 50\){\rm \, \mu eV}}{\Delta^{7/6}}.
	\label{eq:axionmass_diluted}
\end{equation}
The range explored by KLASH is reached if the contribution from the
dilution factor is of order $\Delta \approx \(10 - 100\)$.
The value of the quantity $\Delta$ would depend on the details of the
modified cosmological model and, ultimately, on the reheating
temperature. A detailed derivation has been given in
Refs.~\cite{Visinelli:2009kt, Visinelli:2017imh, Visinelli:2018wza,
  Draper:2018tmh, Nelson:2018via, Ramberg:2019dgi, Blinov:2019rhb}.

\subsubsection{B.\ Early non-standard thermal history} \label{sec:non-standard thermal history}

A non-standard cosmological history of the early universe might also
open new pathways to enlarge the allowed QCD axion dark matter mass
range. For example, if around $T\sim 1\,$GeV the universe evolution is
characterised by a period of non-canonical expansion, the onset of
axion oscillations could start earlier (lower number density), or
could be delayed (enhanced number density), opening up the canonical
axion mass window respectively towards smaller or larger masses. This
can occur for example if the evolution of the universe is described by
a scalar-tensor gravity
theory~\cite{jordan1955schwerkraft,Fierz:1956zz,Brans:1961sx} rather
than by general relativity. Scalar-tensor theories benefit from an
attraction mechanism which at late times makes them flow towards
standard general relativity, so that discrepancies with direct
cosmological observations can be avoided. Such a possibility was
already put forth in relation to possible large enhancements of WIMP
dark matter relic density~\cite{Catena:2004ba, Catena:2009tm,
  Meehan:2015cna, Visinelli:2015eka, Dutta:2016htz, Dutta:2017fcn,
  Visinelli:2017qga}, or to lower the scale of leptogenesis down to
the TeV range~\cite{Dutta:2018zkg}, as a consequence of a modified
expansion rate. To our knowledge, no specific studies have been
performed for axion cosmology along this line.

\begin{figure}[h!]
\begin{center}
	\includegraphics[width=0.7\linewidth]{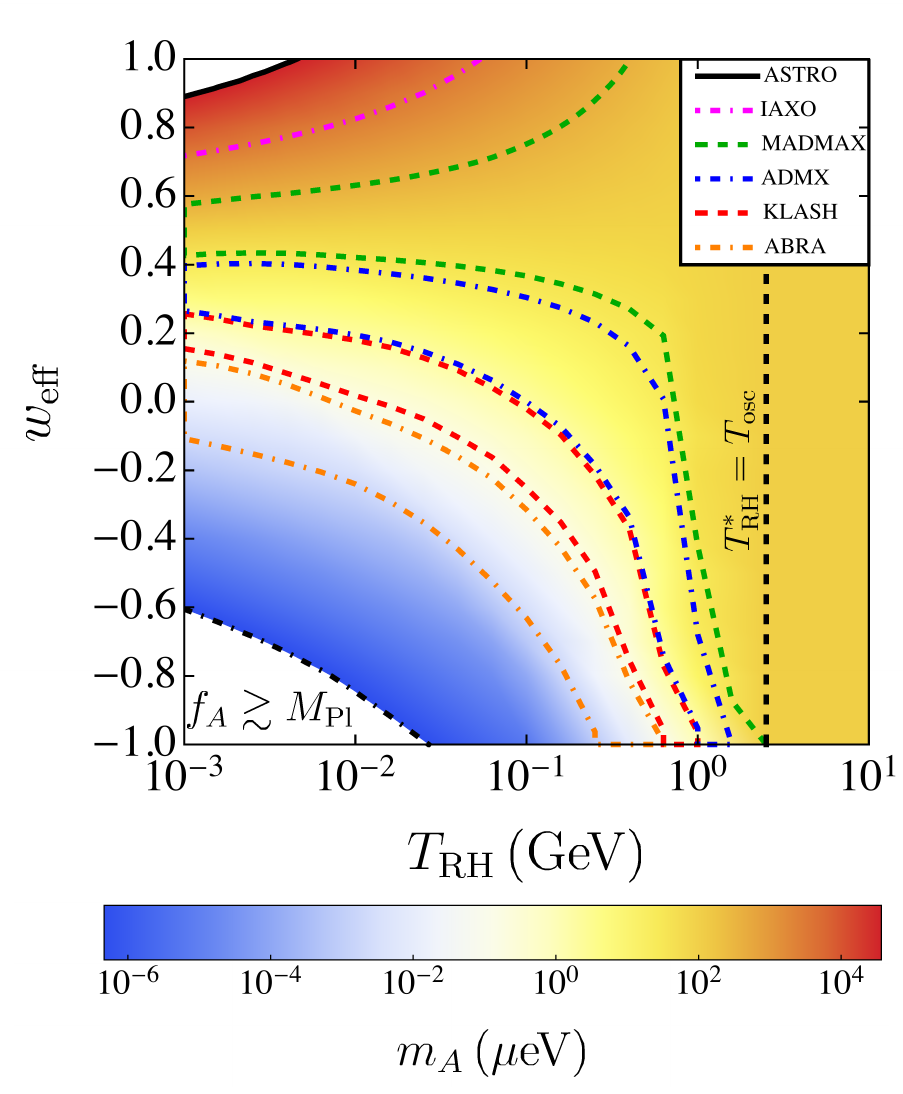}
	\caption{The value of the axion mass (in ${\rm \mu eV})$ that
          yields the observed dark matter abundance for different
          cosmological scenarios. In the vertical axis $w_{\rm eff}$
          gives the effective equation of state, while the reheat
          temperature in given in the horizontal axis. The dashed
          vertical black line marks the reheat temperature $\TRH^*$ at
          which $T_{\rm RH}^* = T_{\rm osc}$. The projected
          sensitivities of ABRACADABRA (``ABRA'', orange dot-dashed
          line), KLASH (red dashed line), ADMX (blue dot-dashed line),
          MADMAX (green dashed line), and IAXO (magenta dot-dashed
          line) are also given. The black dashed line marks the region
          beyond which 
%  the axion decay constant violates the weak gravity conjecture, 
          $f_A \gtrsim \MP$, while the black solid line ``ASTRO''
          gives the bound on $m_A \gtrsim 15{\rm meV}$ excluded by
          astrophysical considerations~\cite{RAFFELT1986402,
            Raffelt2008, Viaux:2013lha, Giannotti:2017hny}.}
	\label{fig:axionmass_nostrings}
\end{center}
\end{figure}

Recently, the authors in Ref.~\cite{Ramberg:2019dgi} have considered a
scenario in which the early universe is characterised by a modified
expansion rate, since the energy content during that period is
dominated by some substance $\psi$ whose equation of state differs
from that of radiation, $w_{\rm eff} \neq 1/3$. This period lasts
until the new substance completely decays into radiation at
temperatures higher than a given reheat temperature
$\TRH \gtrsim 5\,$MeV, with the lower bound on the parameter coming
from requiring that successful BBN is not
spoiled. Fig.~\ref{fig:axionmass_nostrings} shows the value of the
axion mass (in ${\rm \mu eV}$) that is required to saturate the
totality of CDM for given reheating temperature and effective equation
of state of the additional energy component, as a function of the new
parameters $w_{\rm eff}$ and $\TRH$. The value of the misalignment
angle is fixed to $\theta_i = \pi/\sqrt{3}$ and it is assumed that
there is no contribution from topological defects:
$\alpha_{\rm td} = 1$. For each choice of the parameters, the
corresponding value of $m_A$ is the smallest value of the axion mass
attained in the theory, since smaller values would yield an axion
energy density larger than what is observed in dark matter. Higher
values of the axion mass are possible, although the corresponding
energy density would not saturate the totality of the dark matter. The
dashed vertical black line marks the reheating temperature
$T_{\rm RH}^*$ for which $T_{\rm RH}^* = T_{\rm osc}$, which can be
approximated as the solution to the expression
$m_A(T_{\rm RH}^*) = 3H(T_{\rm RH}^*)$. The projected sensitivity to
axion masses from several experiments are shown in
Fig.~\ref{fig:axionmass_nostrings}, including the forecast reach of
KLASH over the mass range $\(0.3 - 1.0\){\rm
  \,\mu eV}$.
We have also demanded that $f_A \lesssim \MP$ (although this
requirement does not take into account the ``axion quality''
problem). Also of interest is the region of the parameter space
already excluded by the non-observation of anomalous cooling of
astrophysical objects due to additional axionic channels labeled as
``ASTRO''~\cite{RAFFELT1986402, Raffelt2008, Viaux:2013lha,
  Giannotti:2017hny}.

\subsection{III.\  Extended QCD axion models} \label{sec:Extended QCD axion models}
\label{mirror}

We have seen that the expected axion abundance can be modified in
cosmological theories that predict a different evolution of the Hubble
parameter during the early universe. On the other hand, it is possible
that the relation between the temperature and the mass of the axion in
Eq.~\eqref{eq:QCDaxion_mass} is modified by the addition of particle
physics content, while the cosmological evolution is left essentially
unchanged. These models have the advantage of insuring the
preservation of well-tested cosmological predictions, in particular
BBN.

In general, suppose that a particular model predicts a non-standard
mass-temperature relation that allows the condition
$ m_A(T_{\rm osc}') \approx 3H(T_{\rm osc}')$, see
Eq.~\eqref{eq:condition_oscillations}, to be satisfied at a higher
temperature $T_{\rm osc}' > T_{\rm osc}$ so that the abundance of
cosmic axions would be different from what expected in the standard
scenario. Since the axion number density dilutes during the expansion
of the universe as
\begin{equation}
	n_A^{\rm mis}(T_0) \propto \frac{H(T_{\rm osc}')}{(T_{\rm osc}')^{3}} \propto \frac{1}{T_{\rm osc}'},
	\label{eq:numberdensitydilution}
\end{equation}
in this type of scenario we expect
$n_A^{\rm mis}(T_{\rm osc}') < n_A^{\rm mis}(T_{\rm osc})$ for a
higher temperature $T_{\rm osc}' > T_{\rm osc}$. The global effect is
that of lowering the expected axion energy density in the present
universe, so that the condition $\Omega_A \leq \Omega_{\rm CDM} $
leads to possible smaller values of the axion mass. We now discuss
some explicit models that employ such a mechanism.

\subsubsection{A.\ Axions in mirror world scenarios}

The cosmological scenario predicted by the theory of the
\textit{mirror world}~\cite{Berezhiani:2000gw}, extended to include
the axion~\cite{Giannotti:2005eb}, allows for a change in the
axion-temperature relation while leaving the standard cosmological
predictions unchanged.  Ultimately, this leads to a reduction of the
expected axion abundance for a fixed axion mass.

The mirror world idea is very
old~\cite{Lee:1956qn,Holdom:1985ag,Glashow:1985ud,Khlopov:1989fj} and
based on the assumption that the gauge group is the product of two
identical groups, $ G\times G^{\prime} $.  In the simplest possible
model, $ G $ is the standard model gauge group and $ G^{\prime} $ an
identical copy of it.  Standard particles are singlets of
$ G^{\prime} $ and mirror particles are singlets of $ G $.  This
implies the existence of mirror particles, identical to ours and
interacting with our sector only through gravity.  Since the
gravitational interaction is very weak, mirror particles are not
expected to thermalise with ordinary particles and so there is not
need to expect that the two universe have the same temperature.  In
fact, cosmological observations require a lower mirror temperature to
reduce the radiation energy density at the time of the
BBN~\cite{Berezhiani:2000gw}.  This energy density is traditionally
parameterised with the effective number of extra neutrino species,
$ \Delta N_{\rm eff}$.  The most recent combined analysis of Planck of
the Cosmic Microwave Background (CMB) and observations of the Baryon
Acoustic Oscillations (BAO) give
$ N_{\rm eff} =2.99\pm 0.17 $~\cite{Aghanim:2018eyx}, while the
standard model value is
$ N_{\rm eff, SM} =3.046$~\cite{PhysRevD.46.3372}.
Denoting by $ x $ the ratio $ T^{\prime}/T $ of the mirror and standard temperature,
one finds $ \Delta N_{\rm eff} =6.14 \, x^4$~\cite{Berezhiani:2000gw}.
Therefore, the current bounds on dark radiation can be accommodated requiring $ x\lesssim 0.4 $.

% \subsubsection*{Axions in mirror world scenarios}

The possibility to implement the PQ mechanism in the mirror world
scenario was proposed in
Refs.~\cite{Rubakov:1997vp,Berezhiani:2000gh,Gianfagna:2004je}.  The
general feature is that the total Lagrangian must be of the form
$\mL+\mL^\prime+\lambda \mL_{\rm int}$, where $\mL$ represents the
ordinary Lagrangian, $\mL^\prime$ is the Lagrangian describing the
mirror world content, and $\mL_{\rm int}$ is an interaction term with
a coupling $\lambda$ which is taken to be small enough to insure that
the two sectors do not reach thermal equilibrium.  A simple
realisation of the mechanism restricts the interaction to the Higgs
sector. Ordinary and mirror world have each two Higgses, which
interact with each other. The axion emerges as a combination of their
phases in a generalisation of the Weinberg-Wilczek mechanism.
For $\lambda=0$, the total Lagrangian contains two identical
$U(1)_{\rm axial}$ symmetries, while the $\mL_{\rm int}$ term breaks
these in just the usual $U(1)_{\rm PQ}$, so that only one axion field
results.

\begin{figure}[!ht]
\begin{center}
	\includegraphics[width=0.75\linewidth]{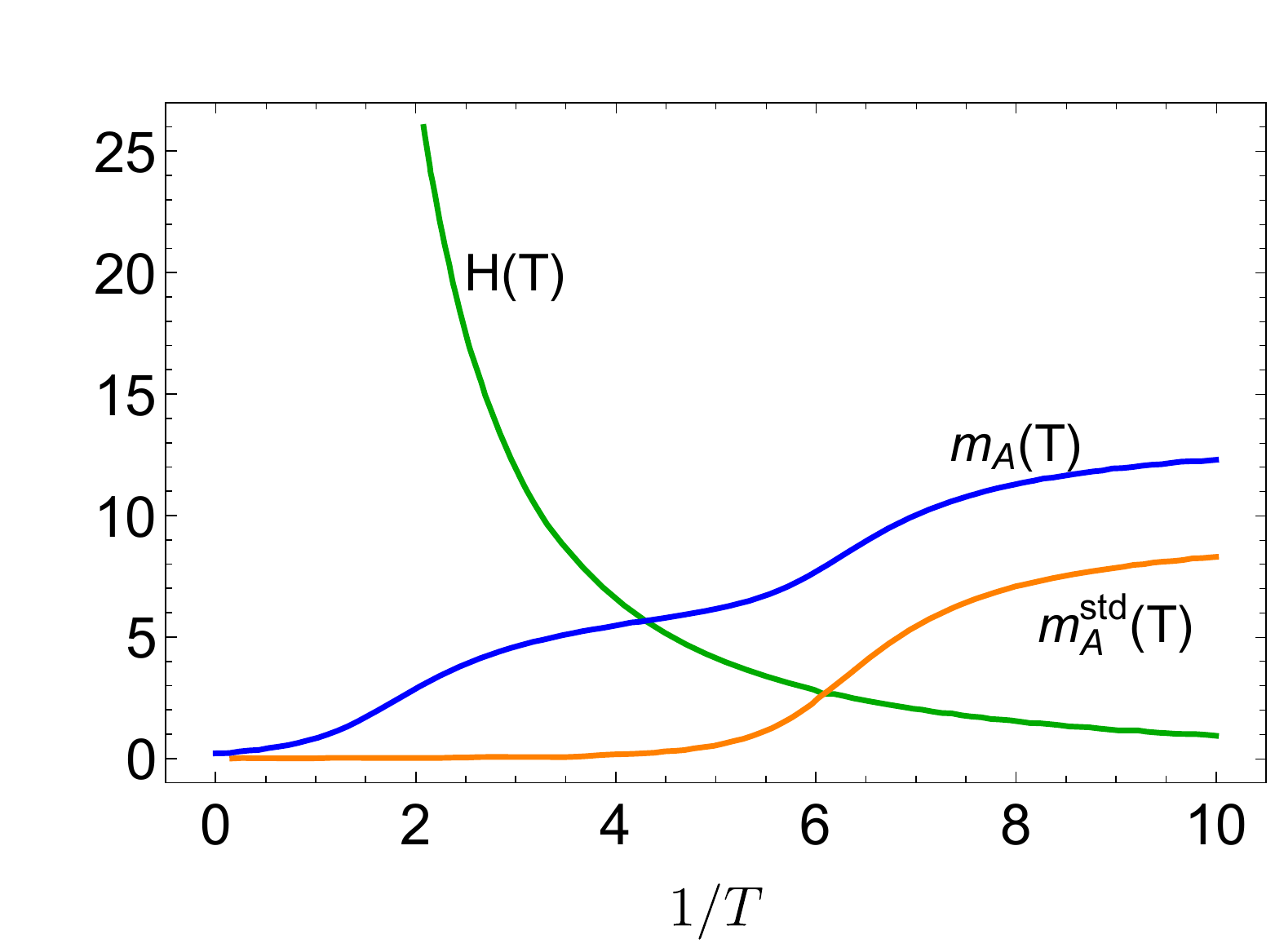}
	\caption{ Dependence of the axion mass on the temperature
          (in arbitrary units) in the standard case $m_A(T) $
          and in mirror world models $m_{A, m}(T)$. The second rise in
          $m_{A, m}(T)$ around $T=T_{\rm osc}$ is due to the standard QCD
          contribution to the axion mass.  The green curve labeled
          $H(T)$ represents the evolution of the Hubble parameter. }
	\label{fig:MassTemperature}
\end{center}
\end{figure}

As long as the mirror-parity is an exact symmetry, the particle
physics is exactly the same in the two worlds, and so the strong CP
problem is simultaneously solved in both sectors.  In particular, the
axion couples to both sectors in the same way and their
non-perturbative QCD dynamics produces the same contribution to the
axion effective potential.  Hence, defining the mass of the axion with
the mirror world contributions as $m_{A,m}(T)$, we have at zero
temperature $m_{0, m} \equiv m_{A,m}(0) = \sqrt{2}\, m_0$, with $m_0$
given in Eq.~\eqref{eq:m0}. This is depicted in
Fig.~\ref{fig:MassTemperature} by the asymptotic value of
$m_{A, m}(T)$ for $T\to 0$.
However, at temperature $T \sim 1\,$GeV the axion mass could be
considerably larger than its standard value. Assuming that the
confinement temperature $T_C$ is the same between both the mirror
world and the visible sector, and neglecting a possible dependence of
the exponent $\gamma$ of the topological susceptibility on the
temperature, $\gamma(T)\approx \gamma(xT) = \gamma \approx 4$, the
expression in Eq.~\eqref{eq:QCDaxion_mass} for $T \gtrsim T_C$ gives
\begin{align}
\label{Eq:maMirror}
m_{A, m}^2(T)=
m_0^{2}\left[ \left( \dfrac{T_C}{T}\right)^{2\gamma} 
+ \left( \dfrac{T_C}{x\,T}\right)^{2\gamma} \right] = m_A^{2}(T)\( 1 + \frac{1}{x^{2\gamma}} \)\,.
\end{align}
Given the number density in Eq.~\eqref{eq:numberdensitydilution}, we
expect the present energy density in cold axions in the mirror world
model to be
\begin{equation}
\rho_{A, m}(T_{\rm osc,m}) = \frac{\sqrt{2} T_{\rm osc}}{T_{\rm
    osc,m}}\rho_{A}(T_{\rm osc}) = \frac{\rho_{A}(T_{\rm
    osc})}{\Delta}, 
\end{equation}
where we have defined the quantity
$\Delta = T_{\rm osc,m}/ \sqrt{2} T_{\rm osc}$ that acts as an
effective dilution factor, with $ T_{\rm osc,m}$ the temperature in
the visible sector at which the axion field starts oscillating in the
presence of the mirror world contribution, while $T_{\rm osc}$ is the
oscillation temperature in the standard case.  Given the results in
Eq.~\eqref{eq:oscillation_onset}, we have
$T_{\rm osc,m} = T_{\rm osc}\( 1 + x^{-2\gamma} \)^{1/2(2+\gamma)}$,
so that
\begin{equation} 
\Delta =\frac{1}{\sqrt{2}}\( 1 + \frac{1}{x^{2\gamma}}
\)^{\frac{1}{2(2+\gamma)}} \approx
\frac{1}{\sqrt{2}}\(\frac{1}{x}\)^{\frac{\gamma}{2+\gamma}}\,,
\end{equation} 
where in the last step we have taken the limit $x \ll 1$. As discussed
in Sec.~\ref{sec:entropygeneration}, a dilution factor of the order of
$\Delta \approx \(10 - 100\)$
sets the energy density in the correct range for the search of KLASH
with an initial misalignment angle of order unity. For the mirror
world model, in order that the value of the axion mass which accounts
for the totality of the CDM will fall in the range accessible to
KLASH, this translates into a ratio between the temperatures of the
two sectors in the range $x \approx \(5 - 200\)\times 10^{-4}$.

\subsubsection*{B.\ Dark matter axion in models with nonstandard mass-axion
  decay constant relation}
\label{massdecay}

As we have seen above, adding a mirror-QCD sector has the effect of
increasing the zero temperature axion mass by a factor $\sqrt{2}$ and,
along this line, it is not difficult to imagine how the axion could be
made heavier than expected~\cite{Berezhiani:2000gh, Gianfagna:2004je}.
It is instead difficult to make the axion {\it lighter} than expected.
A particle physics model that realises this scenario has been proposed
in \cite{Hook:2018jle}. The model relies on a $Z_N$ symmetry under
which $a \to a + \frac{2 \pi f_A}{N}$, and furthermore the axion
interacts with $N$ copies of QCD whose fermions transform under $Z_N$
as $\psi_k \to \psi_{k+1}$.  Surprisingly, adding up the contributions
of all the sectors one finds that cancelations occur in the axion
potential with a high degree of accuracy, and as a result, for even
$N$ the axion mass gets exponentially suppressed:
\begin{align}
  \label{eq:ZN}
m_0 (N) = \frac {4 \, m_0^{\rm std}}{2^{N/2}} \,, 
\end{align}
while, if $N$ is odd, the axion potential retains the minimum in
$\bar\theta =0$. The mass range accessible to KLASH corresponds to 
$9 \lesssim N \lesssim 13$. 

%\subsubsection{C.\ Axion abundance suppressed by hidden monopoles}
%\label{monopoles} 

%In the presence of a CP violating $\theta$-term, monopoles acquire a
%non-zero electric charge and become dyons due to the Witten
%effect~\cite{Witten:1979ey}.  When a dynamical axion field replaces
%$\theta$, its potential receives additional contributions from
%interactions with the monopoles~\cite{Fischler:1983sc} and the field
%oscillations begin much before the epoch of the QCD phase transition.
%This scenario has been implemented exploiting monopoles of a hidden
%$U(1)$ symmetry~\cite{Kawasaki:2015lpf, Kawasaki:2017xwt}.  The final
%result is that the axion abundance turns out to be inversely
%proportional to the abundance of hidden monopoles, and the axion dark
%matter window can thus be extended to a range that lies within the
%reach of KLASH, see Fig.~\ref{fig:Haloscopes}.

 \chapter{Physics Reach}\label{ch2:sec_phys-reach}
The axion is a pseudoscalar particle predicted by S. Weinberg \cite{Weinberg:1977ma} and F.\,Wilczek \cite{Wilczek:1977pj} as a consequence of the
mechanism introduced by R.D. Peccei and H. Quinn \cite{Peccei:1977ur,Peccei:1977hh} to solve the ``strong CP problem''. Axions are also well motivated 
dark-matter (DM) candidates with expected mass lying in a broad range from peV to few meV \cite{PDG2018}. 
As discussed in chapter \ref{ch1:sec_theory}, even if post-inflationary scenarios favours the mass region $(10-10^3)$ $\mu$eV \cite{PDG2018} 
smaller values are also well theoretically motivated. A rich experimental program will probe the axion existence in the next decade. 
Among the experiments, ADMX \cite{ADMX}, HAYSTAC \cite{HAYSTAC}, ORGAN \cite{ORGAN}, CULTASK \cite{CULTASK}, RADES \cite{RADES}, 
and QUAX \cite{QUAX} will use a haloscope, i.e. a detector composed of a resonant cavity immersed in a strong magnetic field as proposed by 
P. Sikivie \cite{Sikivie}. When the resonant frequency of the cavity $\nu_c$ is tuned to the corresponding axion mass $m_a c^2/h$, the expected power 
deposited by DM axions is given by \cite{HAYSTAC}
\begin{equation}
	\label{eq:power}
	P_{a} = \left( g_{\gamma}^2\frac{\alpha^2}{\pi^2}\frac{\hbar^3 c^3\rho_a}{\Lambda^4} \right) \times
	\left( \frac{\beta}{1+\beta} \omega_c \frac{1}{\mu_0} B_0^2 V C_{mnl} Q_L \right),
\end{equation}
where $\rho_a=0.45$\,GeV/cm$^3$ is the local DM density, $\alpha$ is the fine-structure constant, $\Lambda = 78$ MeV is a scale parameter related 
to hadronic physics, and $g_{\gamma}$ is a model dependent parameter equal to $-0.97$ $(0.36)$ in the KSVZ (DFSZ) axion model\,\cite{KSVZ,DFSZ}.
It is related to the coupling appearing in the Lagrangian $g_{a\gamma\gamma}=(g_{\gamma}\alpha/\pi\Lambda^2)m_a$. The second parentheses contain 
the vacuum permeability $\mu_0$, the magnetic field strenght $B_0$, the cavity volume $V$, its angular frequency $\omega_c=2\pi\nu_c$, the coupling 
between cavity and receiver $\beta$ and the loaded quality factor $Q_L=Q_0/(1+\beta)$, where $Q_0$ is the unloaded quality factor; here 
$C_{mnl} \simeq O(1)$ is a geometrical factor depending on the cavity mode.

Since $P_{\mbox{sig}}$ can be as low as $10^{-22}$~W, the cavity is cooled to cryogenic temperatures and ultra low noise cryogenic amplifiers are needed 
for the first stage amplification. According to the Dicke radiometer equation \cite{Dicke}, the signal to noise ratio $SNR$ is given by:
\begin{equation}
  \label{eq:snr}
  SNR=\frac{P_{\mbox{sig}}}{k_B T_{sys}}\sqrt{\frac{\tau}{\Delta\nu_a}}
\end{equation}
where $k_\mathrm{B}$ is the Boltzman constant, $T_{sys}$ is the combination of amplifier and thermal noise, $\tau$ is the integration time and $\Delta\nu_a$ the 
intrinsic bandwith of the galactic axion signal ($\Delta\nu_a/\nu_a\simeq10^{-6}$).

\section{Summary of KLASH Haloscope}\label{ch1:sec_sum-halo}
%Geometry

\begin{figure}[htbp]
  \begin{center}
    \includegraphics[totalheight=5cm]{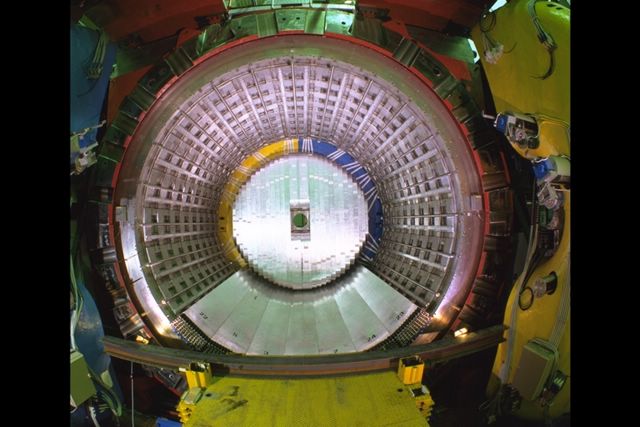}
    \caption{Kloe magnet}
    \label{fig:kloemag}
  \end{center}
\end{figure}

The KLASH haloscope is composed of a large resonant cavity made of copper, inserted in a cryostat cooled down to 4.5\,K. The cryostat is inserted inside 
the KLOE \cite{KLOEreview} magnet \cite{KLOEMAG,MODENA} (Fig. \ref{fig:kloemag}), an iron shielded solenoid coil made from an aluminium-stabilised 
niobium titanium superconductor, providing an homogeneus axial field of 0.6\,T. 

\begin{figure}[htbp]
  \begin{center}
    \includegraphics[totalheight=6cm]{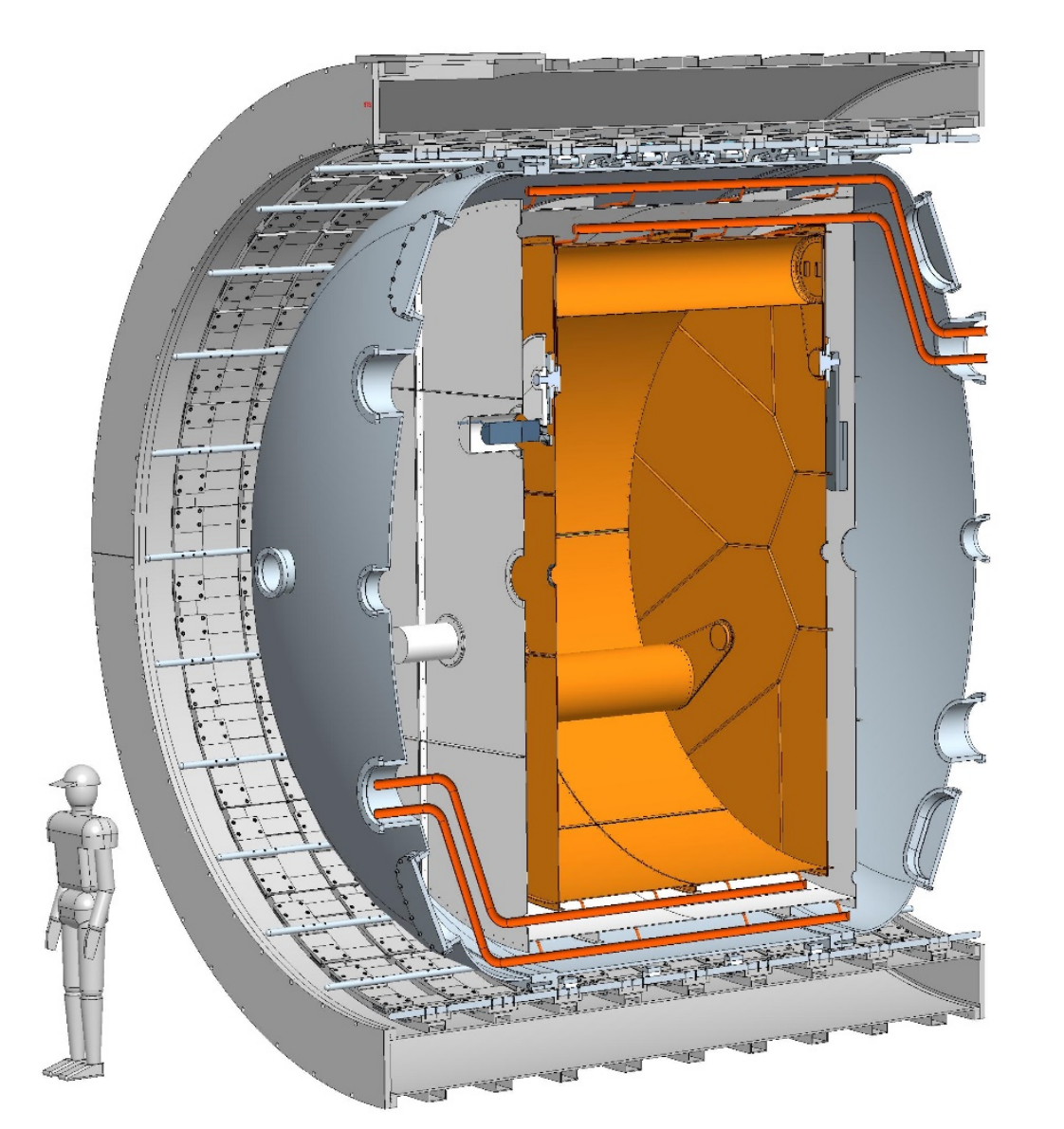}\hspace{3cm}
    \includegraphics[totalheight=6cm]{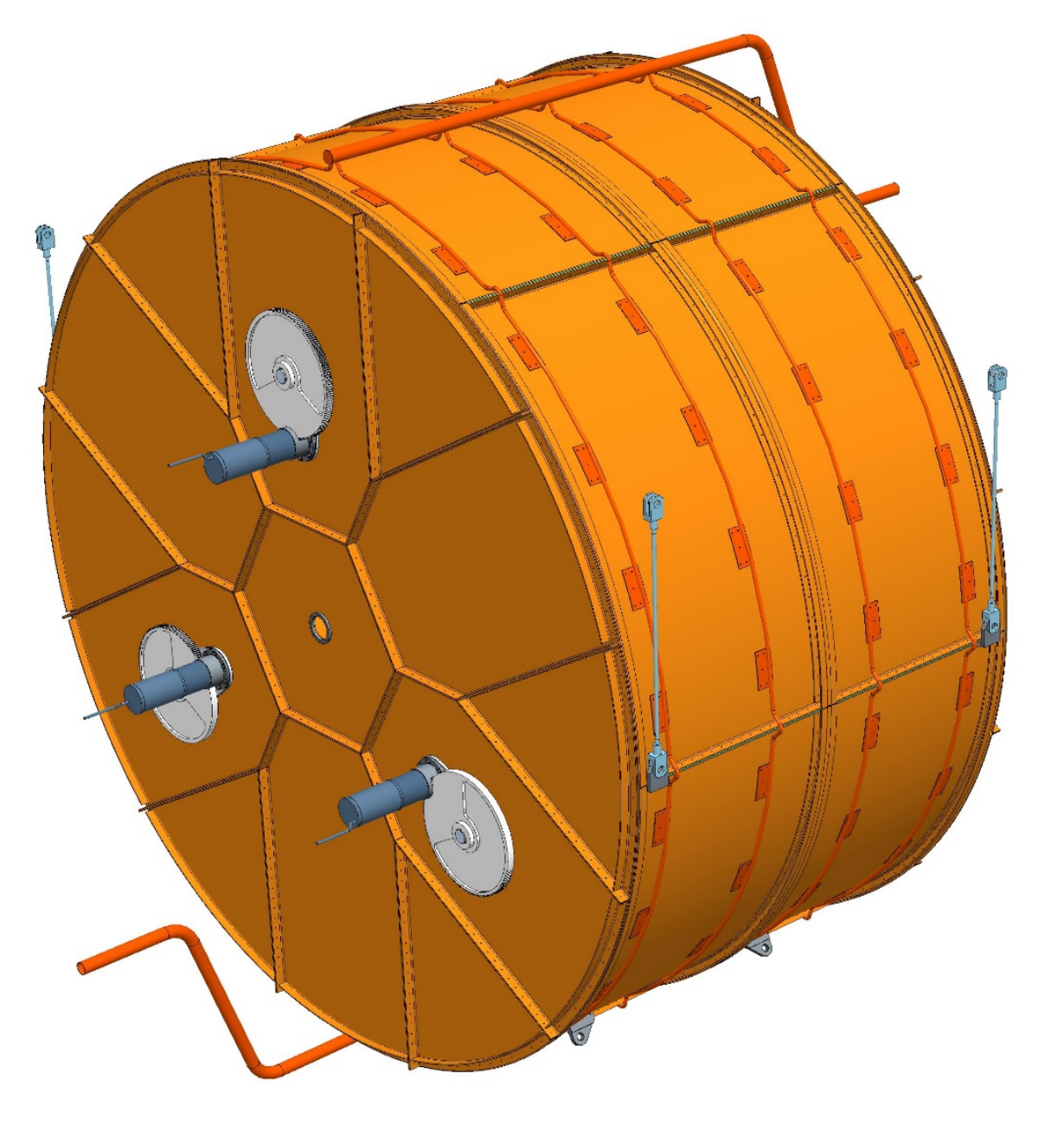}
    \caption{Left: Cutaway picture of the KLASH haloscope. Right: The KLASH resonant cavity.}
    \label{fig:summaryGeo}
  \end{center}
\end{figure}

A detailed description of the mechanical design and cryogenics in discussed 
in section (add reference to section on mechanical design).The cutaway of the haloscope is shown in Fig. \ref{fig:summaryGeo}.
Two different resonant cavities are foreseen to investigate the axion mass region between 0.3 and 1 $\mu$eV. They are cylindrical with length 2041 mm and 
radius 1,860 mm and 900 mm.
\begin{table}[h!]
  \begin{center}
    \caption{The KLASH geometry}
    \label{tab:geometry}
  \vspace*{0.5cm}
    \begin{tabular}{c|c}
      Parameter & Value \\\hline
      L$_{Large Cavity}$ [mm] & 2042 \\
      R$_{Large Cavity}$ [mm] & 1860 \\
      R$_{Large Rod}$ [mm] & 200\\
      L$_{Small Cavity}$ [mm] & 2042 \\
      R$_{Small Cavity}$ [mm] & 900 \\
      R$_{Small Rod}$ [mm] & 100\\
      \hline\hline
    \end{tabular}
  \end{center}
\end{table}

\begin{figure}[htbp]
  \begin{center}
    \includegraphics[totalheight=4cm]{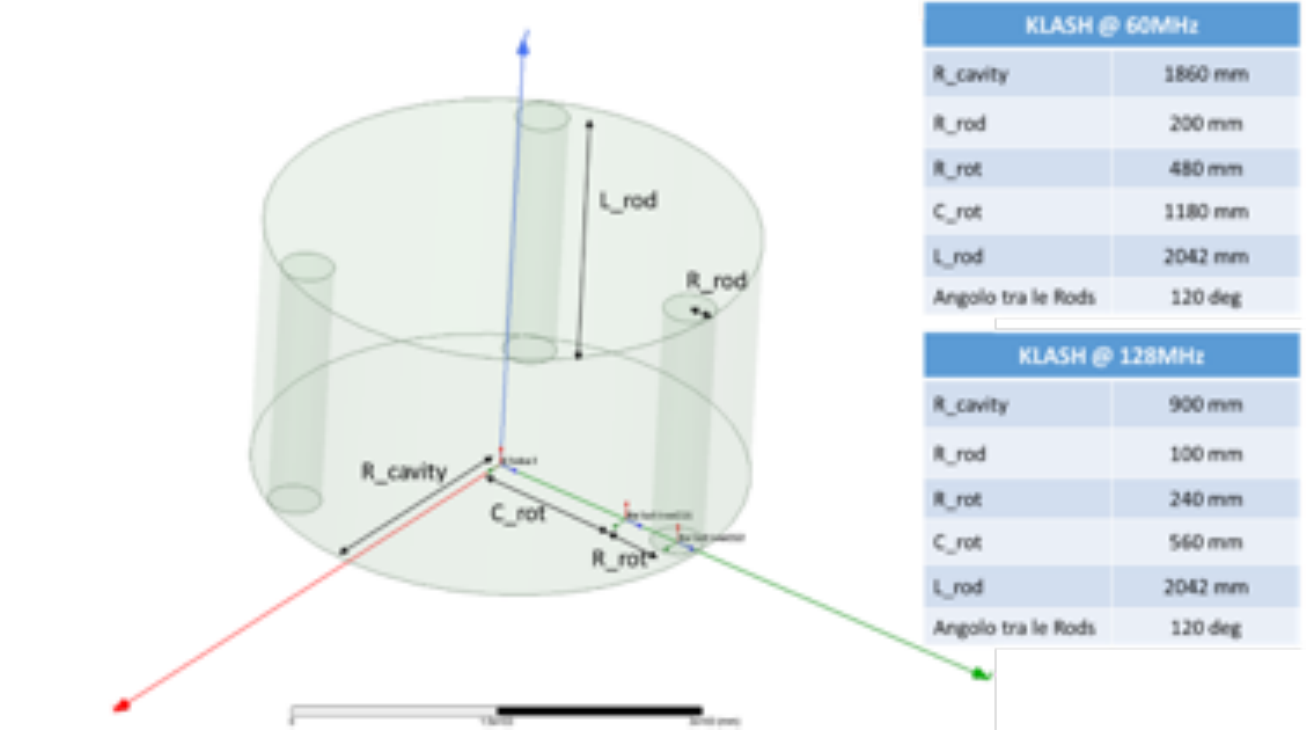}\hspace{3cm}
    \includegraphics[totalheight=4cm]{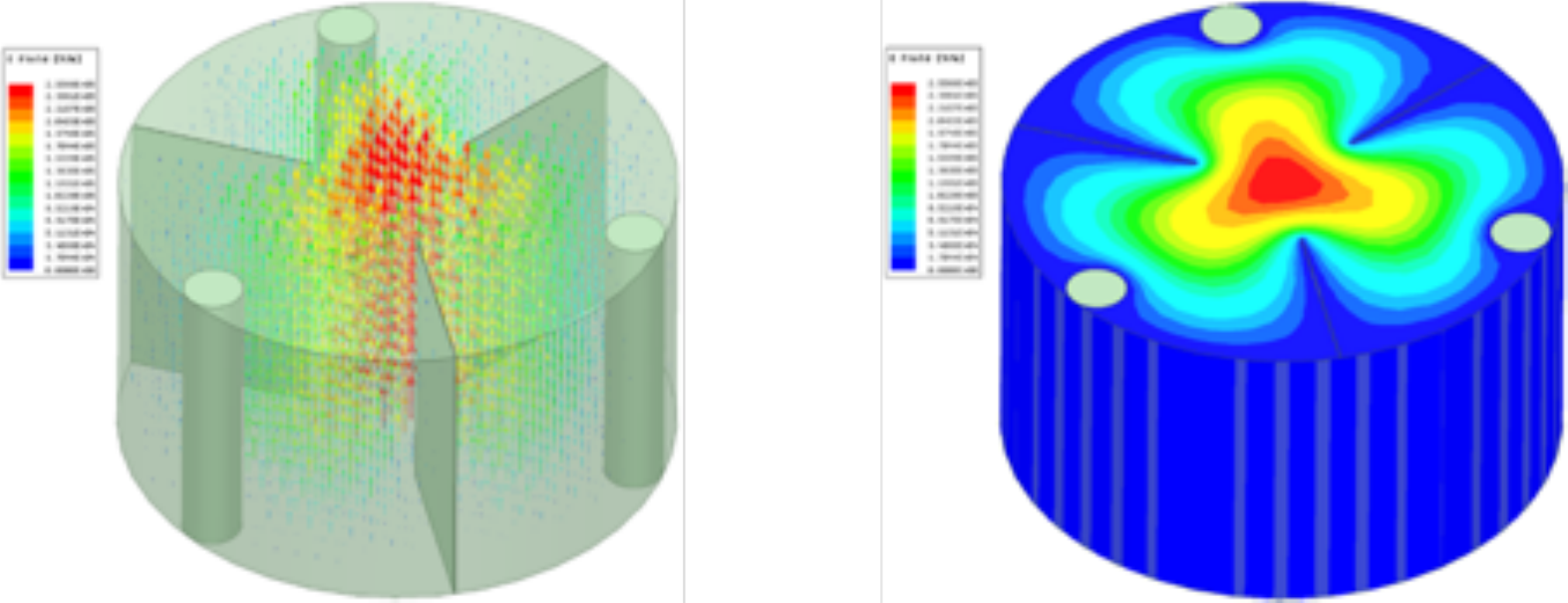}
    \caption{Upper: geometry of the "rotisserie" tuning mechanism. Lower: Tuning by means of rods and fins.}
    \label{fig:summaryTuning}
  \end{center}
\end{figure}

%Tuning
Frequency tuning is obtained by means of three metallic rods and fins as shown in Fig. \ref{fig:summaryTuning} and by replacement of the larger cavity with 
the a smaller one. A detailed description of the tuning procedure and of simulation results are discussed in Chap. \ref{cha:rf}.
Fine frequency tuning, between rod rotations, is obtained by the insertion of a dielectric rod. A similar technique will be used to avoid mode crossing. A summary 
of the cavity and tuning rod geometry is listed in Tab. \ref{tab:geometry}.

\begin{figure}[htbp]
  \begin{center}
    \includegraphics[totalheight=5cm]{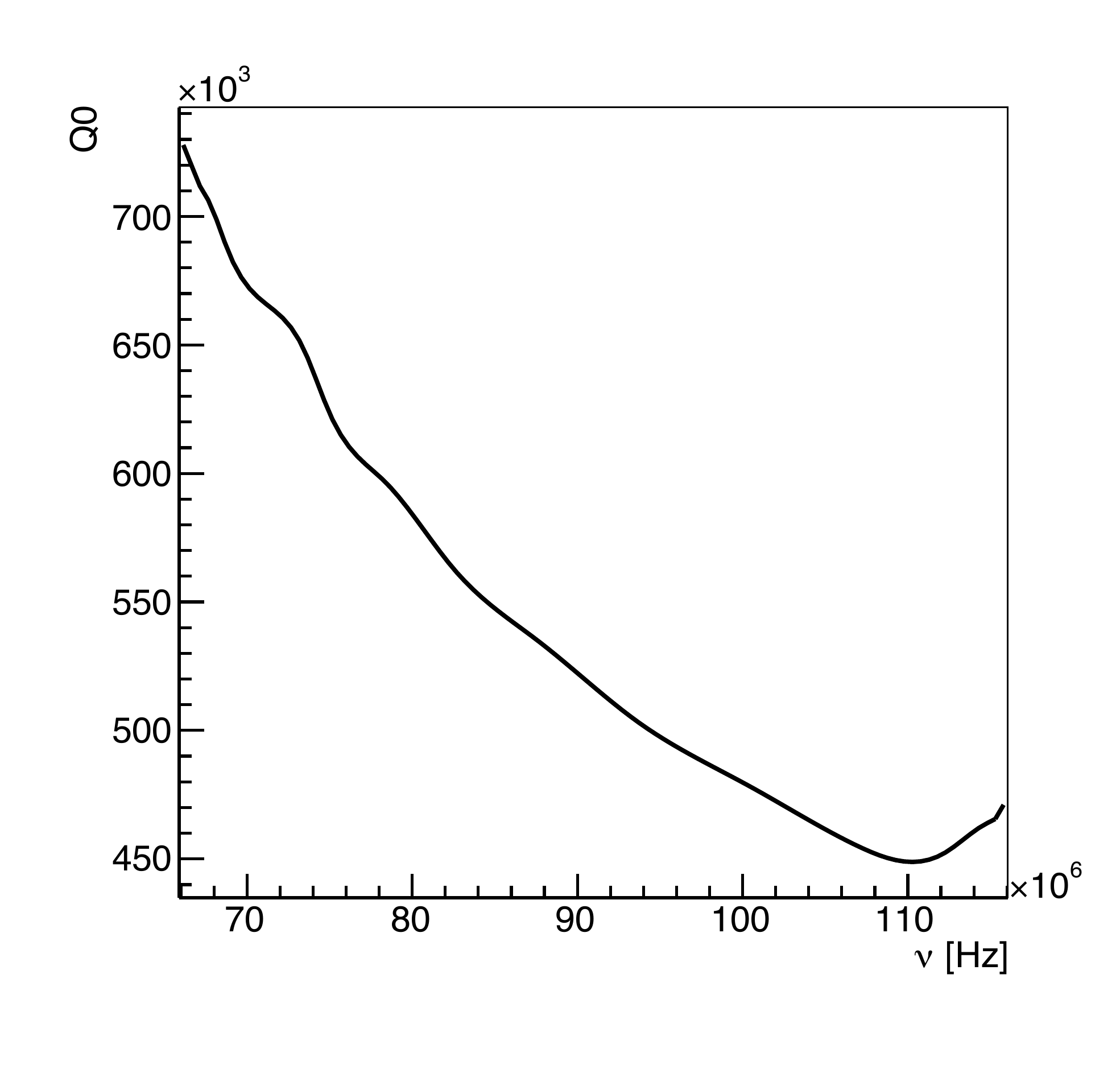}
    \includegraphics[totalheight=5cm]{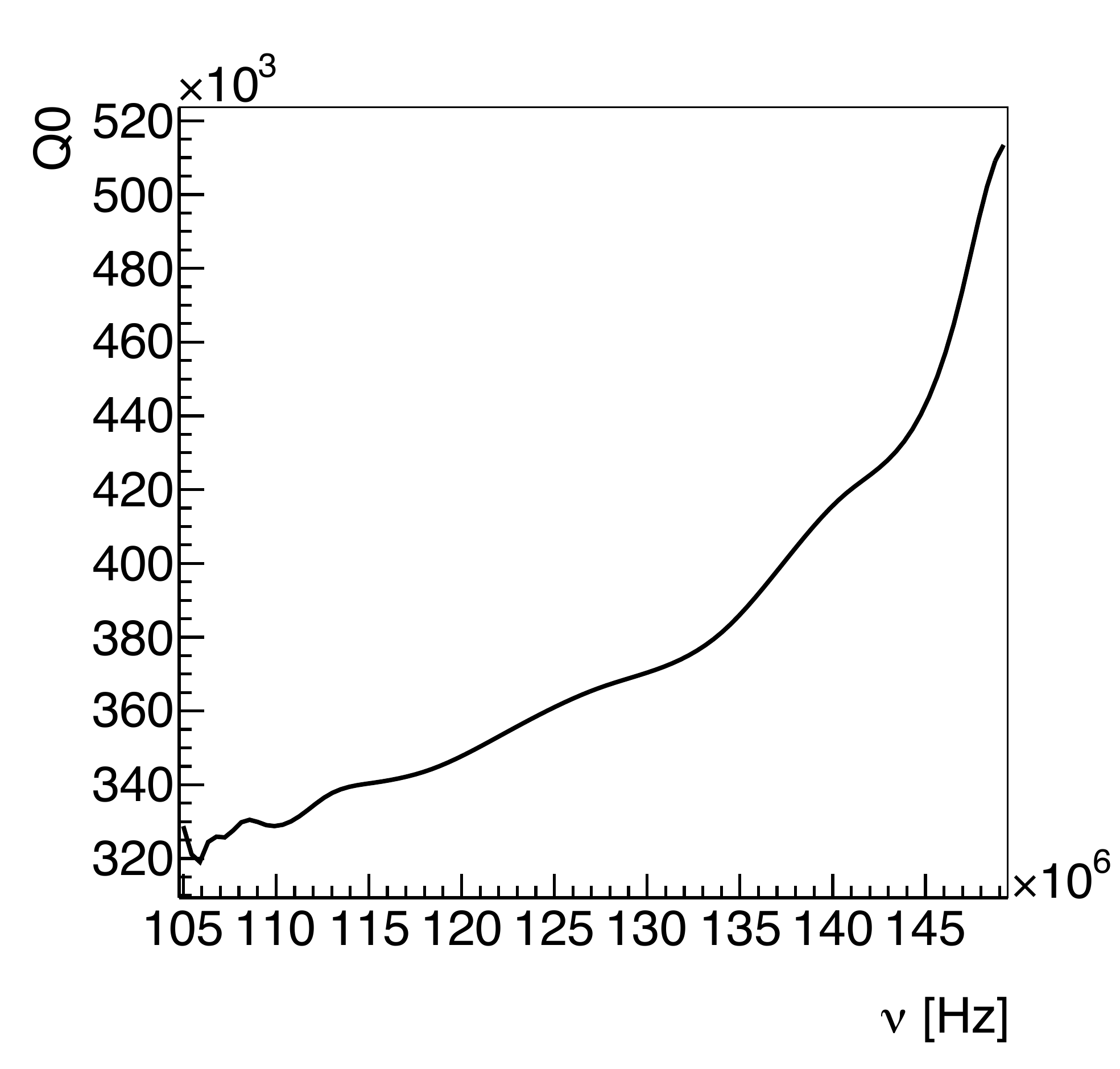}
    \includegraphics[totalheight=5cm]{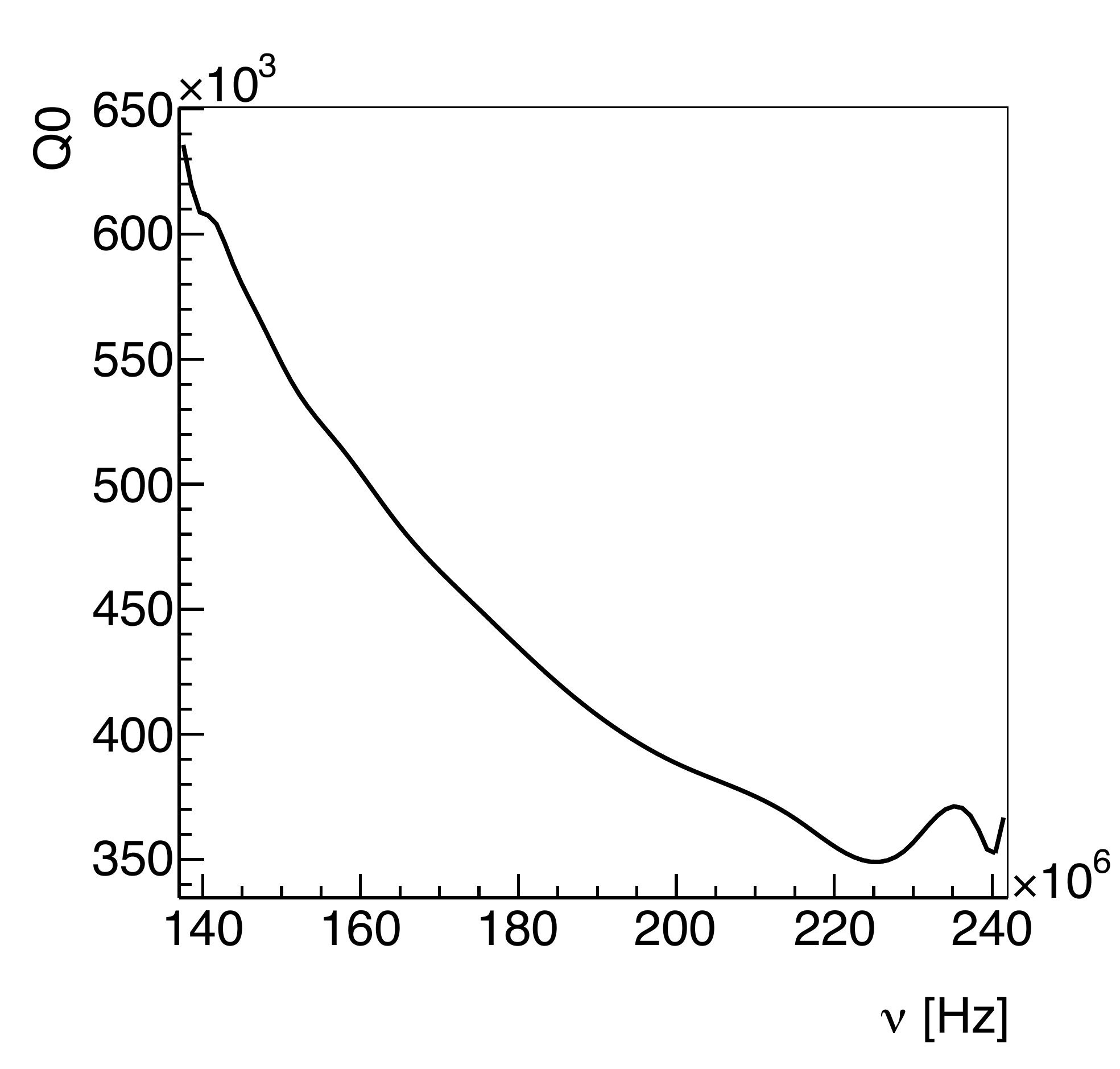}
    \caption{Assuming copper with RRR=50}
    \label{fig:summaryQ}
  \end{center}
\end{figure}

\begin{figure}[htbp]
  \begin{center}
    \includegraphics[totalheight=5cm]{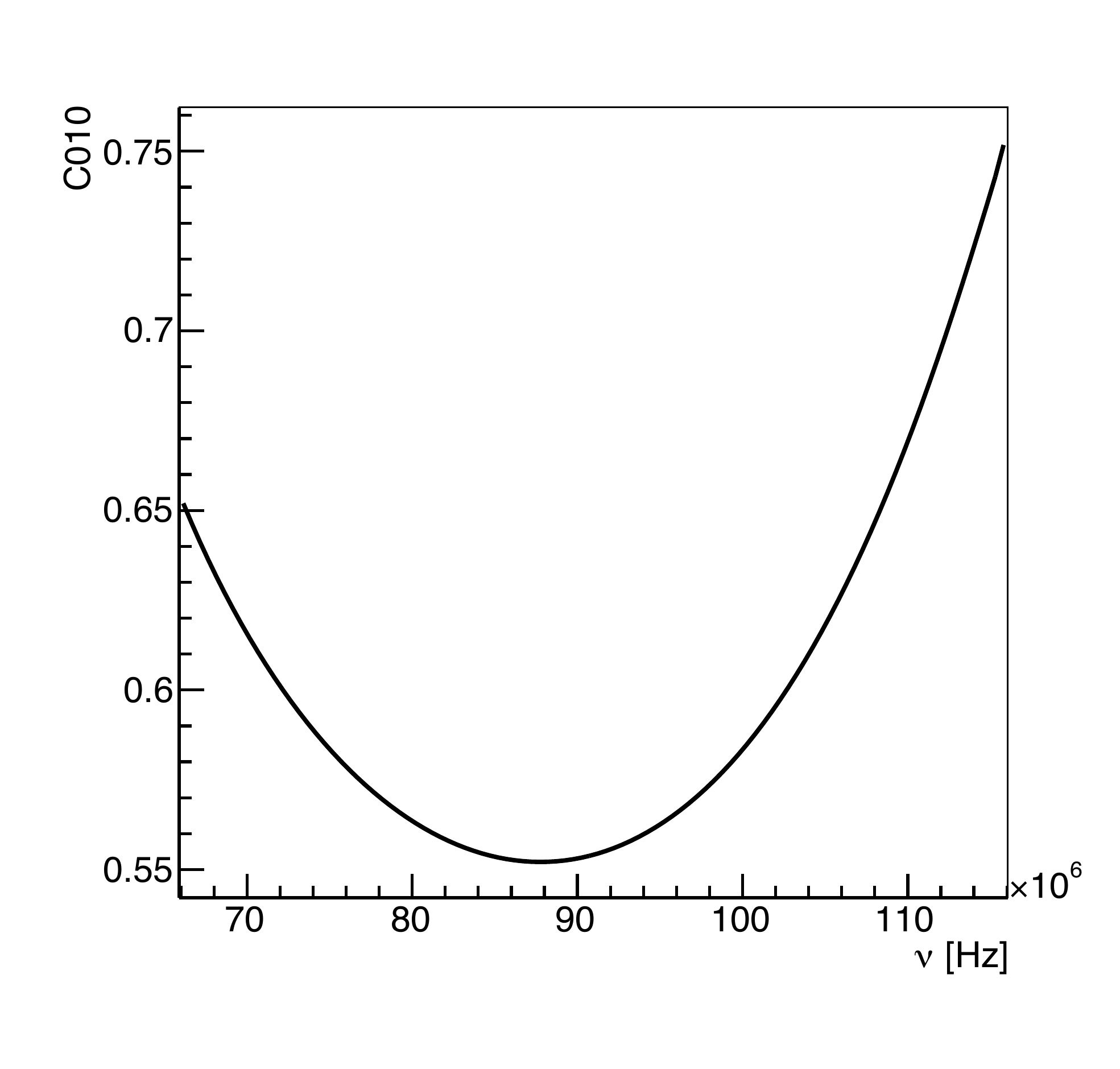}
    \includegraphics[totalheight=5cm]{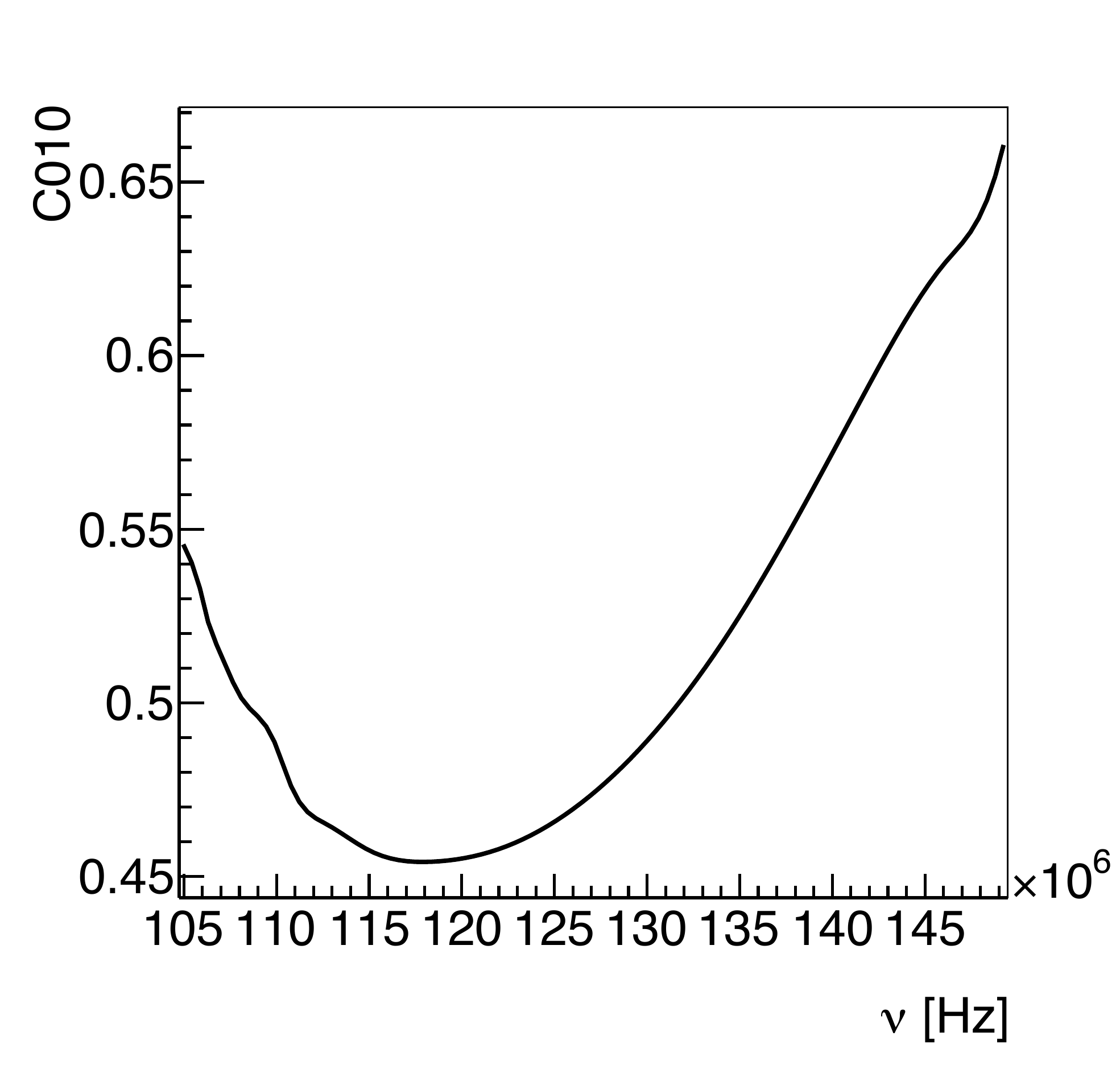}
    \includegraphics[totalheight=5cm]{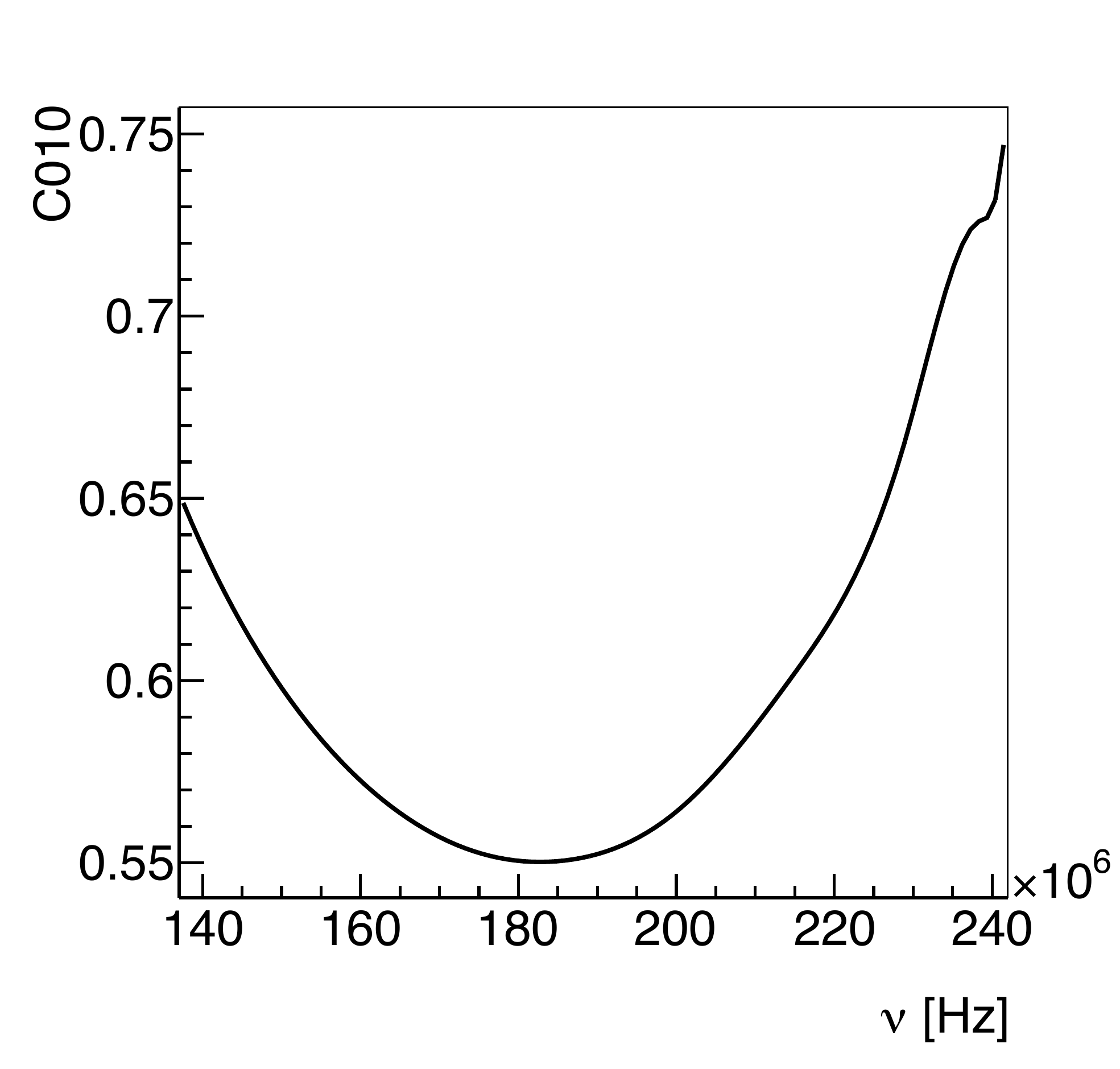}
    \caption{TM010 geometrical factors $C_{010}$.}
    \label{fig:summaryC010}
  \end{center}
\end{figure}

% Q e C010
Three different phases are foreseen. In the first phase, the large cavity will be tuned with three rods changing the frequency from 65 to 115 MHz. In the second 
phase three fins will be inserted to allow frequency tuning up to 150 MHz. Finally, the large cavity will be replaced with a smaller one to allow frequency tuning 
from 140 to 225 MHz. The resonant cavity must be build in oxygen-free high thermal conductivity copper (OFHC). This type of copper may show residual resistance 
ratios (RRR) that vary from 50 to 700. We assume RRR = 50 in the following. With this value we simulated with ANSYS-HFSS code \cite{ansys} the quality factor and form factor 
$C_{010}$ of TM$_{010}$ mode. The results are shown in Fig. \ref{fig:summaryQ} and~\ref{fig:summaryC010} as a function of the mode frequency for the three 
different cavity configurations. These results are used in the following section to determine the KLASH sensitivity to QCD galactic axions and Dark Photons.
In Tab. \ref{tab:tuning} we summarize for the three different resonant cavities (Large, Large with fins, and small) the frequency range and the number of frequency steps.

\begin{table}[h!]
  \begin{center}
    \caption{Summary of frequency tuning}
    \label{tab:tuning}
%  \vspace*{0.5cm}
    \begin{tabular}{c|c|c}
      Cavity &  frequency range & number of steps \\ \hline
      Large & 65 MHz - 115 MHz & 104,000 \\
      Large with fins &  115 MHz - 150 MHz & 44,000 \\
      Small & 140 MHz - 225 MHz & 85,000 \\
      \hline\hline
    \end{tabular}
  \end{center}
\end{table}

% elettronica e noise

As discussed in Sec. \ref{sec:cha5-squid} a Microstrip SQUID Amplifier (MSA) is an optimal solution, in terms of low noise, frequency band and gain, for the first stage of signal 
amplification. The summary of amplification steps and equivalent temperature noise is shown in Tab. \ref{tab:noise}.
\begin{table}[h!]
  \begin{center}
    \caption{Summary of amplification steps and equivalente noise temperature}
    \label{tab:noise}
%  \vspace*{0.1cm}
    \begin{tabular}{c|c|c|c}
      Device &  gain & Noise Temperature & operating temperature \\ \hline
      MSA & 20 dB & 0.4\,K & 4.5\,K \\
      HFET & 15 dB & 5\,K & 4.5\,K \\
      Secondary Amplification & 60 dB & 150\,K & 300\,K \\
      \hline\hline
    \end{tabular}
  \end{center}
\end{table}

\section{The KLASH Sensitivity}\label{ch1:sec_sens}

\begin{figure}[!ht]
  \begin{center}
    \includegraphics[width=0.85\linewidth]{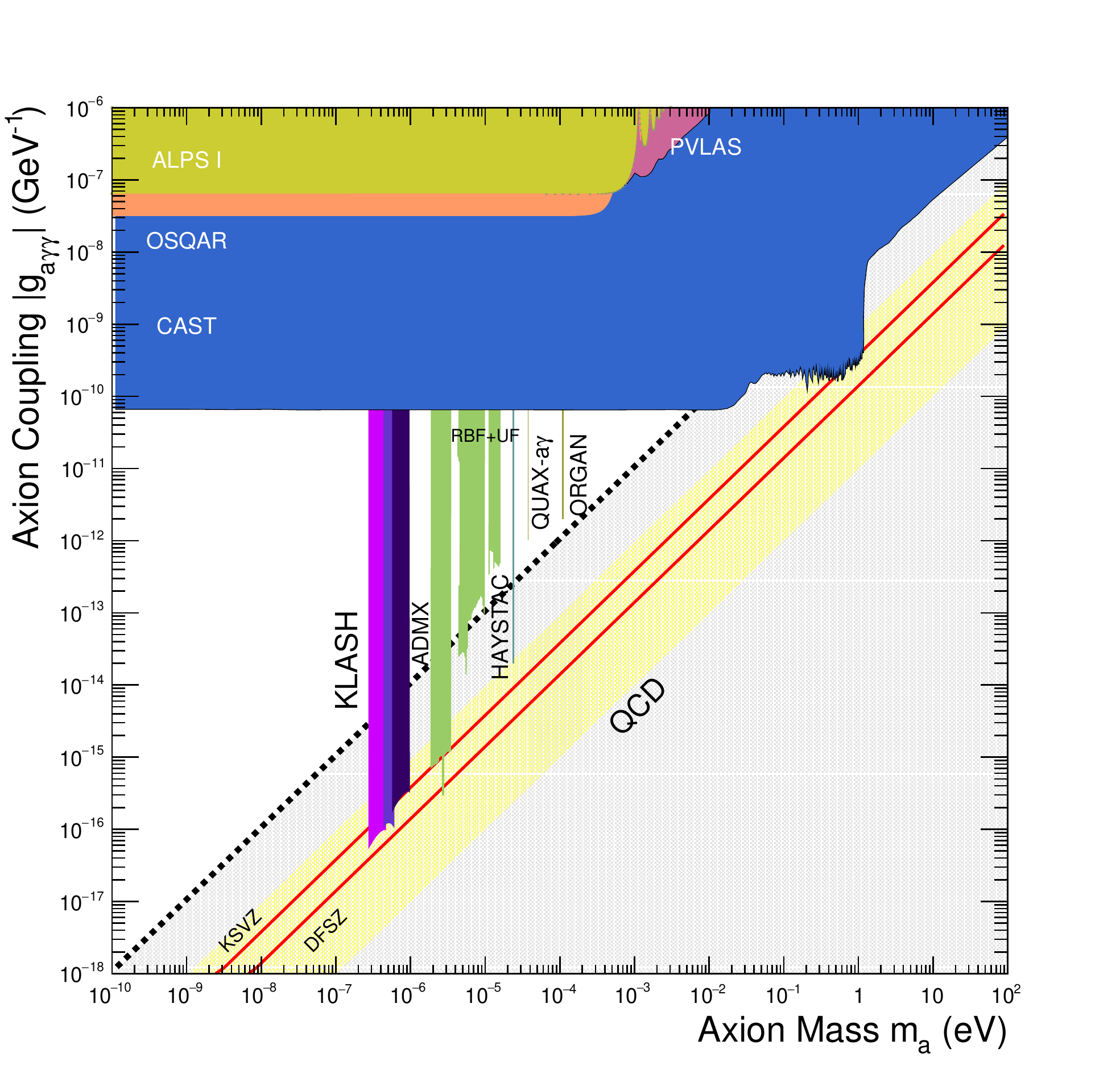}
    \caption{The KLASH discovery potential (90\% c.l.) compared to existing experimental limits. 
    The red lines with yellow error-band show the theoretical predictions for the KSVZ and DFSZ 
    axions \cite{KSVZ,DFSZ}. The grey area shows the prediction from other hadronic axions 
    models \cite{DiLuzio:2016sbl}. The experimental limits are measured with light shining through 
    a wall experiments \cite{ALPS,OSQAR}, from changes in laser polarization \cite{PVLAS}, 
    helioscopes \cite{CAST}, and haloscopes \cite{ADMX,HAYSTAC,ORGAN,BNL,UF}.}
    \label{fig:sensitivity}
  \end{center}
\end{figure}

The discovery potential in the coupling-mass plane is shown in Fig. \ref{fig:sensitivity}. The sensitivity band reaches the band predicted for QCD axions of the 
KSVZ and DFSZ models.
We considered here an integration time of 5 minutes for a single measurement with the large Cavity and 10 minutes for measurements with large cavity with fins 
and small cavity. According to the estimated number of measurements in Tab. \ref{tab:tuning} the total integrated time will be about 3.5 years.
\begin{table}[h!]
  \begin{center}
    \caption{The KLASH discovery potential for KSVZ axions. $\beta$ is chosen equal to 2 to optimize the scan rate \cite{HAYSTAC}.}
    \label{tab:sensitivity}
  \vspace*{0.5cm}
    \begin{tabular}{c|c}
      Parameter & Value \\\hline
      $m_a$ [$\mu$eV] & 0.27\\
      $g_{a\gamma\gamma}^{KSVZ}$ [GeV$^{-1}$] & $1.0\times10^{-16}$  \\
      $P_{\mbox{sig}}$ [W] & $1.33\times10^{-22}$  \\
      Rate [Hz] & 3,050 \\
      $B_{max}$ [T]  & 0.6 \\
      $\beta$ & 2 \\
      $\tau$ [min] & 5\\
      $T_{sys}$ [K] & 4.9 \\
      $g_{a\gamma\gamma}$ 90\% c.l. [GeV$^{-1}$] & $5.3\times10^{-17}$  \\
      \hline\hline
    \end{tabular}
  \end{center}
\end{table}

\section{Dark Photons sensitivity}\label{ch:sec_dark-pho}

The \emph{hidden photon} (HP) field $\chi_\mu$ describes a hidden $U(1)$ symmetry group that mixes with the photon through a Lagrange density of the form
\begin{equation}
\mathcal{L}_{\chi_\gamma} = - \frac{1}{4 \mu_0} F_{\mu\nu}\,F^{\mu\nu} - \frac{1}{4 \mu_0} X_{\mu\nu}\,X^{\mu\nu} + 
\frac{\sin \chi}{2 \mu_0} X_{\mu\nu}\,F^{\mu\nu} + 
\frac{\cos^2 \chi}{2 \mu_0} m^2_{\gamma^{\,\prime}}X_\nu\,X^\nu- j_{\mathrm{em}}^\mu\,A_\mu\,,
\label{hp:eqL1}
\end{equation}
where $X_{\mu\nu}$ is the field strength tensor, $m_{\gamma^{\,\prime}}$ the hidden photon mass, $j_{\mathrm{em}}^\mu$ is the electromagnetic current, 
$\mu_0$ is the permeability of the vacuum, and $\chi$ is the dimensionless mixing parameter we want to determine experimentally. 

The kinetic term in Eq. \eqref{hp:eqL1} can be diagonalised by a linear transformation of the fields $X_\mu$ and $A_\mu$, and it can be chosen such that only 
one of the gauge fields couples to electric charges. This is an interaction eigenstate and can be then identified with the Standard Model photon. 
By rescaling $X_\mu$ and $A_\mu$ the kinetic terms can be put in the form with the standard normalisation, and after diagonalising the mass term, one obtains
\begin{equation}
\mathcal{L}_{\chi_\gamma} = - \frac{1}{4 \mu_0} F_{\mu\nu}\,F^{\mu\nu}\; - \frac{1}{4\mu_0} X_{\mu\nu}\,X^{\mu\nu} + 
\frac{1}{2\mu_0} m^2_{\gamma'}X_\mu\,X^\mu\, + j_{\mathrm{em}}^\mu\,( A_\mu + \tan \chi X_\mu)\,,
\label{hp:eqL2}
\end{equation}
so that the HP couples to e.m. current with strength $e \tan \chi$. Eq. \eqref{hp:eqL2} implies that for both $A_\mu$ and $X_\mu$ the Maxwell equations 
for the potentials are modified, and the Lorentz force becomes
$$
f^\mu = q\,(F^{\mu\nu} + \tan \chi\,X^{\mu\nu})\,u_\nu.
$$
This implies that an electric current also produces HP, but suppressed by a factor tan$\chi$ relative to the ordinary photon field. 

The mass term can be obtained through the \emph{St\"uckelberg mechanism} \cite{Stuck}. Taking $A_{\mu}$ to be real one can introduce a 
new real scalar field $\phi$, with the Lagrange density becoming
\begin{equation}
\mathcal{L}_{\chi_{\gamma}} = - \frac{1}{4\mu_0} F_{\mu\nu}\,F^{\mu\nu}\, + \frac{1}{2}m^2(A_{\mu} + \frac{1}{m}\partial_{\mu}\phi)\,(A^{\mu} + 
\frac{1}{m}\partial^{\mu}\phi).
\label{hp:eq3}
\end{equation}
This contains the term $\partial_{\mu}\phi^{\mu}/2$ which is indeed the kinetic term for a real scalar field and, furthermore, it has the mass term for a vector field. 

The microwave resonant cavity experiments searching for axion can also be used to probe photon-HP mixing when operated without the magnetic field \cite{Arias}. 
The equation of motion for the photon field $A$ then reads,
\begin{equation}
\partial_{\mu}\partial^{\mu} A^{\nu} = \chi\,m^2_{\gamma{\,\prime}}\,X^{\nu},
\label{hp:eqM}
\end{equation}
from which the HP field acts as a source for the ordinary photon. Expanding the e.m. field inside the cavity in terms of the cavity modes
$$
\textbf{A}(x) = \sum_i \alpha_i \textbf{A}_i^{\mathrm{cav}}(\textbf{x}),
$$
we can obtain the expression for the expansion coefficients. The asymptotic solution is found to be
\begin{equation}
\alpha_i(t) = \frac{b_i}{\omega_0^2 - \omega^2 - i \displaystyle{\frac{\omega\omega_0}{Q}}}\,\mathrm{e}^{-i \omega t}
\label{hp:eqalpha}
\end{equation}
where $b_i$ is the driving force and the frequency is related to the HP energy through
\begin{equation}
\omega = E_{\gamma{\,\prime}} \simeq m_{\gamma{\,\prime}}.
\label{hp:eqE}
\end{equation}
The power emission of the cavity is related to the energy stored and the quality factor of the cavity, and it is given by
\begin{equation}
P_{\mathrm{sig}} = \kappa \chi^2 m_{\gamma{\,\prime}} \rho\, Q\, V\, G,
\label{hp:eq4}
\end{equation}
where $\kappa$ is the coupling of the cavity to the detector, $Q$ the quality factor, $V$ the volume of the cavity and the geometric factor $G$ is defined as
\begin{equation}
G = G_{\mathrm{axion}} \cos^2 \theta.
\label{hp:eq5}
\end{equation}
Assuming the energy density of the HP to be equal to the dark matter density (0.45 GeV$^2$ cm$^{-3}$), and the same quality factor used for the axion search, 
and the volume of the cavity equal to 22 m$^3$ (for the large phase of the experiment) and 5.2 m$^3$ (for the small phase), we can use this formula to constrain the 
HP CDM with the KLASH microwave cavity. It is already evident that the large dimension of the cavity significantly enhance the sensitivity for HP, also thanks to the fact 
that no magnetic field is required for this kind of physics analysis.
\begin{figure}[!ht]
\begin{center}
\includegraphics[width=0.85\linewidth]{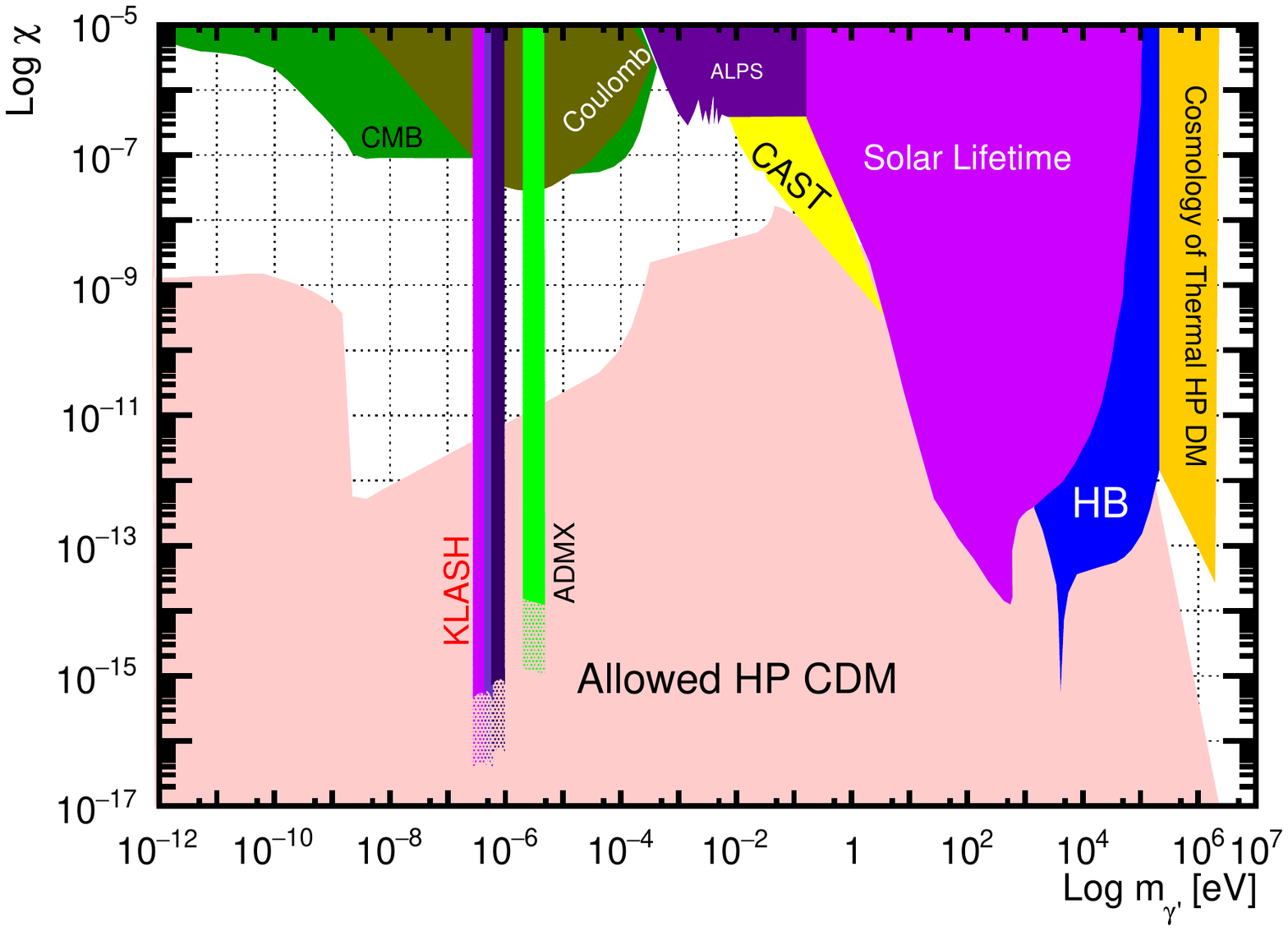}
\caption{\emph{Top panel} - Compilation of current constraints on and sensitivities of planned experiments to photon-HP 
mixing in the $m\gamma^{\,\prime} - \chi$ plane. With bars in blue (red) the conservative (equally likely HP distribution) estimates for 
the ADMX-like haloscope search experiments \cite{Arias} and the KLASH predicted sensitivity. More details are given in the text.}
\label{fig:chivsm}
\end{center}
\end{figure}

{\noindent}
Typical values for the mixing parameter $\chi$ range from $\sim10^{-2}$ down to $\sim 10^{-16}$, KLASH can further improve this value. In top panel of Fig. \ref{fig:chivsm}  
is showed a compilation of current constraints on and sensitivities of planned experiments to photon-HP mixing in the $m_{\gamma^{\,\prime}} - \chi$ plane. The bottom plot 
of the same figure shows the predicted sensitivity of KLASH, compared to the sensitivity of an ADMX-like haloscope. 

The predicted sensitivity for KLASH is computed inserting in Eq. \eqref{eq:snr} the $P_{\mathrm{sig}}$ obtained from Eq. \eqref{hp:eq4}. 
Two scenarios are possible, depending on the relative orientation of the cavity with respect to an a priori unknown direction of the HP field. This is represented by the factor 
$\cos^2 \theta$ in Eq. \eqref{hp:eq5}. As a conservative estimate, assuming that all directions in space are equally likely, we can consider $\cos^2 \theta$ 
such that the real value is bigger with 95\% probability. In this scenario $\cos^2 \theta = 0.0025$ (showed with the histograms in blue and in the full colours in the 
top and in the bottom of Fig. \ref{fig:chivsm}, respectively). The other situation, considers the average over all possible directions (i.e. $\cos^2 \theta = 1/3$) 
and it is more optimistic. This is shown with the histogram in red and with the dashed filling in Fig. \ref{fig:chivsm} (top and bottom).

 \chapterimage{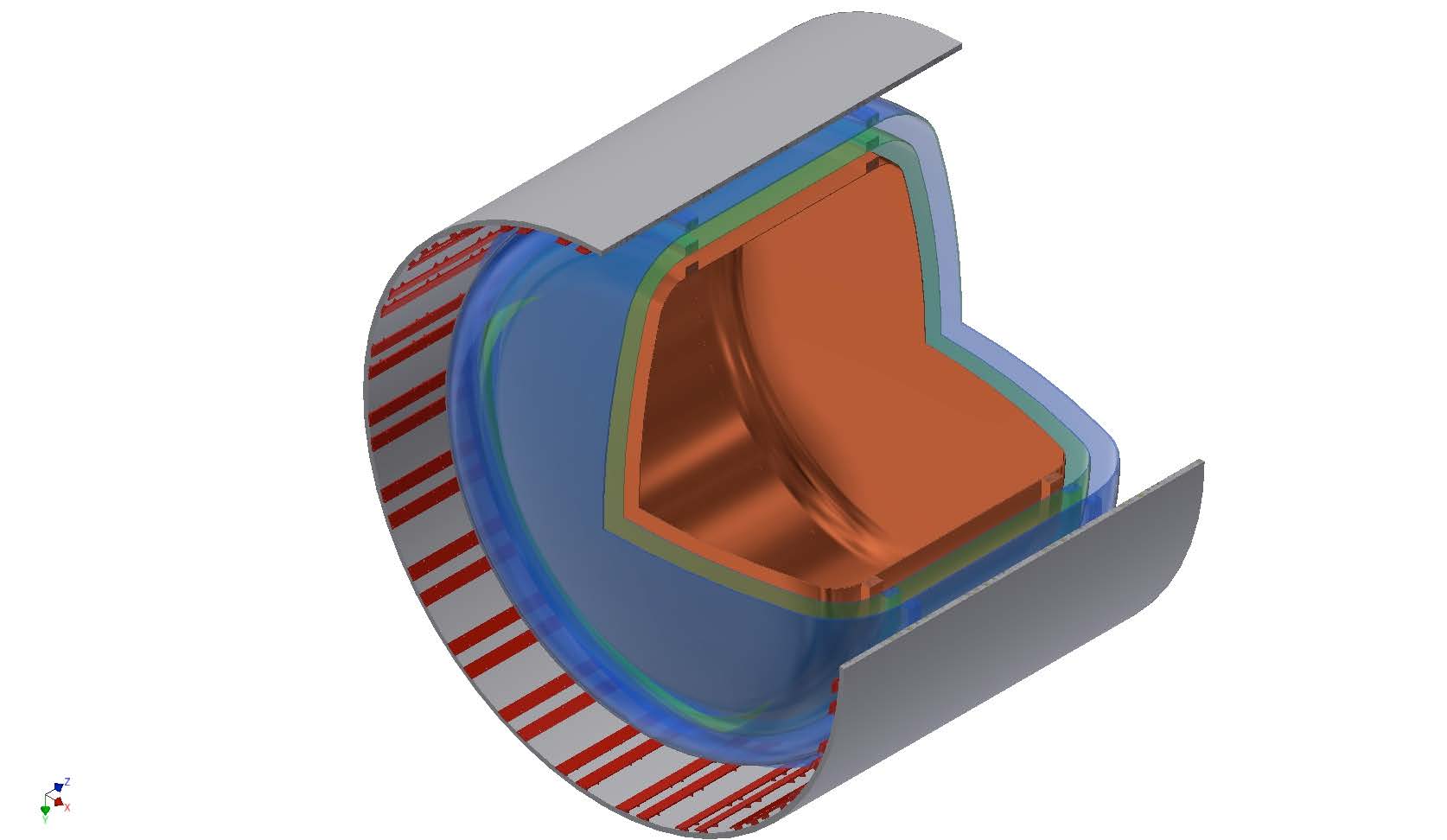}
 \part{Conceptual Design}

\chapter{Mechanical Design}\label{cha:mech}
\section{Cryostat general description}
A design study of the cryostat assembly has been performed by an engineering firm, under our specifications, 
demonstrating its technical and economic feasibility. The whole cryostat has been designed to maximize its 
volume with respect to the available space inside the KLOE magnet (Ø4865 x 4398 mm). The cryostat is 
equipped with the tools for the insertion and mounting inside the KLOE magnet.

The Cryostat includes:
\begin{itemize}
  \item The vacuum vessel, made by a-magnetic stainless steel;
  \item The shield in aluminum alloy, to be cooled to 70 K by gaseous Helium;
  \item The resonant cavity in OFHC copper, with his tuning system, cooled to 4.6 K by 
  saturated liquid Helium; The tuning system includes three cylindrical bars, mounted 
  on three eccentric crank and simultaneously positioned by high precision drives with 
  reduction gearboxes.
\end{itemize}
All the three have the two end caps flanged and are equipped with the mounting and handling tools. 
The shield and the cavity have a cooling circuit.

The cryostat is completed by the superinsulation layers, the cryogenic turret, all the feedthroughs for piping 
and cabling. Included in the design study are the main accessories for the handling. The thermal insulation 
of the cryostat has been analyzed, and a preliminary design of the shield, of the insulation layers and of the 
suspension has been done, to minimize the conduction and radiation heat exchange with the external 
environment. The vessel is designed to stand with the internal overpressure due to an event of helium leak. 

The issues arising from the fabrication of large vessels, especially for the aluminum and copper ones, have 
been analyzed and their feasibility demonstrated at a reasonable cost.

\section{Vacuum vessel}
\begin{figure}[!h]
\begin{center}
	\includegraphics[width=0.6\linewidth]{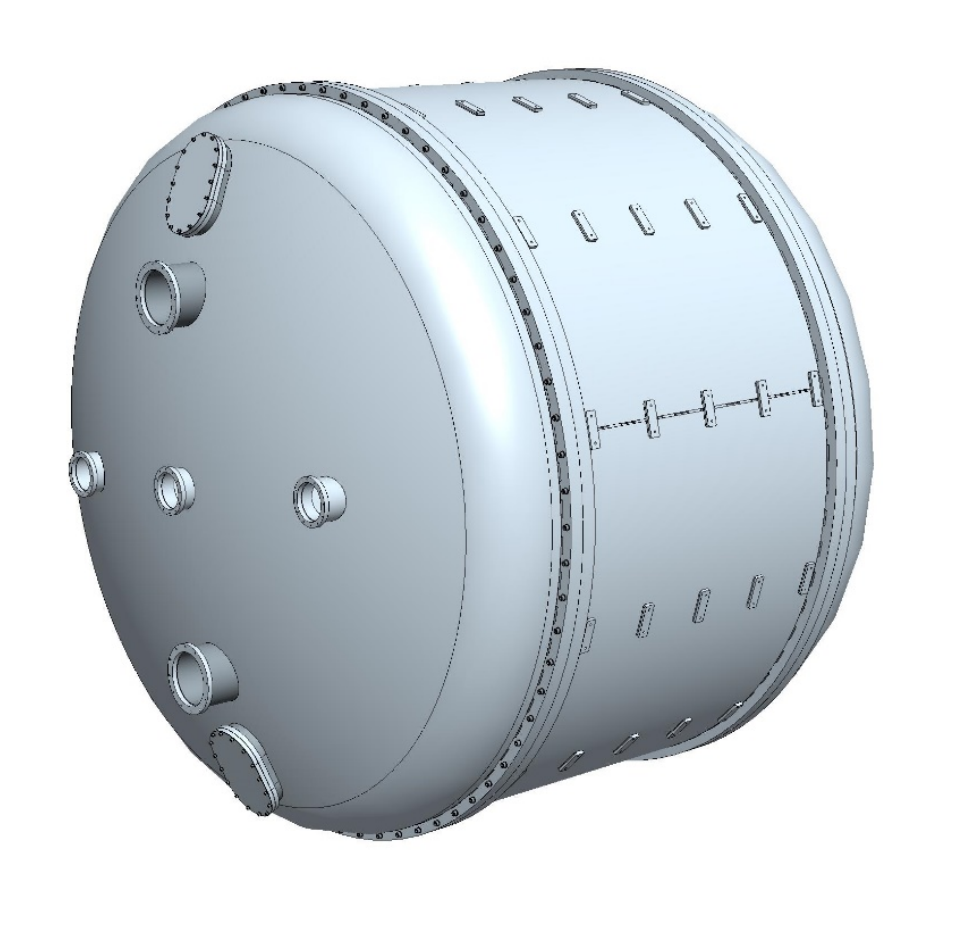}
	\caption{The vacuum vessel.}
	\label{fig:vv}
\end{center}
\end{figure}

The vacuum vessel has an external diameter of 4,660 mm, maximum length of 4,480 mm and a weight of 
14,000 kg (see Fig. \ref{fig:vv}). 
The vessel shell (2 cm thick) is formed by four cold-rolled austenitic stainless steel plates, welded together 
and to the flanges. The heads (1.5 cm thick) will be realized by press-formed dishes with welded flanges; 
the heads are equipped by the openings for the pass-through of pipes and cables and for the mounting 
operation. All the welds and composed by an internal vacuum tight bead and an external structural 
piecewise reinforcement welding. The flanges will be milled with a precision that guarantee the vacuum 
tight coupling. The shell and the heads flanges are bolted together; the sealing is done by VITON O-rings. 

External to the vessel will be realized the interfaces for the anchoring to the magnet, while in the internal face 
will be welded the supports for the suspension of the cavity and the shield.

\section{Cryostat support}
\begin{figure}[!h]
\begin{center}
	\includegraphics[width=0.45\linewidth]{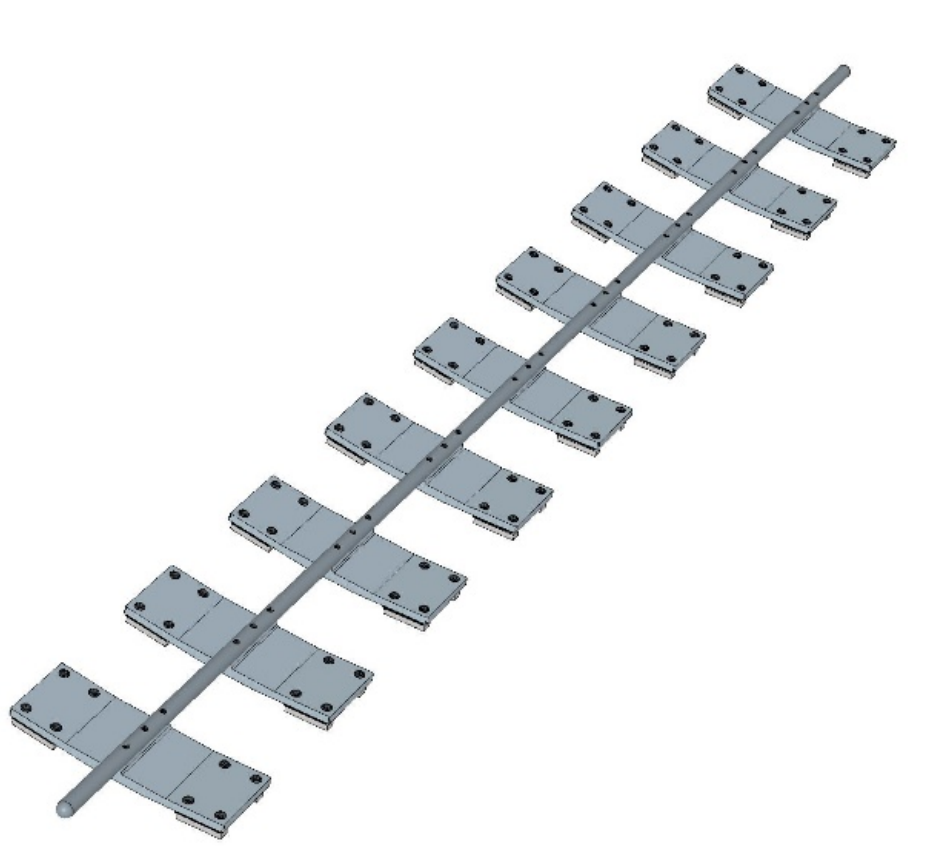} 
	\hspace*{1.0cm}
	\includegraphics[width=0.45\linewidth]{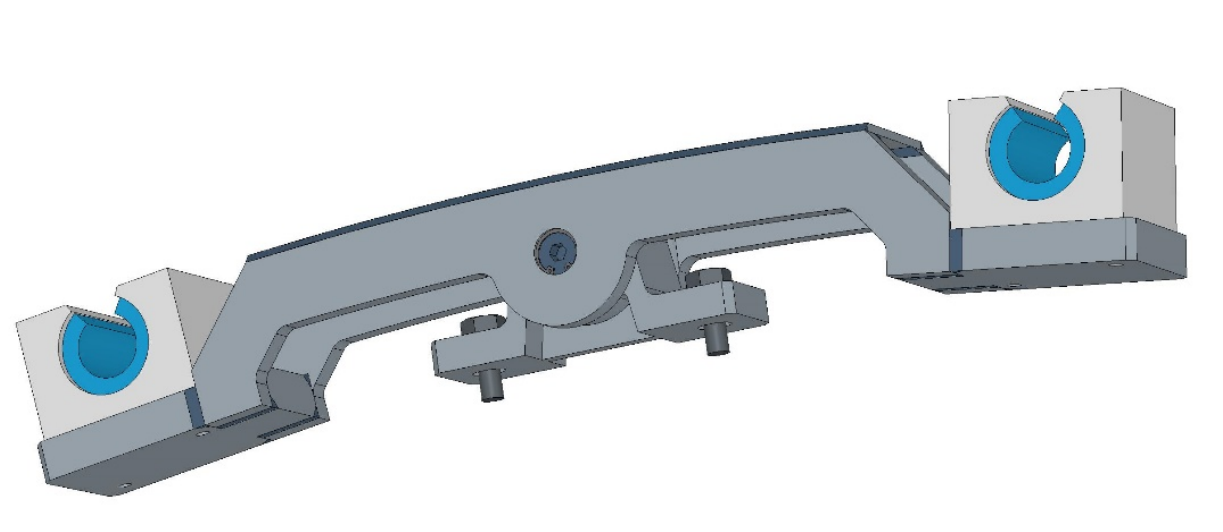}
	\caption{The mounting rail (left) and the support block (right).}
	\label{fig:rail-bloc}
\end{center}
\end{figure}

The cryostat assembly will be supported inside the KLOE magnet by means of the anchoring interfaces 
spaced at 15 degrees that was used for the 24 calorimeter modules, weighing 2 ton each. So, the maximum 
load distributed could be no more than 480 N. 24 longitudinal rails will be mounted by 9 supports each 
(see left panel of Fig. \ref{fig:rail-bloc}), using the existing screwed holes on the internal surface of the magnet, 
and precisely aligned by laser tracker.

Four supports will link each couple of rails to the cryostat. Each support could slide on the couple of rails by 
low friction sleeves (made by Frelon Gold) and will be bolted to the vessel by elastic silent-blocks (see right panel 
of Fig \ref{fig:rail-bloc}). Small alignment errors of each support could be adjusted by the eccentric pin in the middle, 
while the parallelism error between each couple of rails could be accommodated by the elastic blocks.

With this arrangement, the vessel alone, without heads, could be inserted sliding along the rails in the first 
mounting phase.

\section{Cavity}
\begin{figure}[!hb]
\begin{center}
	\includegraphics[width=0.6\linewidth]{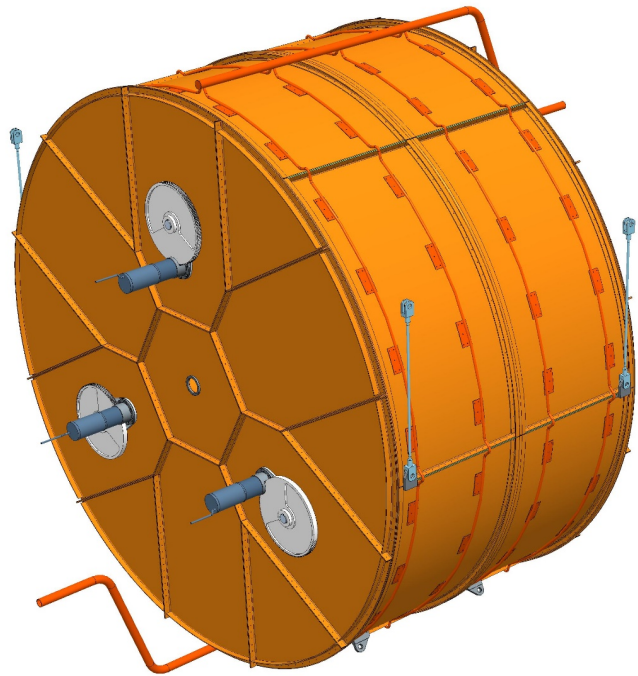}
	\caption{The cavity.}
	\label{fig:cav}
\end{center}
\end{figure}

The cavity will be a cylinder with flat ends (Fig. \ref{fig:cav}), internal dimension Ø3,690 mm x 1,960 mm, external 
Ø3,800 mm x 2,060 mm, weight 2,600 kg.
The shell will be made by 12 tiles, realized by cold formed commercial plates of copper OFHC 
(1,000 x 2, 000 x 5 mm), joint by rivets. Circumferential ribs will enhance the rigidity of the assembly. To minimize 
eddy current loops, the longitudinal joint between tiles will be insulated by NEMA G10 plates and insulating sleeves.
The end plates will be made by 13 tiles, cold formed and joint by rivets and screws. 

The cavity will be suspended to the vessel by four stainless steel (AISI 316L) tie-rods, with a length adjustment 
and a heat link to the 70K shield. Some pins will keep the horizontal position of the cavity. Tie-rods and pins will 
be sized to minimize the heat conduction.

The cooling of the cavity will be done by a 4.6 K liquid helium circuit, made by two manifold (upper and lower) 
connected by Cu-DHP copper tubes in parallel. The tubes will be connected by small plates brazed to the tube and 
riveted to the cavity.

The cavity assembly is such to allow opening of the end plate for servicing.

\section{Tuning system}
In the cavity there three OFHC copper hollow cylinders (``tuners'', Ø400 mm for the whole length of the cavity) 
will be mounted on eccentric positioning devices (three motorized couples of cranks with a counterweight). 
The three cylinders could be positioned simultaneously at a distance from the axis of the cavity from 560 to 
1,640 mm.

The electric continuity of the cavity will be assured by sliding contacts around the tuners shafts. Each tuner will be 
equipped by a vacuum compliant Hybrid Stepper-Motor (200 step/turn, 150 min$^{-1}$) with a three-stage 
epicyclical gear unit (1:200) and a gear couple (1:10). The total angular step will be $9 \time 10^{-4}$ degrees. 
The whole turn will take around 7 min. Each motor will have a thermal probe and an angular resolver, while all the 
three will be driven by a control unit, external to the cryostat. The whole tuning system will weigh 1,050 kg.
\vspace*{1.0cm}
\begin{figure}[!h]
\begin{center}
	\includegraphics[width=0.45\linewidth]{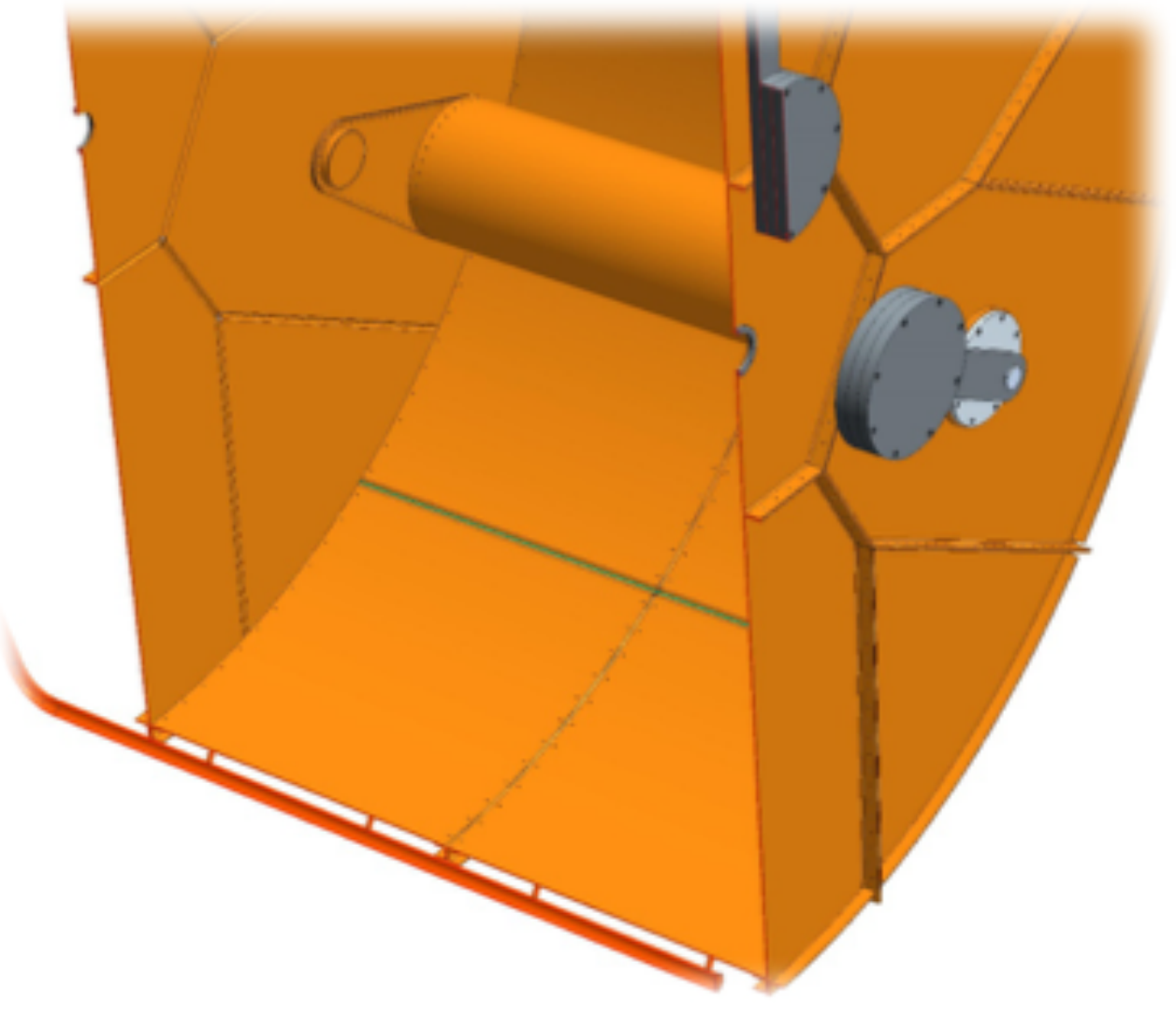}
	\hspace*{1.0cm}
	\includegraphics[width=0.45\linewidth]{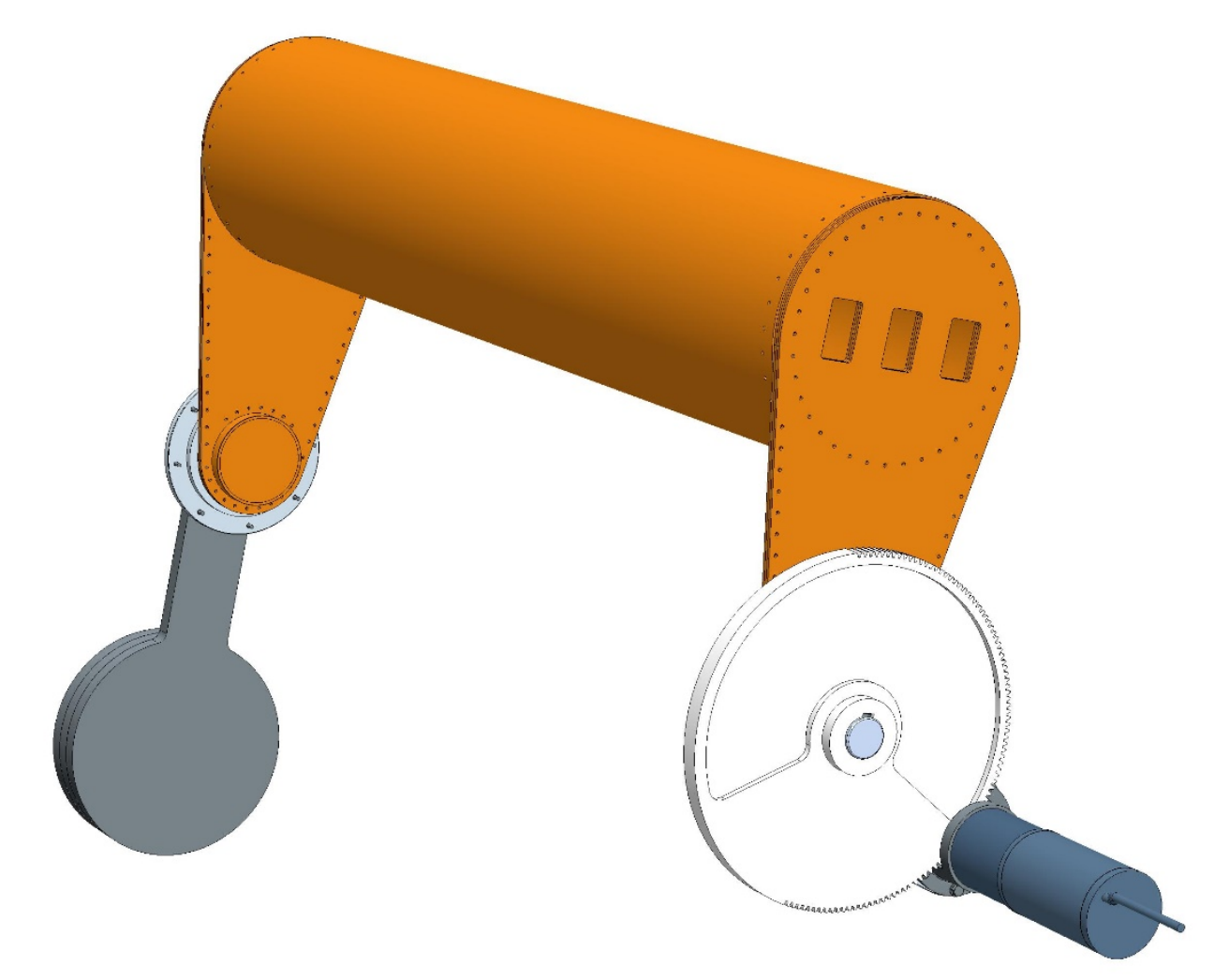}
	\caption{Internal view of the cavity with the tuner (left); tuner assembly (right).}
	\label{fig:tun}
\end{center}
\end{figure}

During the data taking, the motors will be switched off, but their position will be kept in the control memory, to avoid 
any further homing.

\section{Shield}
\begin{figure}[!h]
\begin{center}
	\includegraphics[width=0.5\linewidth]{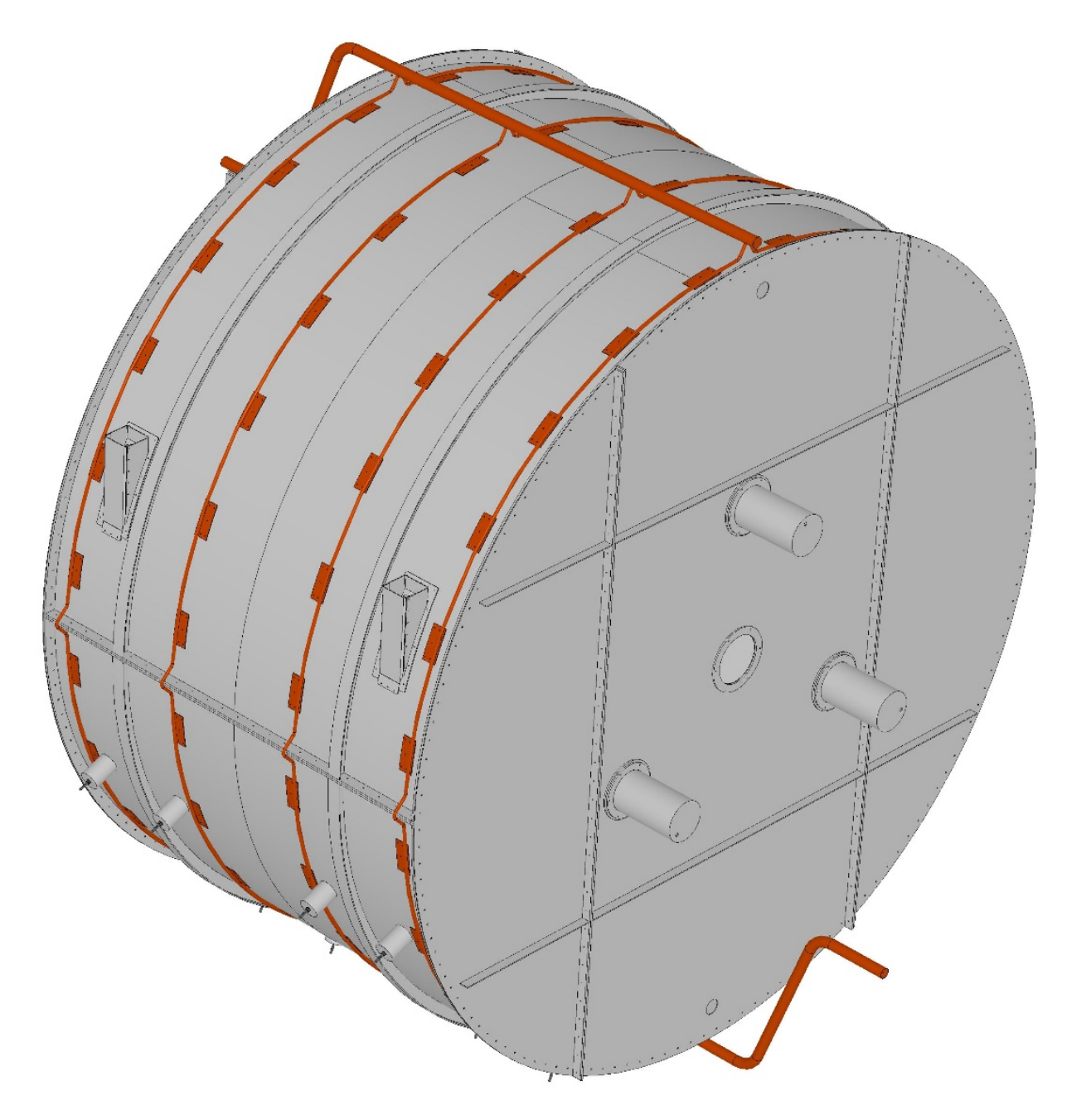}
	\caption{The shield.}
	\label{fig:shie}
\end{center}
\end{figure}

The cylindrical shield (Fig. \ref{fig:shie}) will be made of aluminum alloy AA 5083 H112 (Ø4,220 x 2,500 mm). It will 
cover the cavity without contact, while will lay on the vacuum vessel by means of sixteen adjustable pins. The shell 
of the shield will be made of two halves (4 mm thick) with longitudinal flanges and ribs (10 mm thick), all welded. 
The two flat ends end will be made by three plates bolted together. On the shield there will be bolted the covers for 
the tie-rods and the motors protruding from the cavity, and the shield will be thermally connected with the tie-rods. 
The shield will be cooled by 70 K gaseous helium circulating in a circuit similar to that described for the cavity. The 
total weight of the shield will be 1,000 kg. 

\section{Mounting sequence}
In the KLOE hall, a double crane with a suitable lifting capacity is available.  After mounting the rails inside the 
magnet, the vacuum vessel shell will be inserted by a suitable mounting tool, composed by two stands and a 
trolley running along a rail on ball bearing mountings. The shield and the cavity shells will then be inserted by 
the same tool. Each shell will be aligned and put or hanged on its supports. Then the ends of the cavity will be 
bolted, the tuners put in position and finally the motors and the counterweights. The alignment of the components 
during all the mounting phases should be checked by laser tracker.
\begin{figure}[!hb]
\begin{center}
	\includegraphics[width=0.5\linewidth]{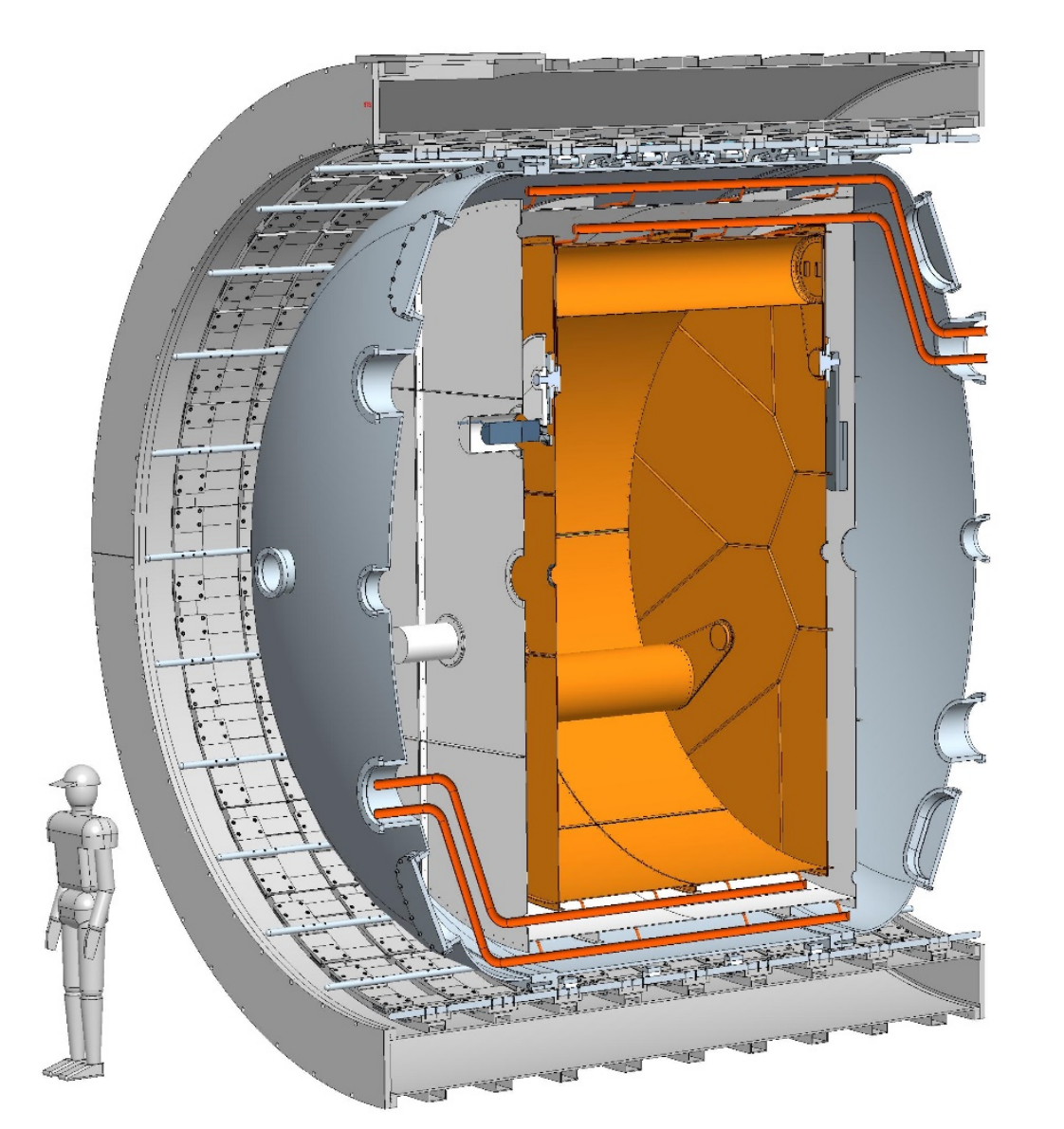}
	\caption{Cryostat assembly section.}
%	\label{}
\end{center}
\end{figure}

After mounting all the accessories, the cryogenic feedthroughs and the insulating layers, the end caps of the shield 
and of the vacuum vessel will be put in position by the crane and a suitable sling.
\begin{figure}[!h]
\begin{center}
	\includegraphics[width=0.45\linewidth]{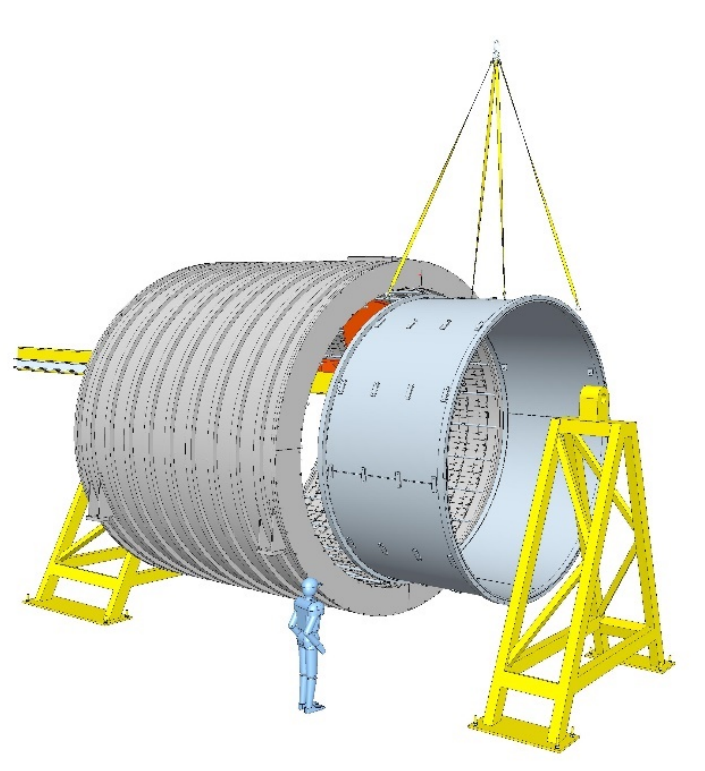}
	\hspace*{1.0cm}
	\includegraphics[width=0.45\linewidth]{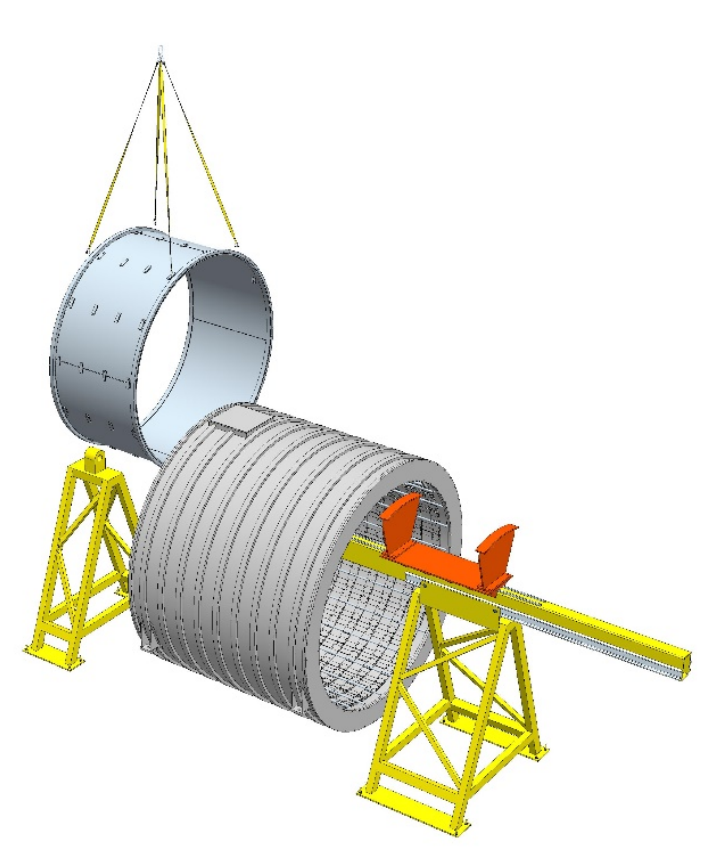}\vspace*{2.0cm}
	\includegraphics[width=0.45\linewidth]{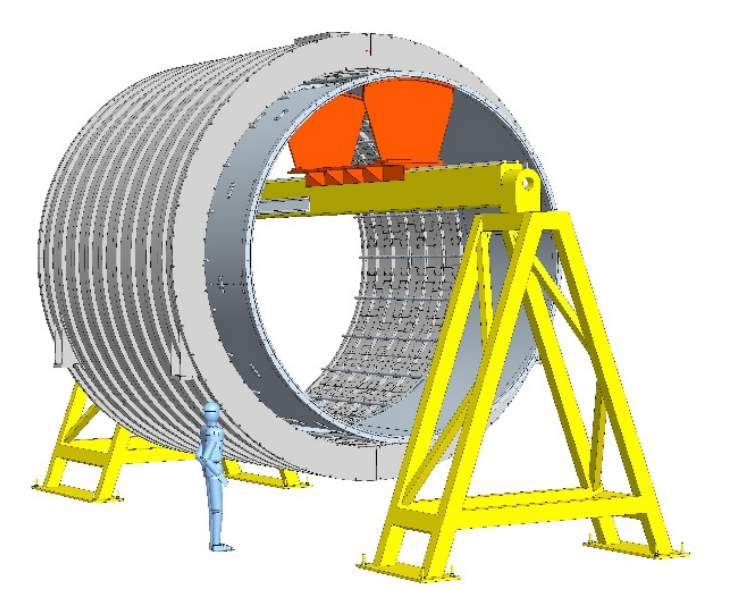}
	\hspace*{1.0cm}
	\includegraphics[width=0.45\linewidth]{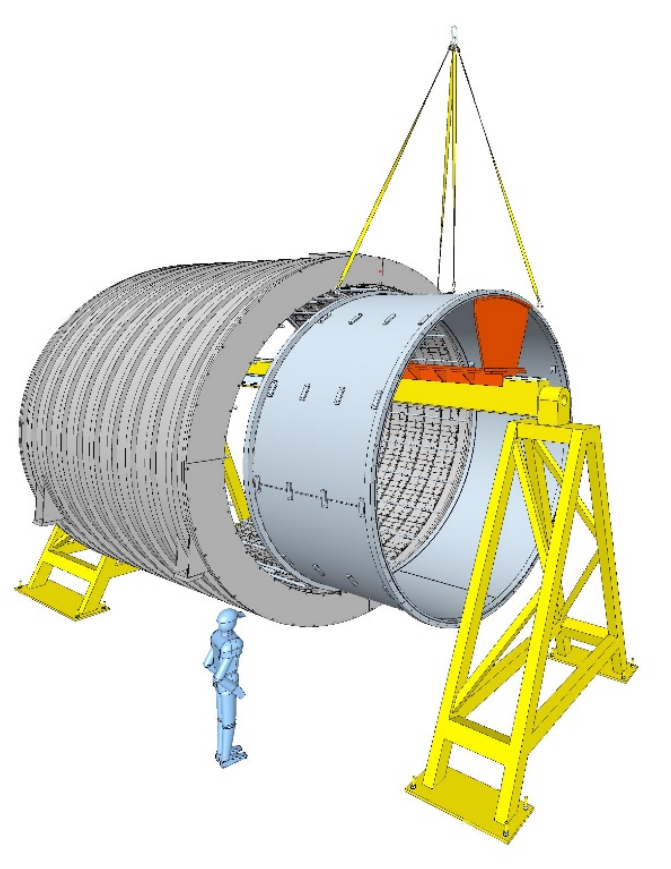}
	\caption{The various phases of the insertion of the vessel: 
	1 (top left); 2 (top right); 3 (bottom left); 4 (bottom right).}
%	\label{}
\end{center}
\end{figure}

\section{Preliminary structural simulations}
Some simulations have been performed to demonstrate the feasibility of the three main components of the cryostat. 
All the results are compliant with the functionality of the cryostat and the safety of the assembly, i.e. the stress values 
are well below the yielding, the displacement values are suitable for the functioning of the components and the assembly 
operations, even when the shells are lifted and mounted without the ends. A special buckling simulation has been 
performed to check the behavior of the vacuum vessel during vacuum leak test at the factory, i.e. without the supports 
inside the magnet.  Finally, a check according to the ASME VIII Div. 2 (4.4.5 Cylindrical Shells) shows a permissible load 
of 0.53 MPa with a safety factor of about 2.

In the figures below are reported the main results of the simulations concerning displacement, stress and buckling of  
cavity, shield and vacuum vessel.
\vspace*{2.0cm}
\begin{figure}[!h]
\begin{center}
	\includegraphics[width=0.45\linewidth]{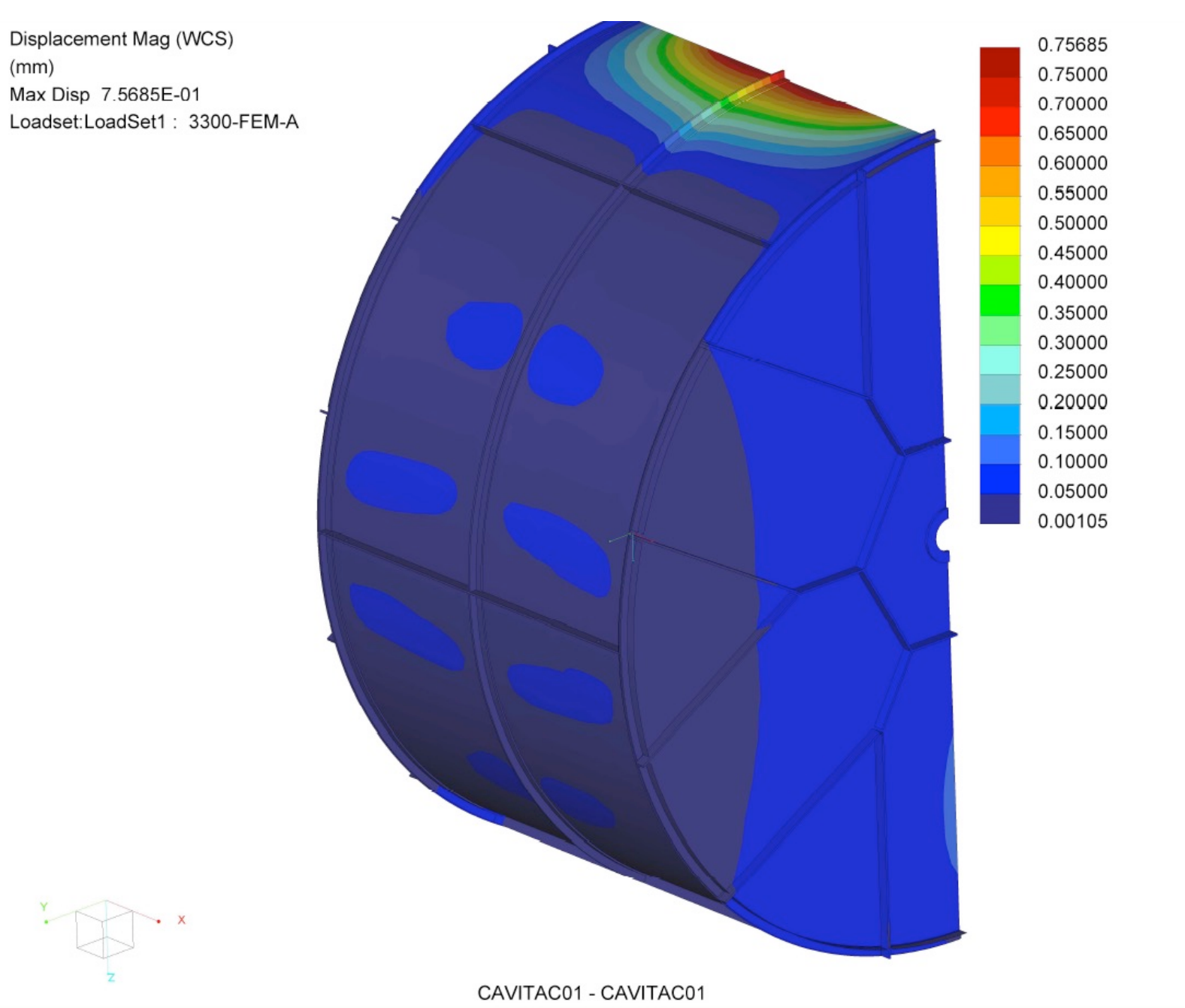}
	\hspace*{1.0cm}
	\includegraphics[width=0.45\linewidth]{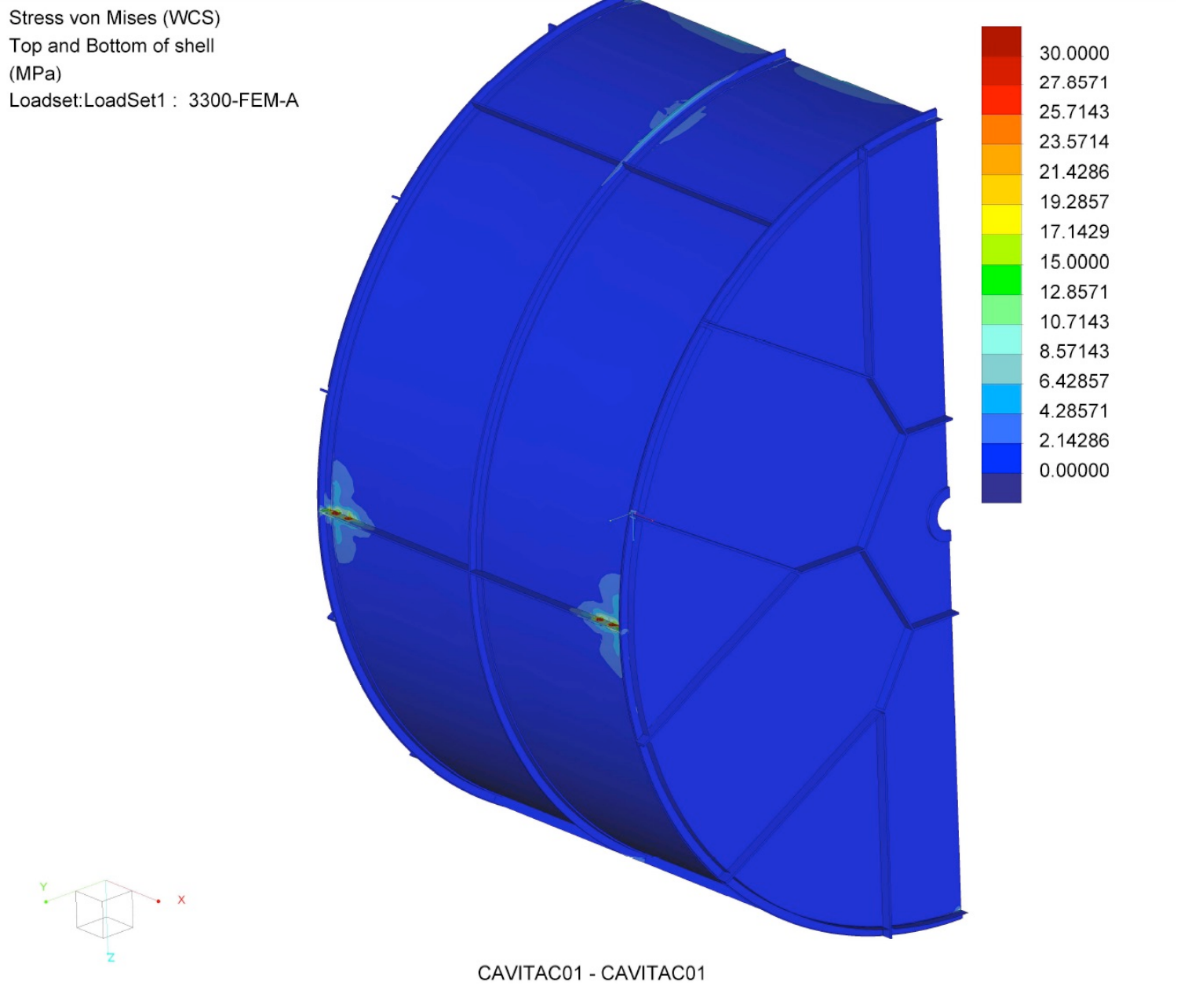}\vspace*{2.5cm}
	\includegraphics[width=0.45\linewidth]{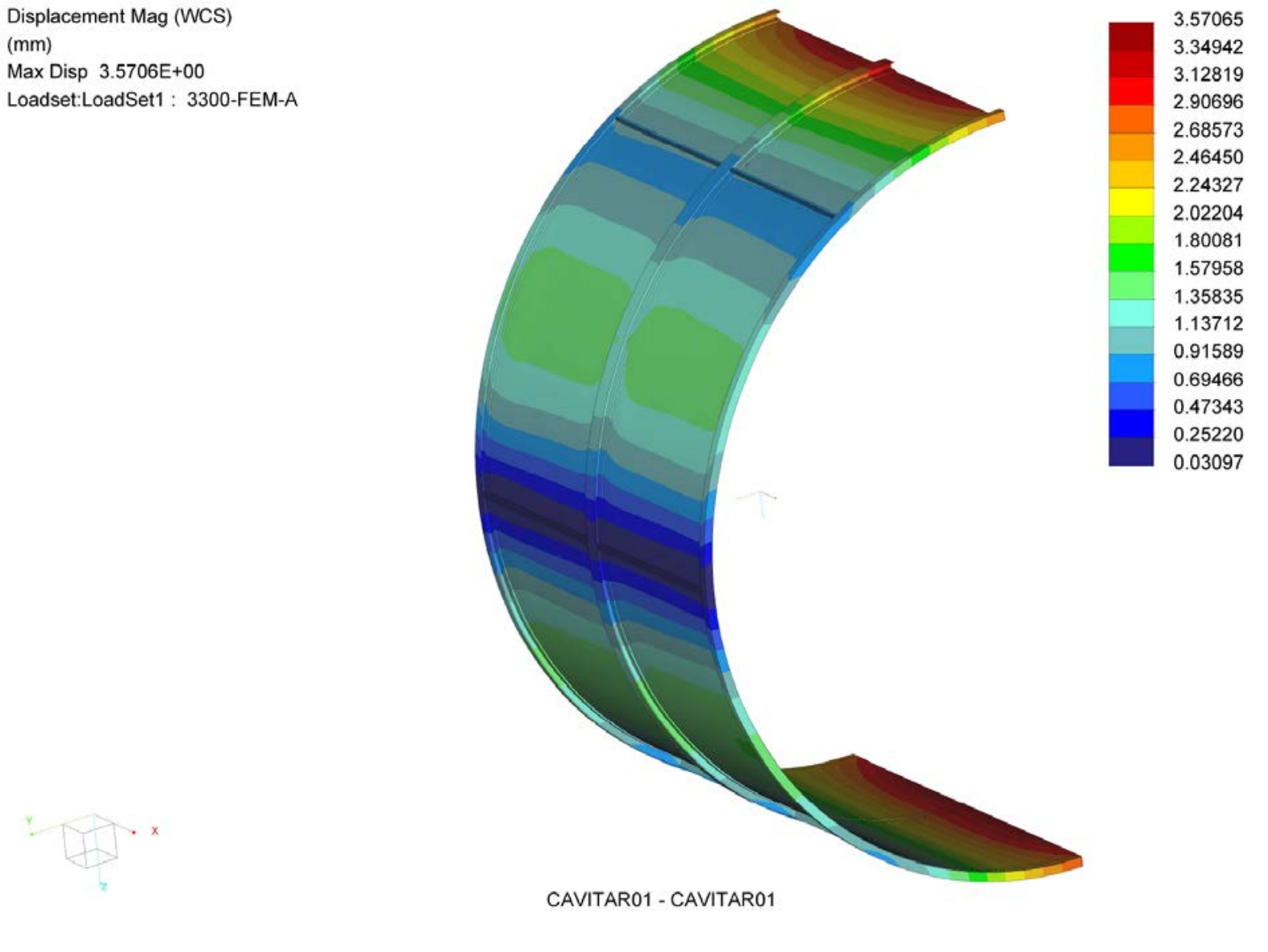}
	\caption{Top: cavity displacement (left) and stress (right) under 
	its own weight. Bottom: cavity shell displacement under its own weight.} 
%	\label{}
\end{center}
\end{figure}

\begin{figure}[!h]
\begin{center}
	\includegraphics[width=0.45\linewidth]{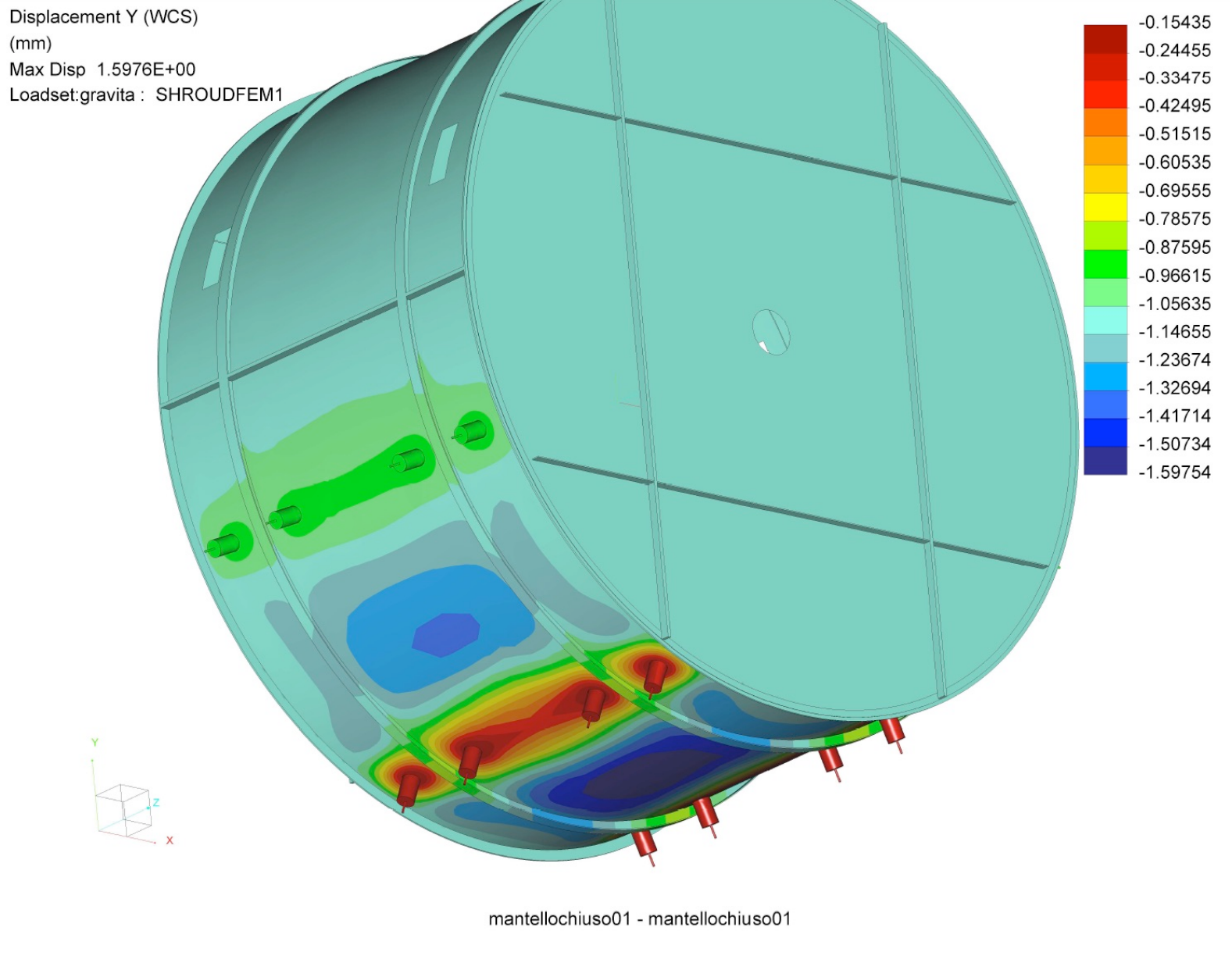}\hspace*{1.0cm}
	\includegraphics[width=0.45\linewidth]{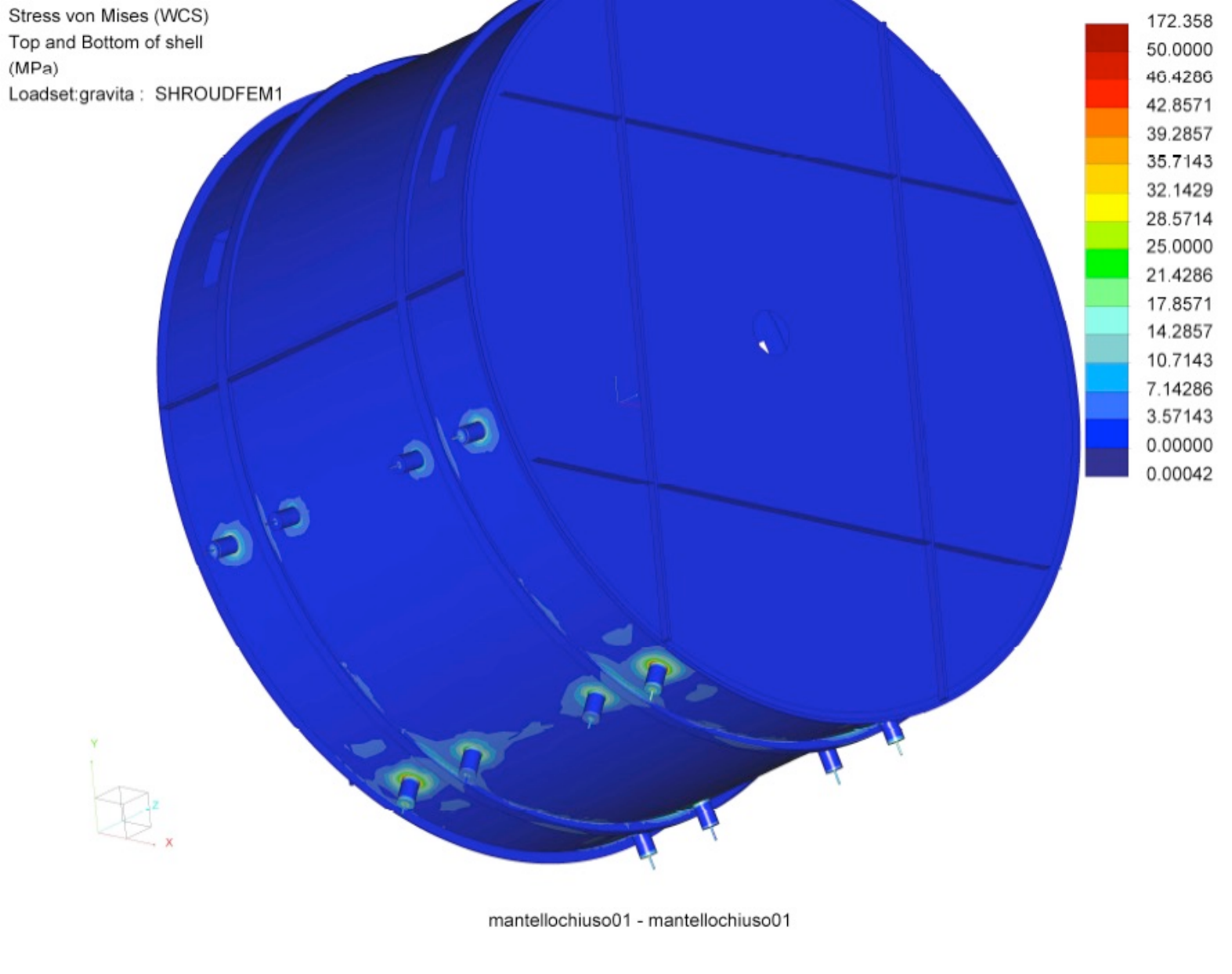}\vspace*{1.0cm}
	\includegraphics[width=0.45\linewidth]{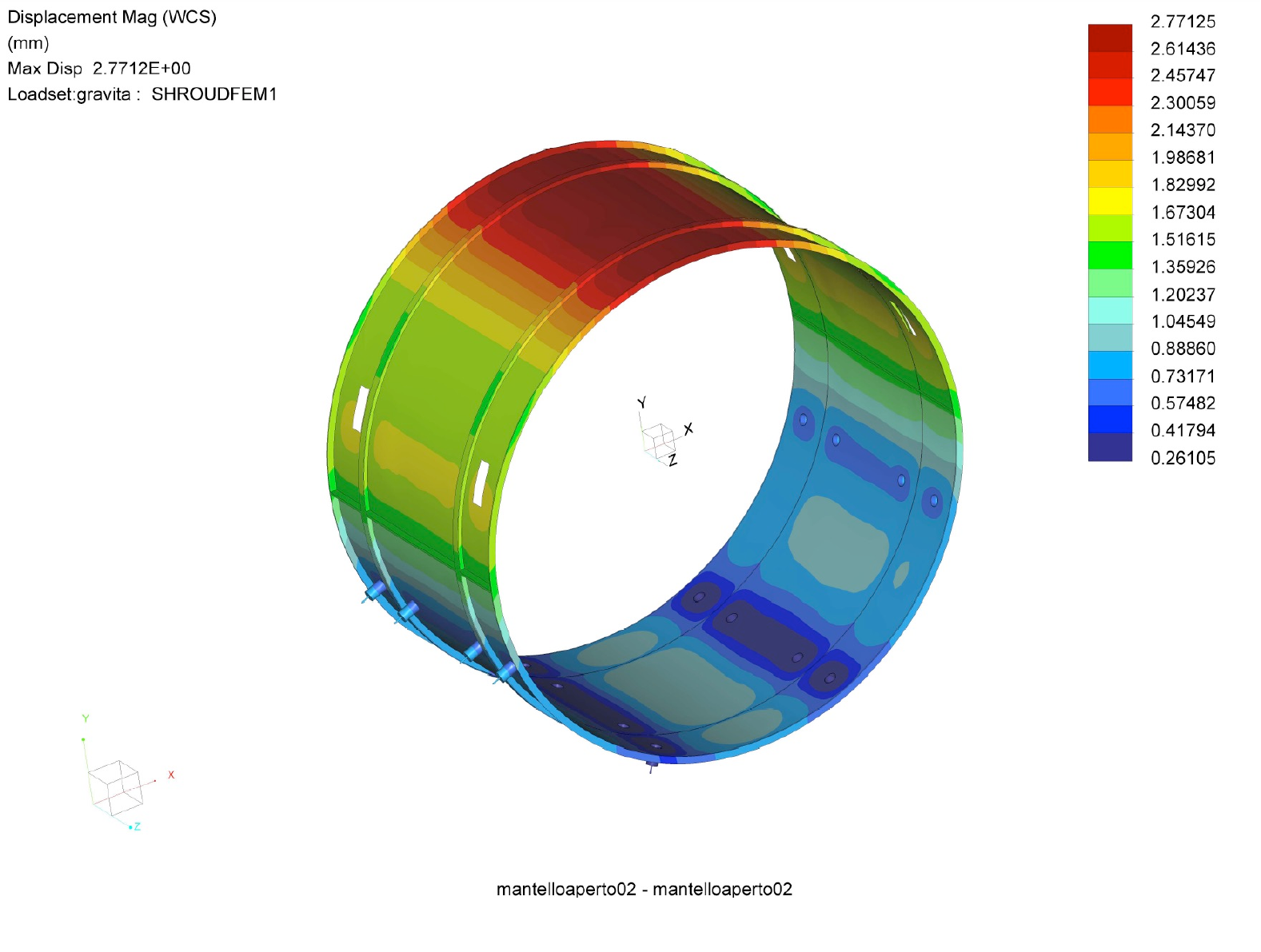}
	\includegraphics[width=0.45\linewidth]{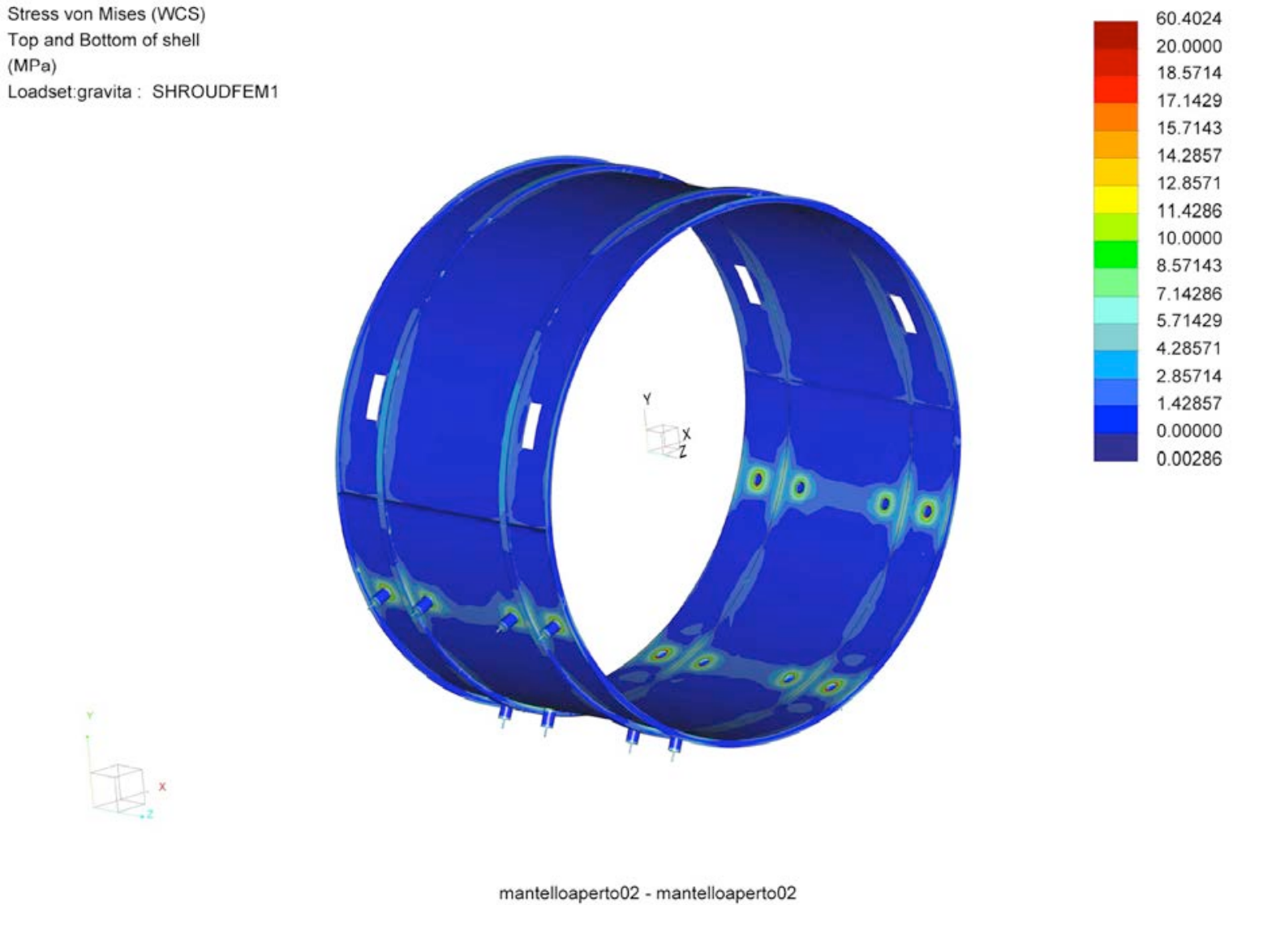}
	\caption{Top: shield displacement (left) and stress (right) under 
	its own weight. Bottom: shield displacement (left) and stress (right) under its own weight 
	without ends.} 
%	\label{}
\end{center}
\end{figure}

\begin{figure}[!h]
\begin{center}
	\includegraphics[width=0.45\linewidth]{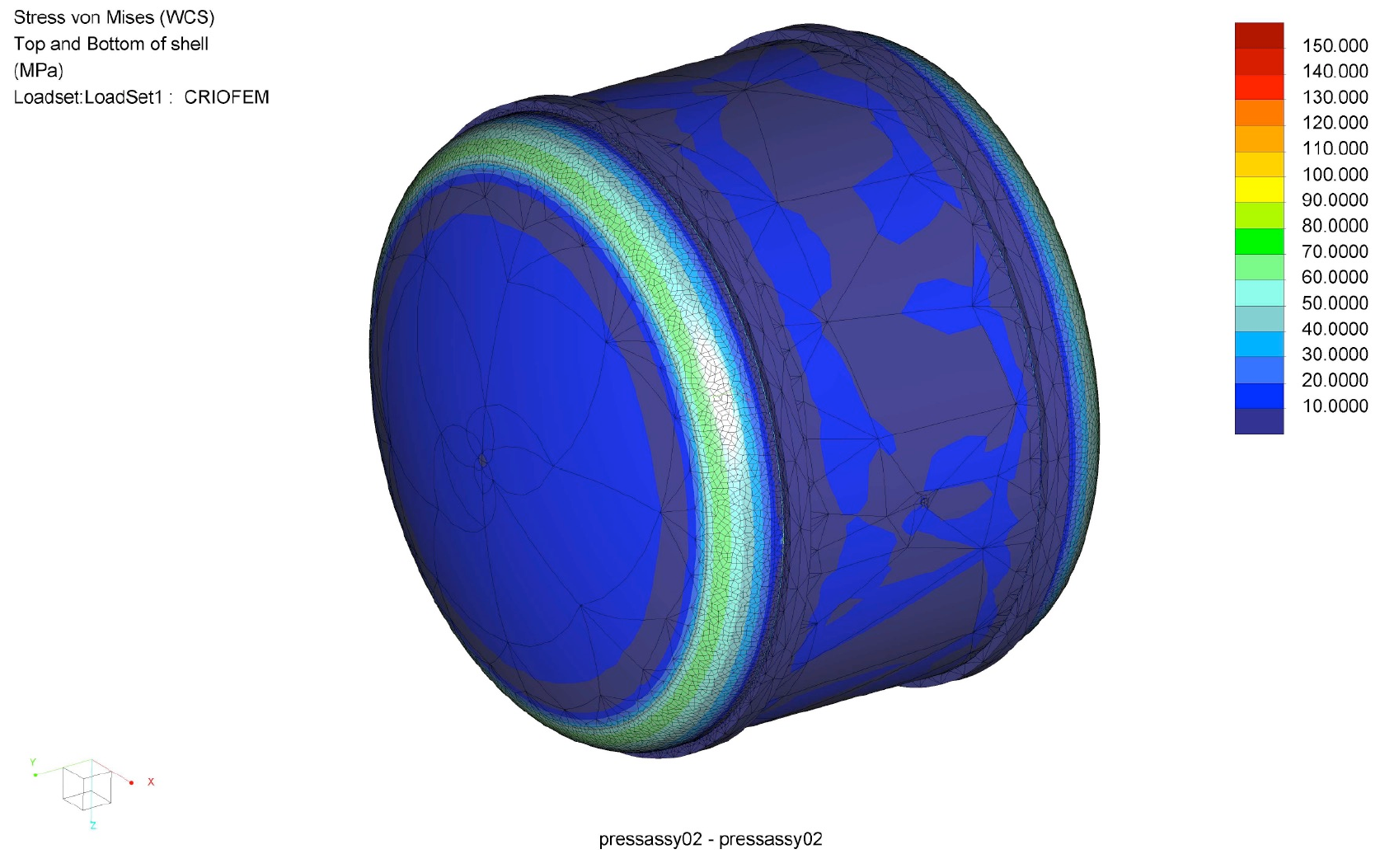}\hspace*{1.0cm}
	\includegraphics[width=0.45\linewidth]{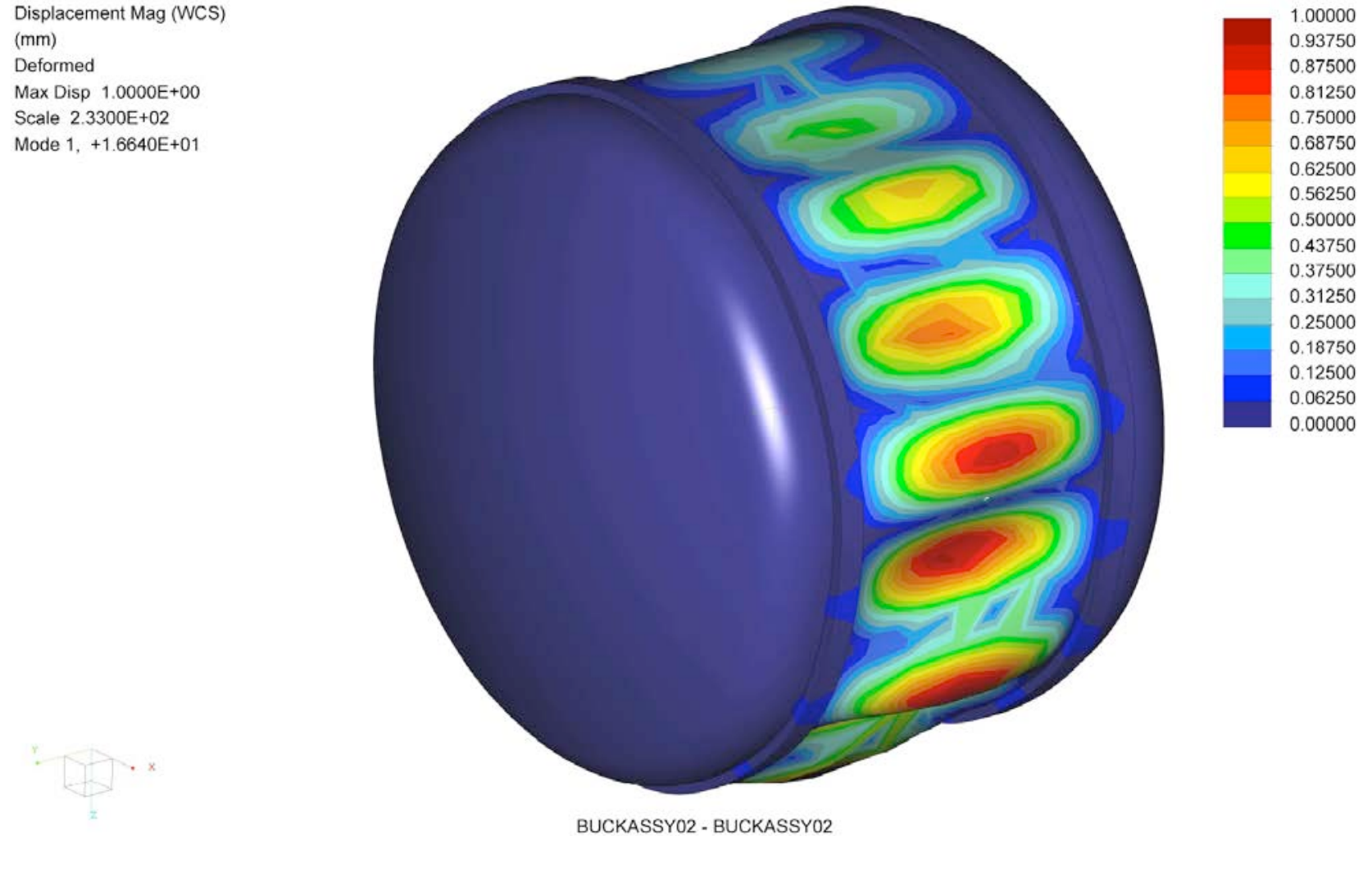}
	\caption{Vacuum vessel stress (left) and buckling (right) under 
	vacuum.} 
%	\label{}
\end{center}
\end{figure}

\chapter{Cryogenics}\label{cha:cryo}

\section{The cryogenic plant}
The cryogenic plant consists in LINDE TCF 50 Cold Box, a LINDE valve box and a KAESER ESD442 compressor, all 
connected together using a transfer line system. This plant operated at DA$\phi$NE since 1997 as a cold He supply 
for the KLOE and FINUDA detectors and for other four OXFORD superconducting anti-solenoids. The former KAESER 
compressor was replaced in 2015 with the present ESD442, which so far operated for 15000 hours.

The plant has the following nominal performance:
\begin{itemize}
 \item Liquefaction rate = 1.14 g/s;
 \item Refrigeration capacity at 4.45 K/1.22 bar = 99 W;
 \item Shield cooling capacity below 80 K = 800 W.
\end{itemize}
The plant is able to take cold the KLOE magnet and the KLASH cryostat at the same time. The capacity request for KLOE, 
cold at steady state, is:
\begin{itemize}
  \item Liquefaction rate (KLOE) = 0.3 g/s;
  \item Refrigeration capacity at 4.45 K/1.22 bar (KLOE) = 55 W
  \item Shield cooling capacity below 80 K (KLOE) = 530 W
\end{itemize}
So, the cooling capacity available for KLASH, cold at steady state, is
\begin{itemize}
  \item Liquefaction rate (for KLASH) = 0.84 g/s;
  \item Refrigeration capacity at 4.45 K/1.22 bar (for KLASH) = 44 W
  \item Shield cooling capacity below 80 K (for KLASH) = 270 W
\end{itemize}

KLASH does not need liquefaction, so the liquefaction capacity can be roughly converted to a refrigeration value, using a conversion 
factor of 100 J/g \cite{Green}. In this case 84 W (0.84 g/s $\times$ 100 J/g), to be added to the real refrigeration capacity. The effective total 
availability become:
\begin{itemize}
  \item Refrigeration capacity at 4.45 K/1.22 bar (for KLASH) = 128 W
  \item Shield cooling capacity below 80 K (for KLASH) = 270 W
\end{itemize}
If necessary, a Liquid Nitrogen line connected to the Cold Box is available to speed-up the KLOE/KLASH cool down.

A Siemens SIMATIC S7-400 PLC provide a fully automatic operation of the cooling process. A PC with a dedicated synoptic enables 
a complete remote view and control of the operations.

\section{The cryogenics layout}
\begin{figure}[!h]
\begin{center}
	\includegraphics[width=1.0\linewidth]{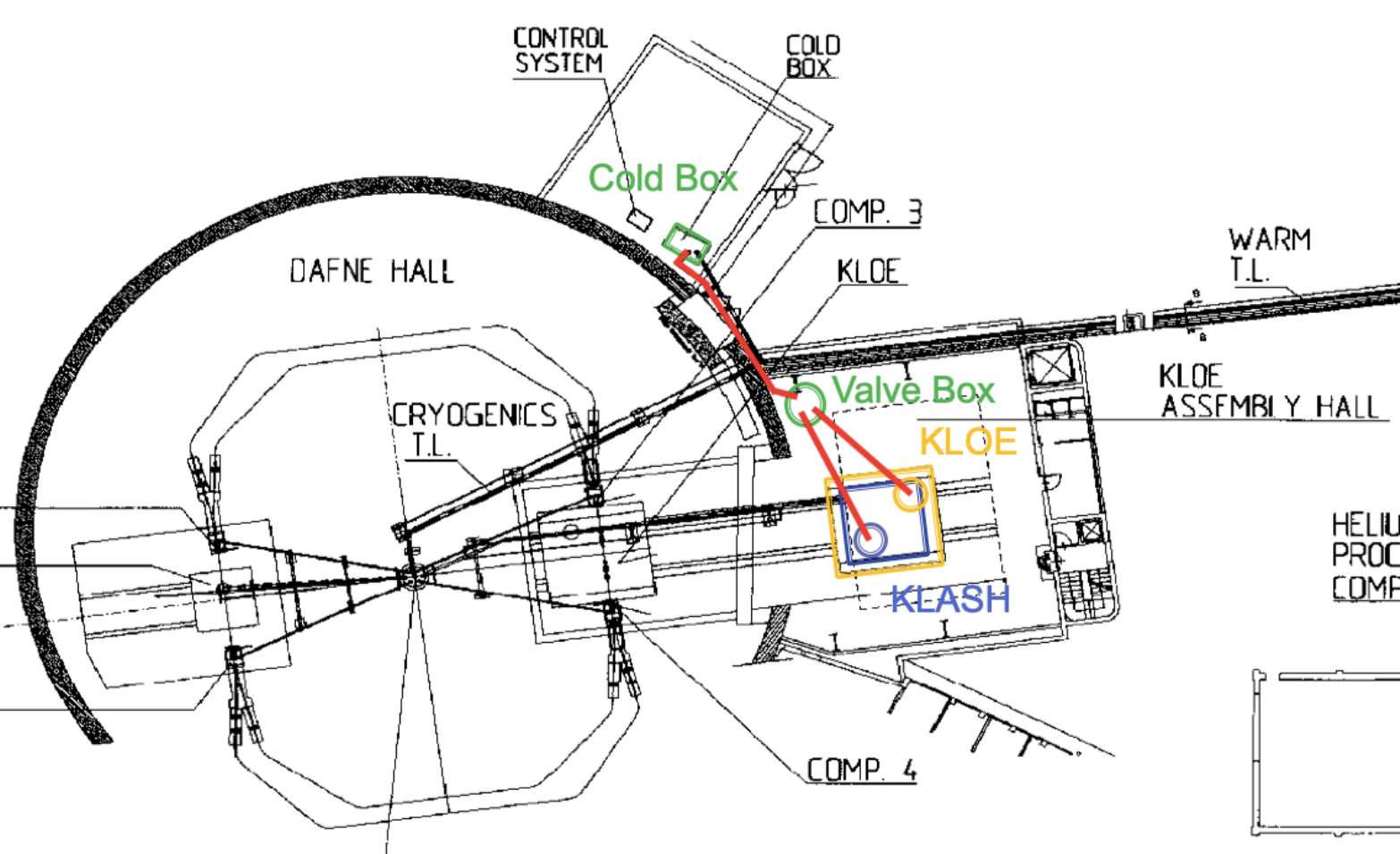}
	\caption{The cryogenics layout.}
	\label{fig:crylay}
\end{center}
\end{figure}

In Fig. \ref{fig:crylay} a view of the proposed cryogenic layout is shown. The cryoplant Cold Box is located in the cryogenic lab. 
KLOE, KLASH and the Valve Box are in the KLOE hall. In red the transfer lines are depicted. The plant compressor is out of 
the picture. The Cold Box-Valve Box transfer line consists of 5.2 SHe/70 K GHe send/return line, all contained in a vacuum 
tight tube. The Valve Box-KLOE and Valve Box-KLASH lines are similar.

The Valve Box is a distribution box in which the coming 70/5.2 K He coming from the Cold Box is splitted in different lines to the 
users by means of cryogenic valves. At present the Valve Box is placed at the center of the DA$\phi$NE hall, so several modifications 
in the plant are foreseen to adapt the plant for KLASH. The Valve Box should be moved inside the KLOE hall to minimize the 
transfer lines path. It can be placed in the platform in place of the KLOE management gas. It must be checked if the platform 
can support the Valve Box and transfer lines weights
Following the valve box repositioning, the transfer line path on between the cold and valve boxes should be modified.

New transfer lines should be made from the Valve Box to KLOE and KLASH. Their total length is of the order of 30 meters. 
Their cost is estimated in 2.5 k\euro~per meter.

\section{The KLASH service turret}
Cold Helium from the LINDE plant flows inside the transfer lines (TL) as a supercritical gas (SHe) at 5.2-5.4 K and 
about 3 bars. This is preferable respect to exiting the plant directly with the liquid, because the heating generated 
in the TLs could produce liquid-to-gas transition that will result in larger pressure drop, worsening the heat transfer 
performance. To reduce the thermal noise, the cavity must be operated at the lowest possible temperature, so it will 
be cooled at LHe temperature (4.4 K/1.2 bar), using a service turret where the SHe will be expanded through a 
Joule-Thomson valve, then collected in a vessel as a liquid and distributed to the user.

The same turret will permit the gas management during the cryostat cool down and warmup, by means of mixing 
valves and heat exchangers. 

\begin{figure}[!h]
\begin{center}
	\includegraphics[width=1.0\linewidth]{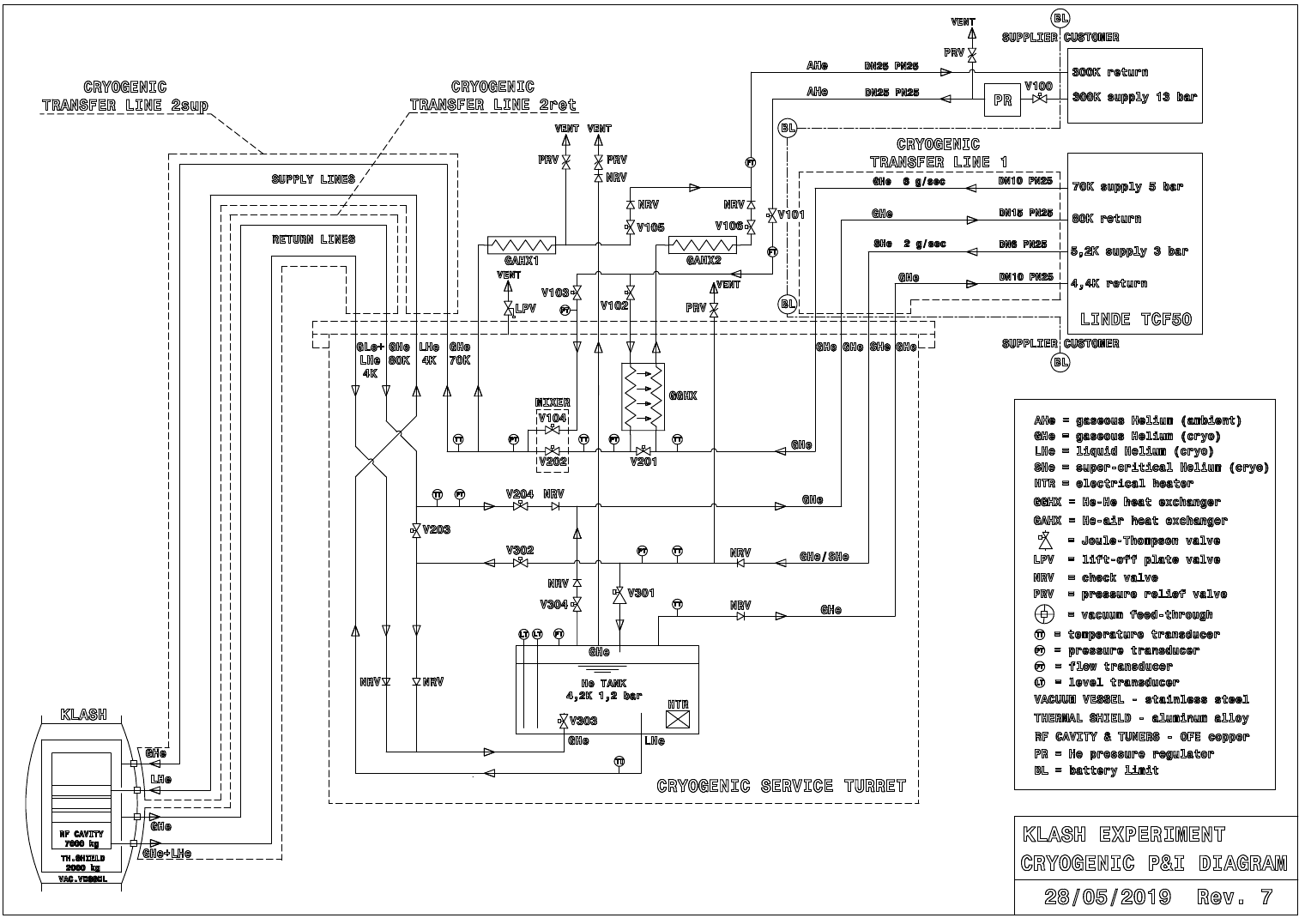}
	\caption{Service Turret layout.}
	\label{fig:turret}
\end{center}
\end{figure}

In Fig. \ref{fig:turret} the service turret scheme is depicted.

During the first part of the cool down (from ambient temperature down to about 100 K), care must be taken to avoid excessive 
thermal stresses to cavity and 70 K shield. For this purpose, the cooling gas should than controlled in temperature to values 
no more than 50 K below the users’ average temperature. To this purpose, warm (300 K) He line is splitted in two. In the first 
line the gas flows unchanged while in the other it is cooled by the 70 K GHe coming from the cryoplant by means a counterflow 
heat exchanger. The two lines are then mixed, in order to get a temperature-controlled gas flow. 

\begin{figure}[!h]
\begin{center}
	\includegraphics[width=0.6\linewidth]{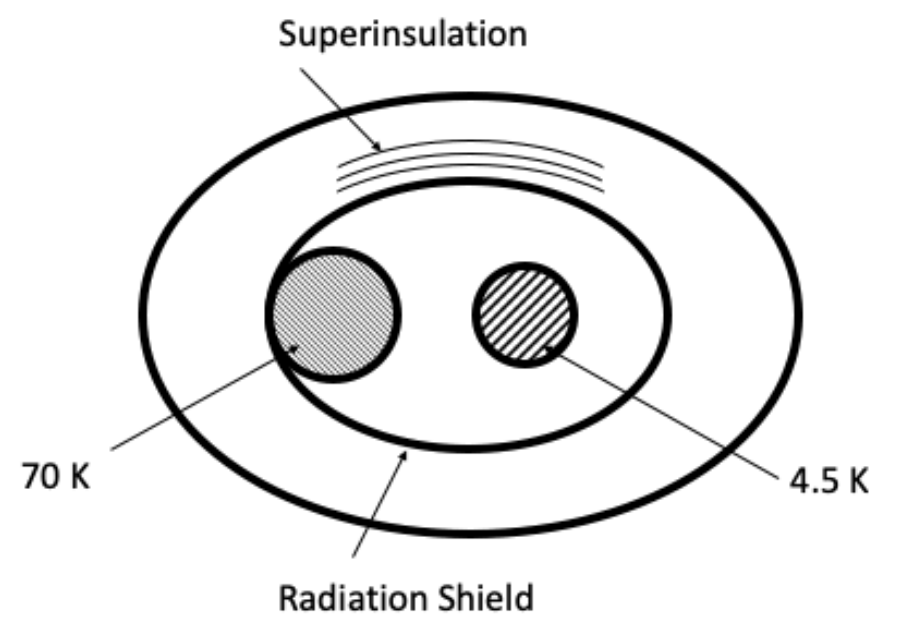}
	\caption{Service Turret/cryostat He transfer line.}
	\label{fig:transf}
\end{center}
\end{figure}

Mixed with this hydraulic scheme allows to choose the temperature of the GHe entering the cryostat at the desired value, in 
order to follow the cavity temperature decreasing. This shall avoid excessive thermal stresses due to temperature gradient.
Due to the lack of available space inside the magnet bore and the needs to access to the valves, the service turret must be 
necessarily positioned outside from the KLOE iron yoke. It will be placed on the platform above the KLOE magnet. The 
connection pipes between the turret and the cryostat should pass through some of the through holes in the iron yoke that 
was formerly used for the signal cables of the KLOE detectors. The holes size (300 x 120 mm each) is such that the four 
He pipes (send/return of 4.5 and 70 K) cannot be hosted in a single vacuum tube. The transfer line will therefore be splitted 
in two smaller tubes, the first housing the 4.5/70 K send pipes end the second the 4.5/80 K return lines. In both pipes, at the 
warmer line will be thermally connected a radiation shield used for the colder one. In Fig. \ref{fig:transf} a sketch of the transfer line 
section is shown.

The turret will be supplied with Helium coming from the LINDE plant valve box, using the exit piping that was formerly used for 
the FINUDA magnet cooling. The turret vacuum will be in connection with the cryostat one by means of the transfer lines, so 
its pumping will be provided by the cryostat pumping system. A number of sensors are required to control the Helium properties. 
In Fig. \ref{fig:crylay} the position of temperature (TT), pressure (PT), liquid Helium level (LT) and flow measurement sensors (FT) are 
indicated. Flow sensors are positioned outside of the cryostat. Pressure sensors are placed at room temperature, connected 
with the vacuum tank. All the cold sensors (thermometers and level probe) shall have a spare, to allow continuous operation 
in case of a sensor fault.

\section{The KLASH cryostat cryogenics}
\begin{figure}[!hb]
\begin{center}
	\includegraphics[width=0.4\linewidth]{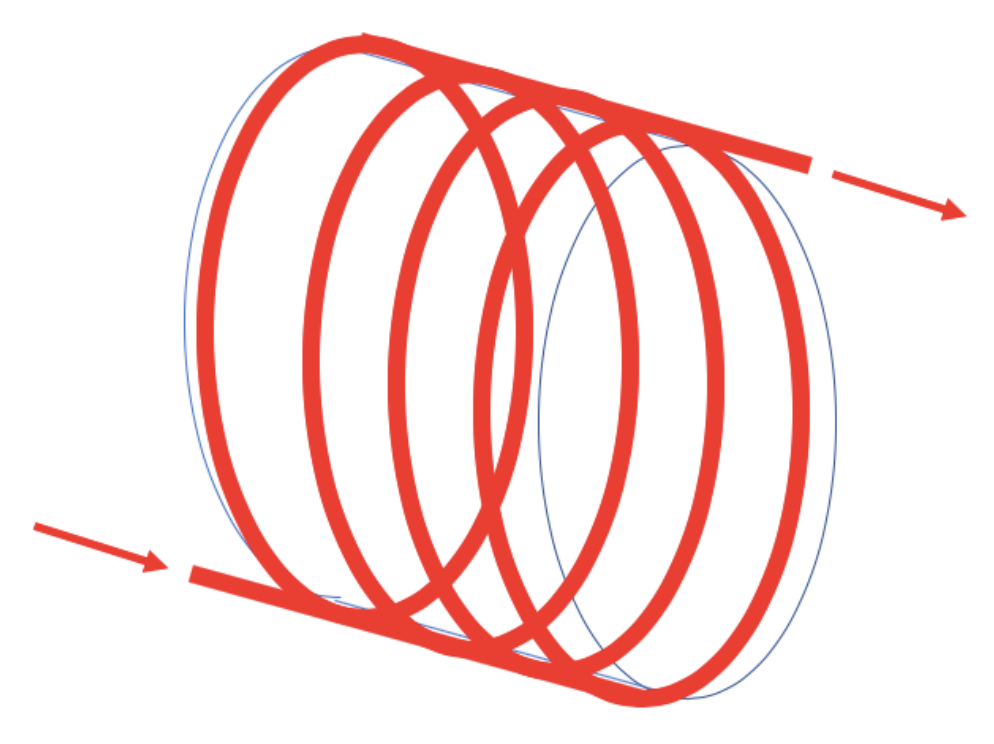}
	\caption{Sketch of the cavity/70 K shield He lines.}
	\label{fig:sketch_cav70K}
\end{center}
\end{figure}

Both cavity and 70 K shield are cooled in thermal contact with lines containing the flowing Helium. The LHe and GHe 
send lines entering the KLASH vacuum tank go on the bottom side of the cavity and 70 K shield, respectively. The 
hydraulic layout is similar on both users. A collector is placed on the bottom, parallel to the user’s longitudinal axis, 
from which start several parallel lines going radially up to the top, where another collector takes the gas to the return lines.
In Figure \ref{fig:sketch_cav70K}, a sketch of the He lines layout is shown.

\section{The KLASH cryogenics diagnostic}
All the sensors (temperature, pressure, flow, Helium level) signals are read by their respective electronics and sent to a PC 
running a customized \texttt{Labview} program. The same software controls the service turret valves, in particular the 
Joule-Thomson valve where the Helium liquefies. For that valve, a software proportional control manages the opening 
of the valve, taking the liquid level constant inside the turret vessel. A set of temperature and pressure ranges must be 
defined in order to create warning/alarm messages in case of anomalous values.

\chapter{RF}\label{cha:rf}

\section{Cavity Design and Tuning}

\begin{figure}[!h]
\begin{center}
	\includegraphics[width=0.8\linewidth]{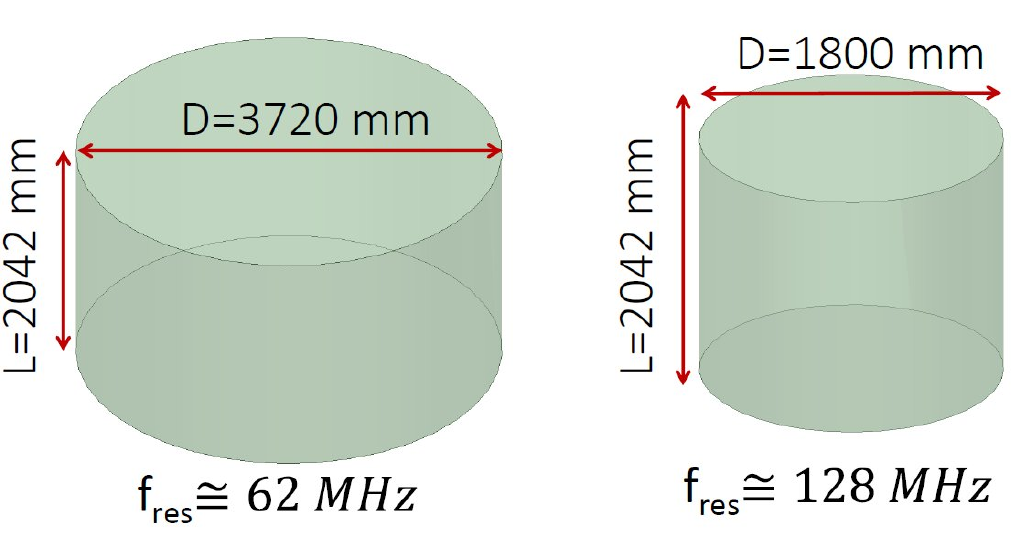}
	\caption{Sketches of the two cavities (without tuning system) proposed
	to cover the frequency range of resonant frequencies 60-250 MHz.}
	\label{fig:cav-sket}
\end{center}
\end{figure}

\begin{figure}[!ht]
\begin{center}
	\includegraphics[width=0.8\linewidth]{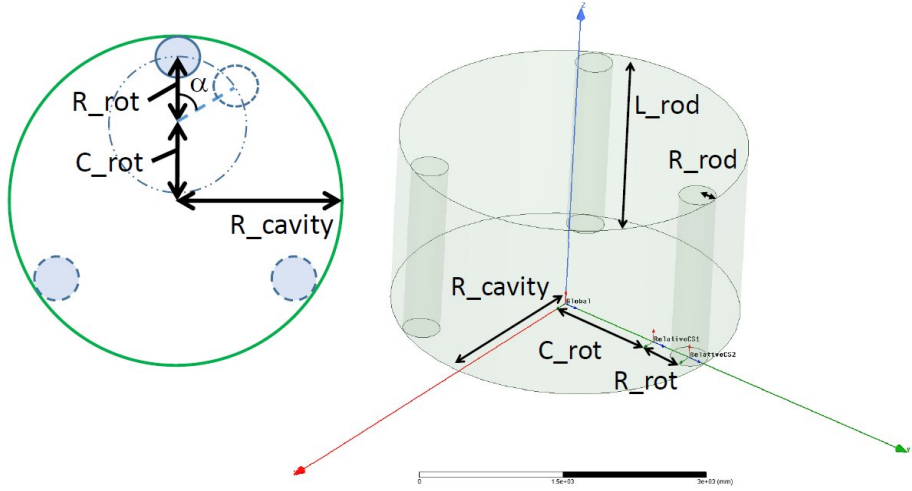}
	\caption{Sketch of the cavity with tuning system.}
	\label{fig:cavity}
\end{center}
\end{figure}

The RF design of the cavity was done using the simulation code Ansys \cite{ansys} with the goal has of designing a resonant cavity working on the mode TM010
and tunable in the range $\sim$ 60-250 MHz. This wide range of tuning cannot be covered with a single cavity  and we decided to implement two cavities with their
own tuning systems. The first one will cover the range $\sim~60-150$ MHz while the second one the range $\sim$ 130-250 MHz with an overlap of few tens of MHz.
The two cavities have different diameter and are schematically represented in Fig. \ref{fig:cav-sket}.
The tuning system is based on the use of metallic movable rods, as schematically represented in Fig. \ref{fig:cavity} where the optimized case of three rods is represented.

A similar tuning system has been already adopted in \cite{ADMX,Lee,Stern}. We have investigated different possible cases in term of number of rods, dimensions and positions
and the final parameters are reported in the Table \ref{tab:param} for the two different cavities. Each rod of radius $R_\mathrm{rod}$ can rotate (angle~$\alpha$) around the
centre $C_\mathrm{rot}$ moving towards the center of the cavity on a circular trajectory of radius $R_\mathrm{rot}$. This allows tuning the mode TM010 at different frequencies
with a rotation of the rods instead of a rigid translation, that is more difficult to implement in a real cavity inserted in a cryogenics system. From Table \ref{tab:param} it is also
possible to note that the frequency range 116-137 MHz is not covered. We will illustrate in the following the solution proposed to cover it.
\begin{table}[!ht]
  \begin{center}
  \vspace*{0.5cm}
    \begin{tabular}{c|c|c}
      Parameter& KLASH Low Frequency & KLASH High Frequency ($S$)\\
      \hline\hline
      $R$ cavity (mm)               & 1860  & 900 \\ \hline
      $R_\mathrm{rod}$ (mm)  &   200  & 100 \\ \hline
      $R_\mathrm{rot}$ (mm)   &   480  &  240 \\ \hline
      $C_\mathrm{rot}$ (mm)   &  1180  & 560 \\ \hline
      $L_\mathrm{rod}$ (mm)   &  2042 & 2042  \\ \hline
      n. tuning rods                    &   3      &    3  \\ \hline
      Frequency range (MHz)   & 66 - 166 & 137 - 242 \\ \hline
      $Q/1000$                         & 733 - 477 & 643 - 386 \\ \hline
      Form factor                      & 0.66 - 0.76 & 0.65 - 0.76 \\ \hline
      BW (Hz) @ $\beta = 1$   &   180 - 490 & 430 - 1250 \\
      \hline\hline
    \end{tabular}
  \caption{Main parameters of the two cavities with tuning system.}
  \label{tab:param}
  \end{center}
\end{table}

\begin{figure}[!h]
\begin{center}
	\includegraphics[width=0.8\linewidth]{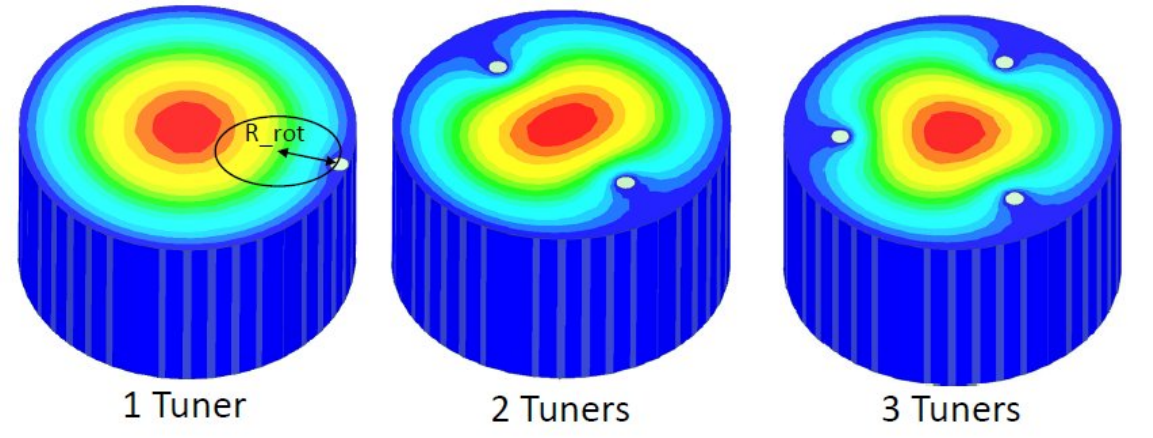}
	\caption{Sketch of the analyzed tuning rods configurations with magnitude
	of the electric field.}
	\label{fig:trod}
\end{center}
\end{figure}

The contact between the rod and the parallel plates of the cavity has to be guaranteed in order to not deteriorate the quality factor. In fact, a non-perfect contact creates, in the
gap, a capacitor and increases losses. Simulations have been performed and indicate that a gap of 1 mm can be tolerable and deteriorated the $Q$ factor by less than 5\%.
This problem has been already addressed in \cite{Lee}.  Different rods configurations have been explored and are schematically represented in Fig. \ref{fig:trod}.

\begin{figure}[!h]
\begin{center}
	\includegraphics[width=1.0\linewidth]{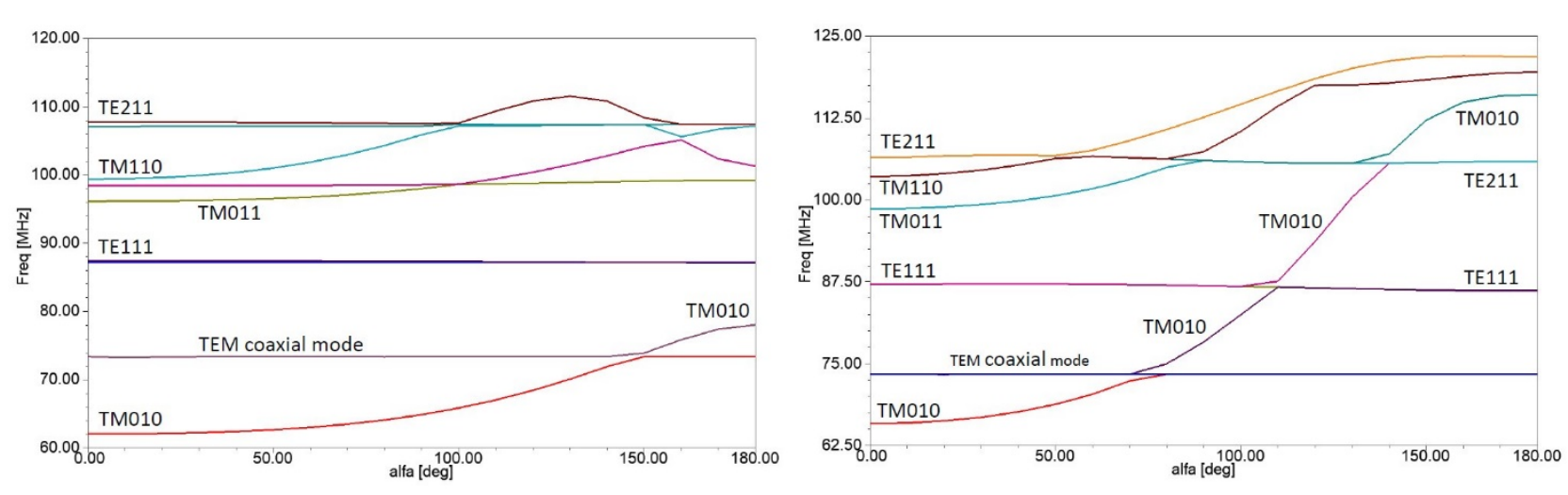}
	\caption{Frequency of the TM$_{010}$ mode and of all other identified modes, as a function of
	the angle of rotation $\alpha$ for the two cases of a single rod with $R_\mathrm{rod} = 100$ mm,
	$R_\mathrm{rot} = 780$ mm, and $C_\mathrm{rot} =980$ mm (left) and the final case
	of three rods with $R_\mathrm{rod} = 200$ mm, $R_\mathrm{rot} = 480$ mm, and
	$C_\mathrm{rot} = 1180$ mm (right).}
	\label{fig:tm010}
\end{center}
\end{figure}

We considered the cases of one, two and three symmetric rods exploring different radius, positions, and centres of rotation. For all configurations the frequency range, the
quality factor and the form factor of the working mode have been calculated and we have finally proposed the configuration with three rods of radius $R_\mathrm{rod} = 200$
mm, $R_\mathrm{rot} = 480$ mm and $C_\mathrm{rot} = 1180$ mm. This choice has been a compromise between the complexity of the system and the tunability range.
Fig. \ref{fig:tm010} shows, as an example, the frequency of the TM$_{010}$ mode and of all other identified modes, as a function of the angle of rotation $\alpha$ for the two
cases of a single rod ($R_\mathrm{rod} = 100$ mm, $R_\mathrm{rot} = 780$ mm and $C_\mathrm{rot} =980$ mm) and the final case of three rods. In the case of
three rods they are moving symmetrically. The rotation angle $\alpha = 0$ corresponds to the case of the rod near the cavity wall.

The second case is, obviously, more effective. From the plots it is quite easy to note that the proposed tuning system does not translate the frequency of all cavity modes
rigidly. There are positions of the rods for which we have an overlap between different modes. We can note, as an example, the overlap between the working mode and TE
modes and also with the modes due to TEM resonances of the coaxial line generated by the rod (inner conductor) and the outer wall (outer conductor) that we indicated as
``TEM coaxial modes''.

\begin{figure}[!ht]
\begin{center}
	\includegraphics[width=0.8\linewidth]{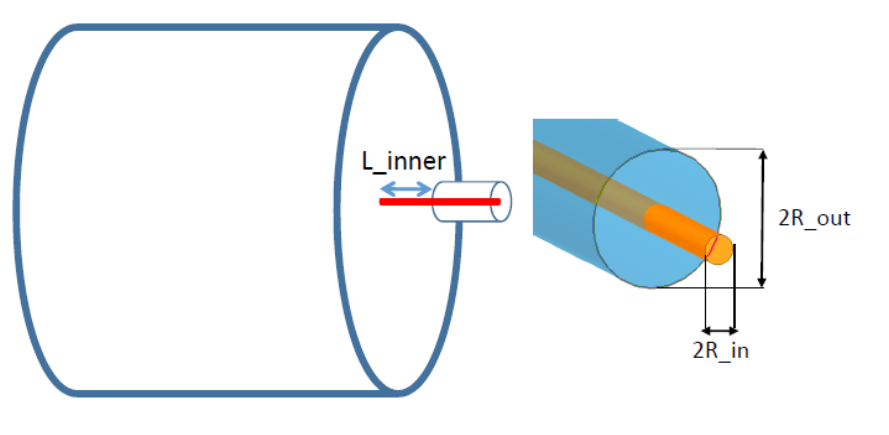}
	\caption{Sketch of the antenna coupled to the cavity modes.}
	\label{fig:ant}
\end{center}
\end{figure}

\begin{figure}[!hb]
\begin{center}
	\includegraphics[width=0.8\linewidth]{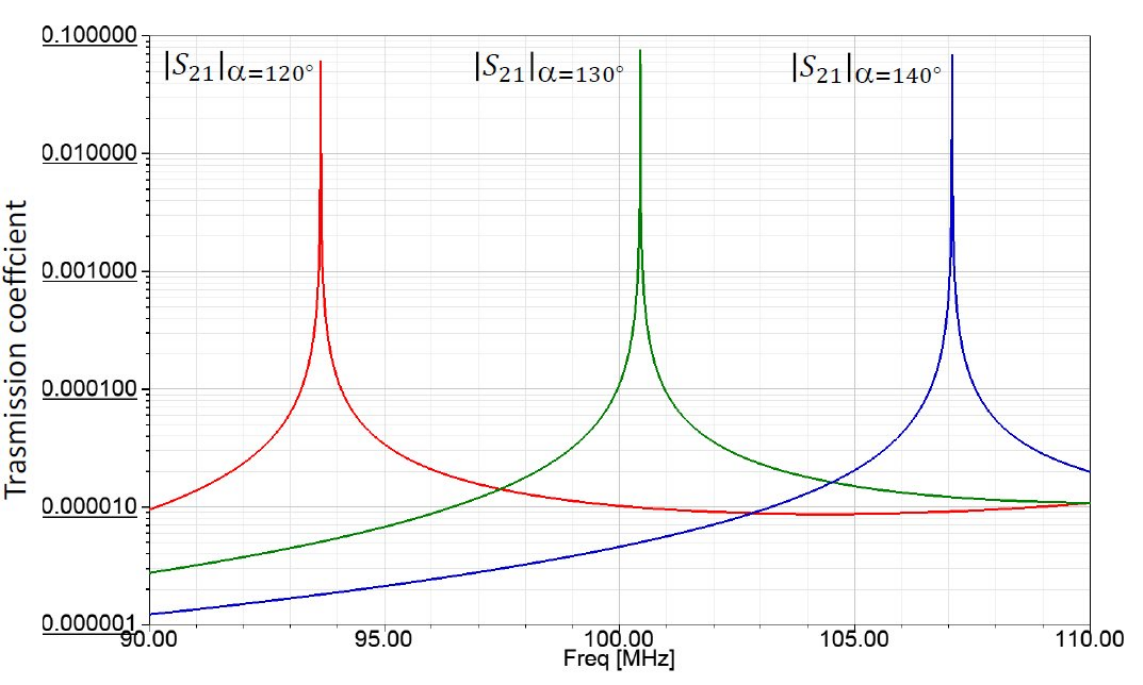}
	\caption{The transmission coefficient between two coaxial probes
	coupled to the cavity is shown for different tuner positions
	that cross the mode mixing region TM$_{010}$-TE$_{211}$.}
	\label{fig:trans}
\end{center}
\end{figure}

The overlap between modes gives a ``mode mixing'' effect that could make the detection of the mode TM$_{010}$ more difficulty in the crossing region between the modes,
as already pointed out in \cite{Lee}. The working mode TM$_{010}$ mode will be detected through a coaxial probe inserted in one of the two end-caps of the cavity and parallel
to the axis of the cavity itself, as sketched in Fig. \ref{fig:ant}.
The coupling of the probe can be varied changing the penetration of the inner conductor
($L_\mathrm{inner}$). We verified, by simulations, that, even with a standard-SMA antenna it is possible to change the coupling $\beta$ of the antenna with the
TM$_{010}$ mode, in a wide range (between 0 and 2) moving the penetration of the inner by few tens of cm.
By simulations we verified that the TM$_{010}$ mode is always very well identifiable from the probes even near the crossing region between TE-TM modes. This is due to
two reasons: first of all because the probe is coupled to the longitudinal electric field of the modes only (that is in principle zero for the TE and TEM modes) and,
secondly, because the mode couplings between the TM$_{010}$ mode and the TE ones is very weak. This is, for instance, visible in Fig. \ref{fig:trans} where the transmission
coefficient between two coaxial probes coupled to the cavity is reported for different tuner positions that crosses the mode mixing region TM$_{010}$-TE$_{211}$.
Also simulations in a smaller frequency range confirmed that the TM$_{010}$ is always very well detectable.
Mode crossing may be resolved by proper rotation of the three tuning rods or by proper insertion of dielectric tuning rods.

\begin{figure}[!h]
\begin{center}
	\includegraphics[width=0.8\linewidth]{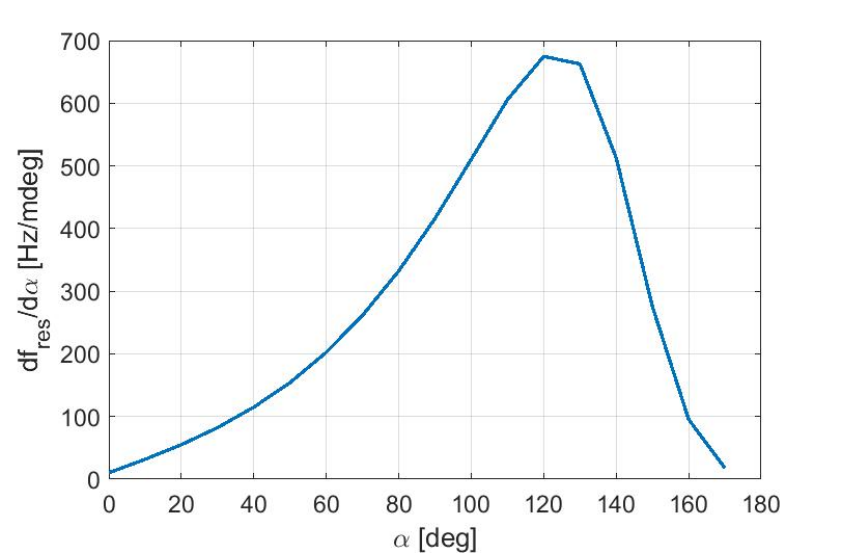}
	\caption{Sensitivity of the resonant frequency as a function
	of the rotation angle is reported.}
	\label{fig:sens}
\end{center}
\end{figure}

The sensitivity of the resonant frequency, as a function of the rotation angle, is reported in Fig. \ref{fig:sens}.
The detection of the induced signals will be done at different resonant frequencies of the cavity. The resolution in term of frequency steps that we would like to achieve is directly
proportional to the bandwidth of the cavity reported in the last line of Table \ref{tab:param}.

\begin{figure}[!h]
\begin{center}
	\includegraphics[width=0.3\linewidth]{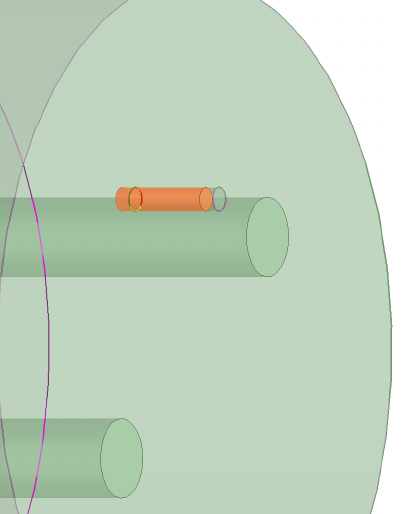}
	\caption{Dielectric tuning system for fine frequency tuning.}
	\label{fig:diele}
\end{center}
\end{figure}

In order to achieve a resolution on the order of the mode bandwidth (or lower), as can be understood from Fig. \ref{fig:sens}, the required sensitivity in term of tuning angle has to 
be much smaller than  17.5 $\mu$rad. If we want to achieve a better sensitivity in the resonant frequency tuning, we have to add another, less sensitive, tuning system. 
One possibility is to insert a small cylinder of dielectric material(alumina or sapphire) as sketched in Fig. \ref{fig:diele}. With a cylinder of 60 mm of radius, in sapphire, we achieved, 
as an example, a sensitivity of 200 Hz/mm that can be further reduced decreasing the diameter of the cylinder itself.

\begin{figure}[!h]
\begin{center}
	\includegraphics[width=0.55\linewidth]{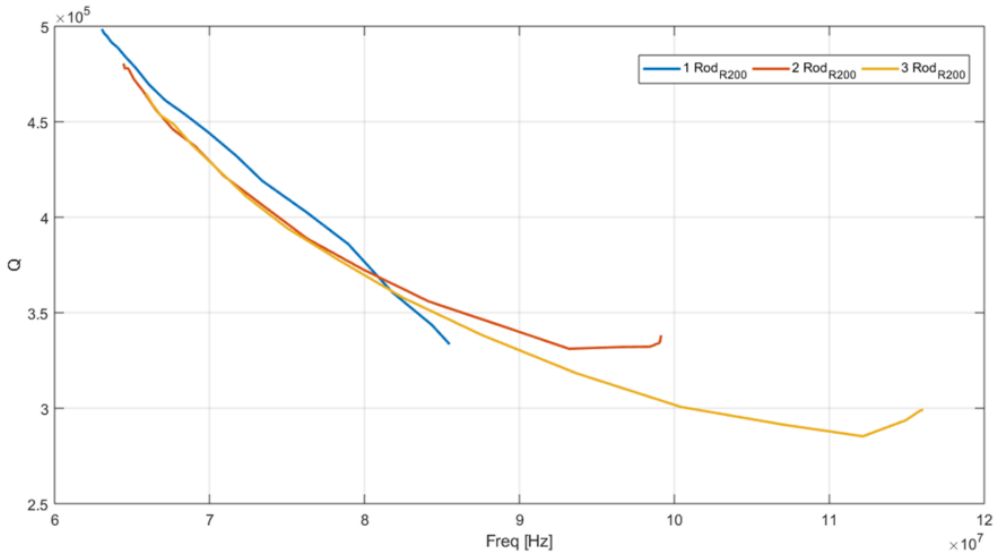}
	\caption{Quality factor and of the TM$_{010}$ mode as a function
	of frequency and for different number of rods.}
	\label{fig:qual}
\end{center}
\end{figure}

\begin{figure}[!ht]
\begin{center}
	\includegraphics[width=0.55\linewidth]{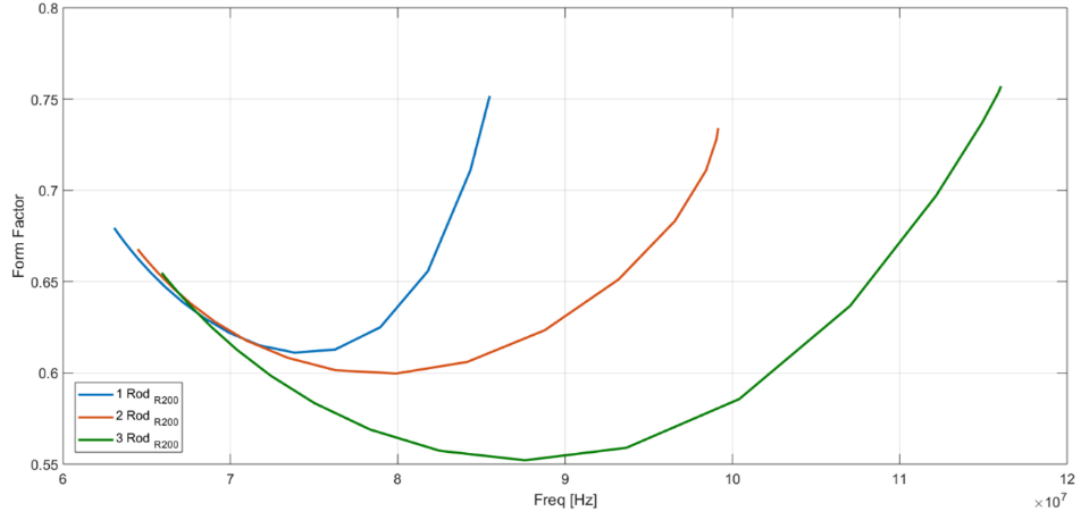}
	\caption{Form factor of TM$_{010}$ mode as a function
	of frequency and  for different number of rods.}
	\label{fig:formf}
\end{center}
\end{figure}

\begin{figure}[!h]
\begin{center}
	\includegraphics[width=0.55\linewidth]{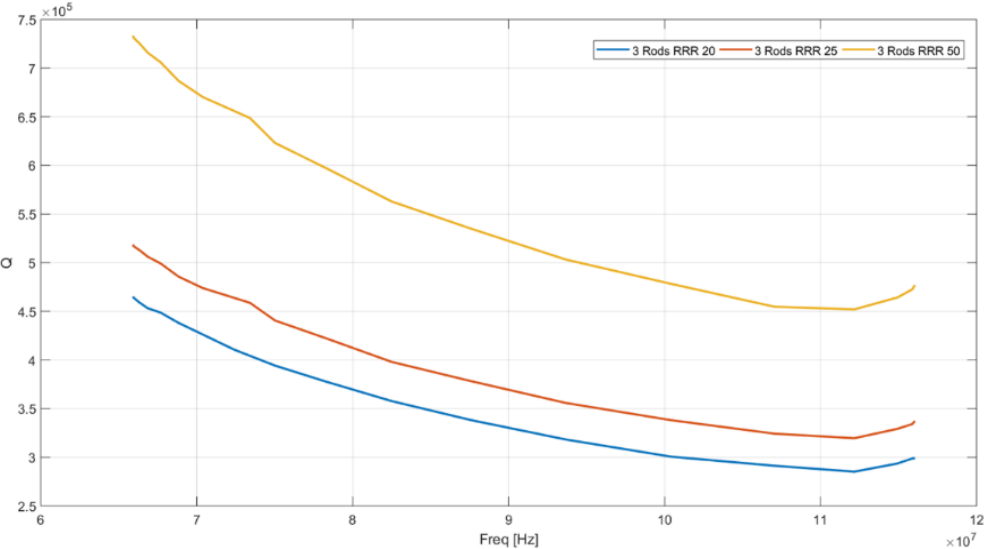}
	\caption{Form factor of TM$_{010}$ mode as a function
	of frequency and for different RRR of copper.}
	\label{fig:formf3R}
\end{center}
\end{figure}

The quality factor and of the TM$_{010}$ mode as a function of frequency and for different number of rods is given in Fig. \ref{fig:qual}. In the plot we supposed that the rods and the cavity
are made of copper with a value of RRR equal to 20 and, as a consequence, with a conductivity at 4 K of $1.16 \times10^9$ S/m \cite{Reuter}. We also calculated the form factor
\cite{Lee} and the results are given in Fig. \ref{fig:formf}.
We also considered the case of different copper RRR (see Fig. \ref{fig:formf3R})

\begin{figure}[!h]
\begin{center}
	\includegraphics[width=0.6\linewidth]{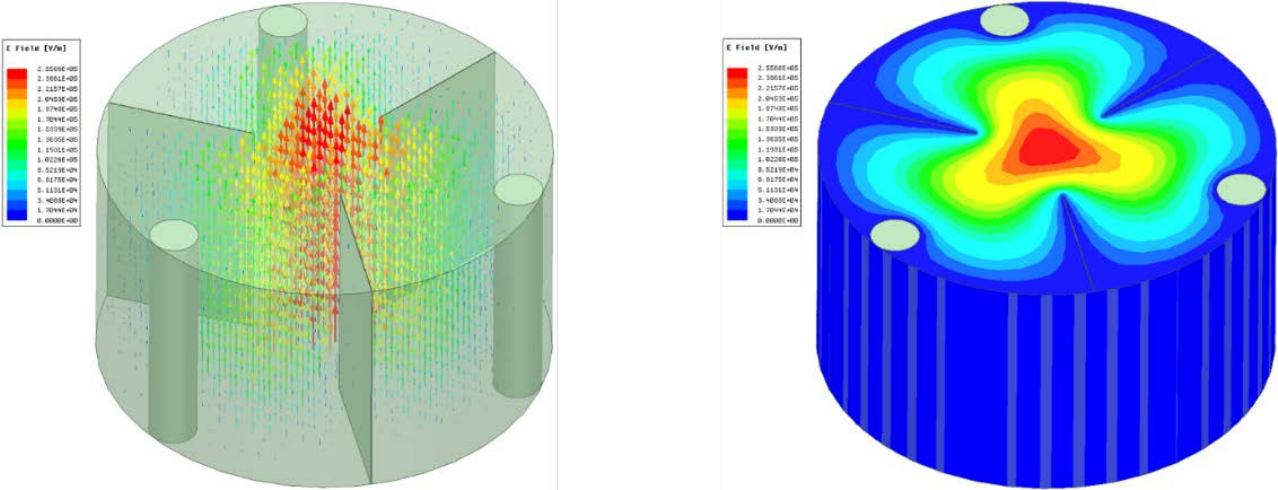}
	\caption{Modification of the cavity with insertion of metallic plates
	to cover the range up to 150 MHz.}
	\label{fig:met}
\end{center}
\end{figure}

\begin{figure}[!h]
\begin{center}
	\includegraphics[width=0.8\linewidth]{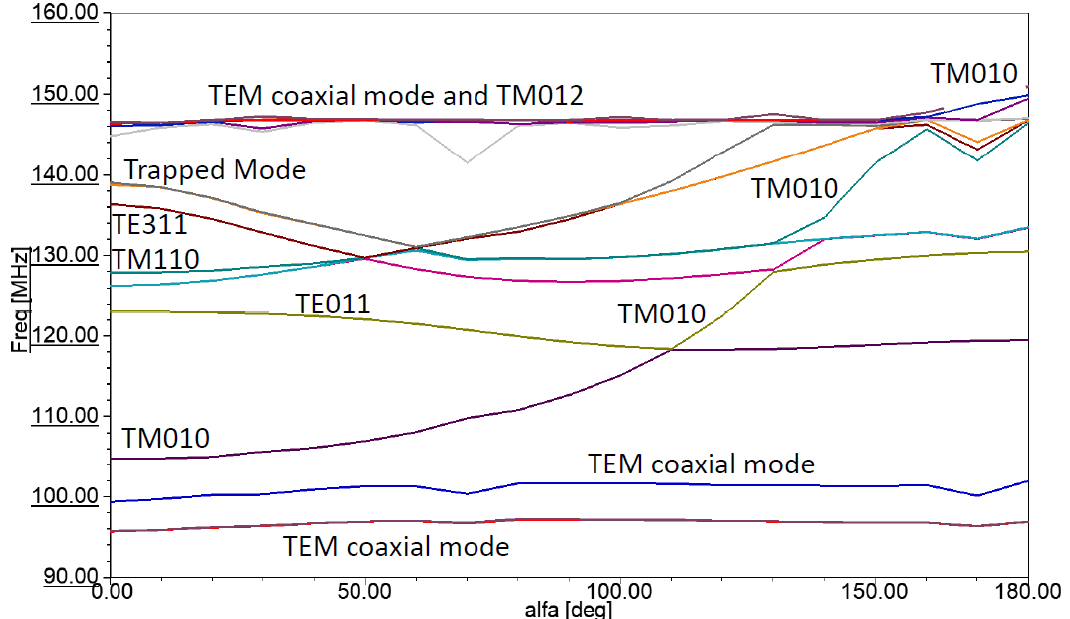}
	\caption{Frequency of the TM$_{010}$ mode and of all other identified modes,
	as a function of the angle of rotation $\alpha$ in the case of cavity with metallic plates.}
	\label{fig:freq}
\end{center}
\end{figure}

\begin{figure}[!h]
\begin{center}
	\includegraphics[width=0.5\linewidth]{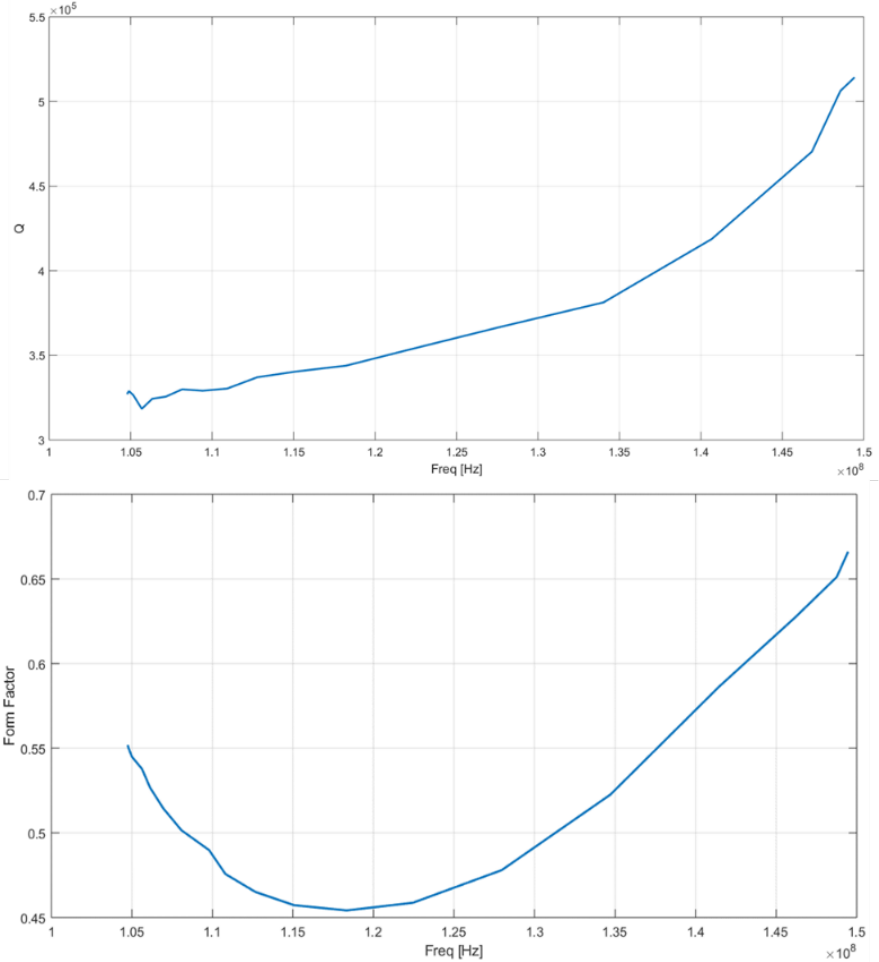}
	\caption{Quality factor and form factor as a function of frequency in the case of cavity
	with metallic plates.}
	\label{fig:quff}
\end{center}
\end{figure}

As already pointed out, to cover the frequency range up to 250 MHz, we will substitute the cavity with another one with a smaller radius. To increase as much as possible, the frequency
range to be explored we decided to choose this cavity radius in order to have a starting frequency higher than the highest frequency reachable with the cavity with larger radius. The
frequency gap of about 20 MHz (see Table \ref{tab:param}) between the two systems can be covered with a modification of the cavity with larger radius. As represented in Fig. \ref{fig:met},
the cavity can be divided in three sectors through metallic plates. A similar solution was proposed in \cite{Jeong}. The metallic rods, also in this case, can be moved on the
same circular trajectory detuning the TM$_{010}$ mode. In this condition the mode will cover a range up to 150 MHz as represented in the mode chart of Fig. \ref{fig:freq}.
In this case, due to the presence of the metallic plates the mode ensemble is more complex but, in any case, the working mode is still very well identified. Fig. \ref{fig:quff} reports the quality
factor and form factor as a function of frequency for the same case.

\begin{figure}[!h]
\begin{center}
	\includegraphics[width=0.4\linewidth]{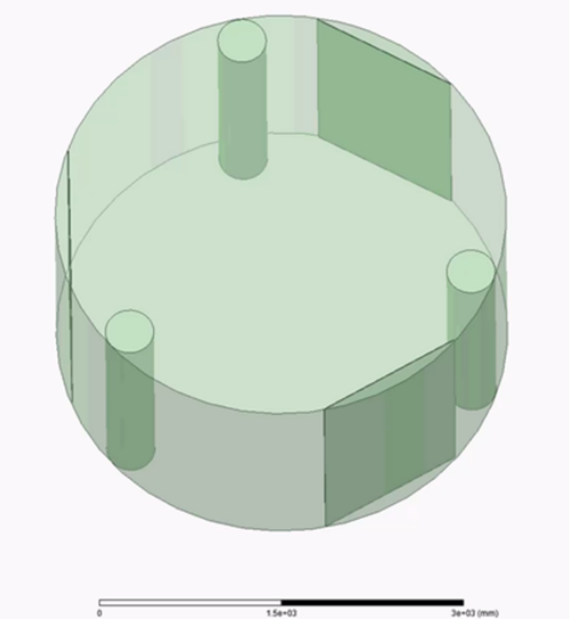}
	\caption{Cavity with the metallic plates inserted but rotated toward the outer wall of
	the cavity (open plates configuration).}
	\label{fig:met2}
\end{center}
\end{figure}

\begin{figure}[!h]
\begin{center}
	\includegraphics[width=0.5\linewidth]{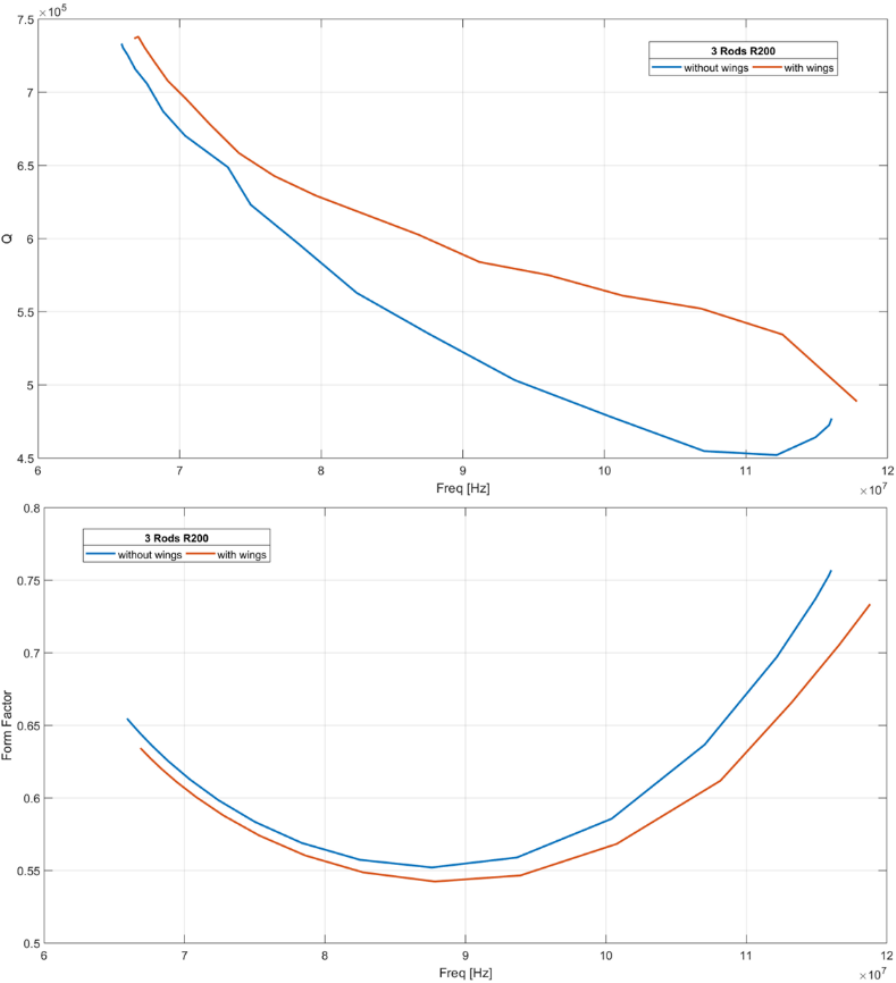}
	\caption{Quality factor and form factor as a function of frequency in the case of
	cavity with metallic plates rotated as represented in Fig. \ref{fig:met2}.}
	\label{fig:quff2}
\end{center}
\end{figure}

An alternative solution we explored is a cavity with the metallic plates inserted but rotated toward the outer wall of the cavity (open plates configuration) as schematically represented in
Fig. \ref{fig:met2}.
In this configuration the first frequency scan will be done covering a frequency interval close to the previous case, without plates, and the second scan will be done closing the plates
(that are free to rotate) as previously illustrated. Fig. \ref{fig:quff2} reports the quality factor and the form factor of the modes as a function of frequency, in the case of open plates. It is
worthwhile to note that these parameters are comparable, or even better, with respect the previous case without plates.

\begin{figure}[!h]
\begin{center}
	\includegraphics[width=0.6\linewidth]{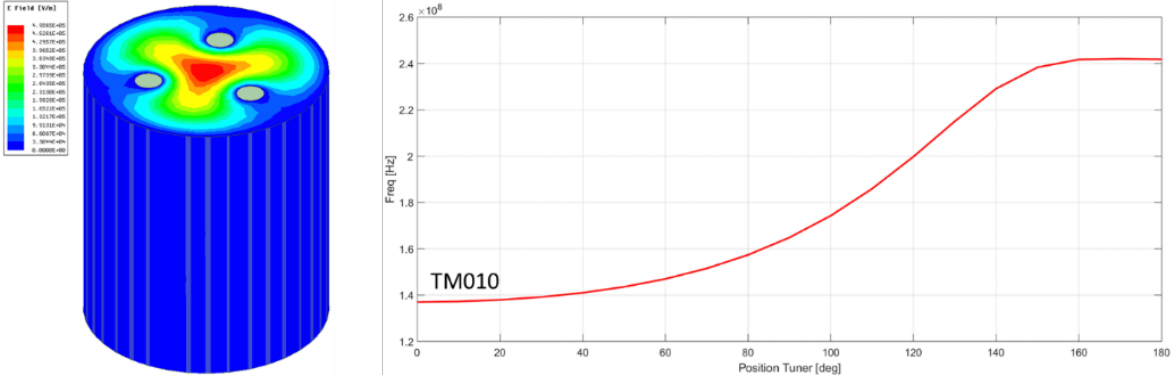}
	\caption{Cavity with smaller radius and tuning system (left); frequency of the
	TM$_{010}$ mode as a function of the tuner position (right).}
	\label{fig:cav2}
\end{center}
\end{figure}

\begin{figure}[!h]
\begin{center}
	\includegraphics[width=0.8\linewidth]{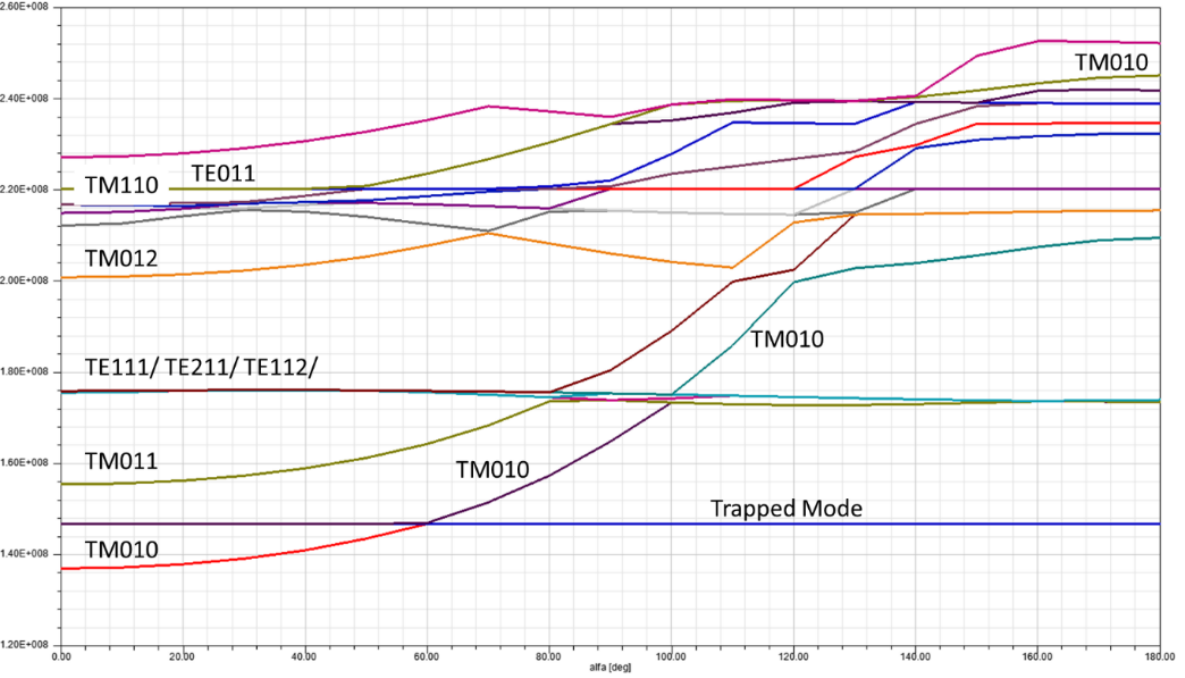}
	\caption{Complete mode mapping in the case of cavity with smaller radius.}
	\label{fig:mod}
\end{center}
\end{figure}

\begin{figure}[!h]
\begin{center}
	\includegraphics[width=0.8\linewidth]{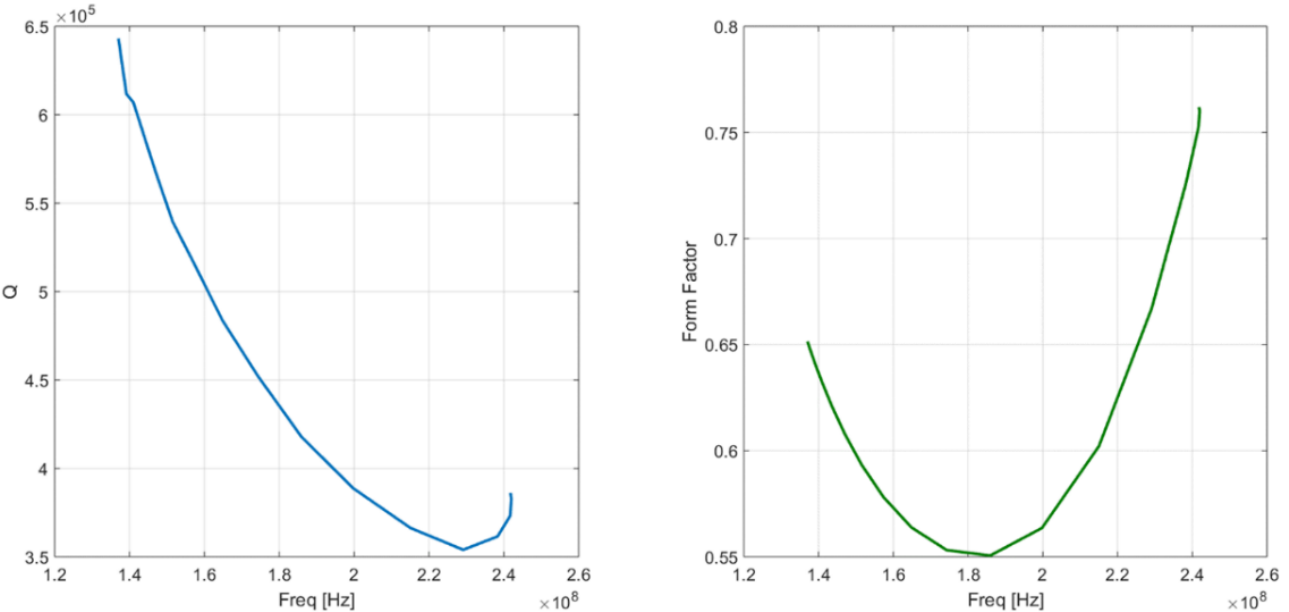}
	\caption{Quality (left) and form (right) factors as a function of frequency.}
	\label{fig:quff3}
\end{center}
\end{figure}

The last frequency range will be covered through another cavity with smaller radius and three tuning rods of radius equal to 100 mm. The cavity is schematically represented in
Fig. \ref{fig:cav2}.
In the same figure we also reported the frequency of the TM$_{010}$ mode as a function of the tuner position. The complete mode mapping is given, in this case, in Fig. \ref{fig:mod} with the
quality and of form factors reported in Fig. \ref{fig:quff3}.

\begin{figure}[!h]
\begin{center}
	\includegraphics[width=0.8\linewidth]{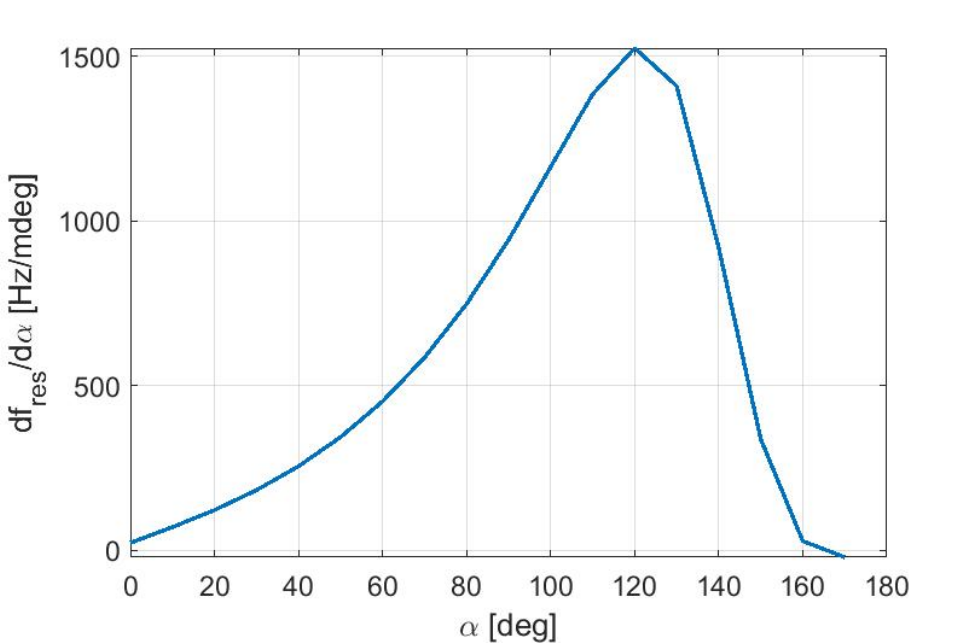}
	\caption{Sensitivity of the tuning system in the case of cavity with small radius.}
	\label{fig:sens2}
\end{center}
\end{figure}

The sensitivity of the tuning system is finally reported in Fig. \ref{fig:sens2}. Also for this case it is possible to add another tuning system for a more accurate frequency scan.

\chapter{Signal Amplification and Acquisition}\label{cha:sig_amp}
\section{SQUID}\label{sec:cha5-squid}  % Falferi

The SQUID (Superconducting QUantum Interference Device) is the most sensitive magnetic flux detector currently available and has long
been used for a wide range of low-frequency applications (\cite{Clarke}): from gravitational wave detection to biomagnetism, from
nondestructive evaluation to magnetic resonance imaging.

\begin{figure}[!h]
\begin{center}
	\includegraphics[width=0.6\linewidth]{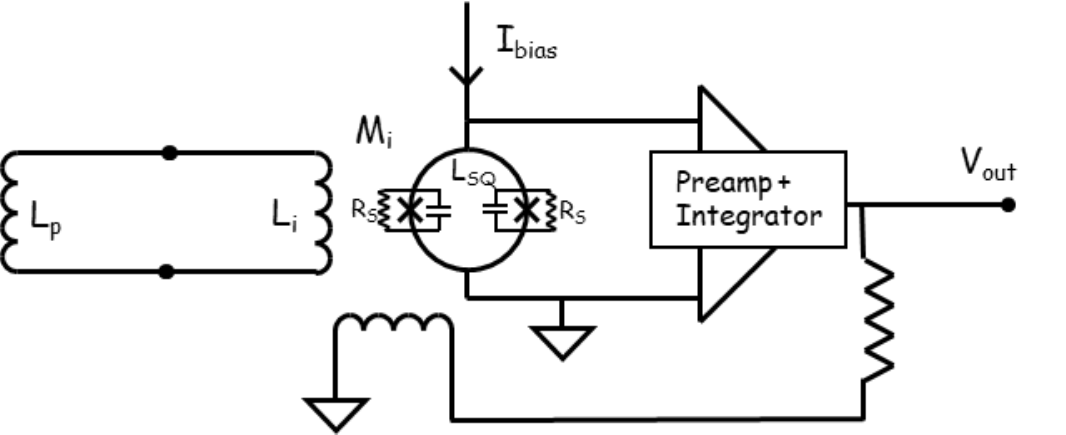}
	\caption{Flux Locked Loop dc SQUID with input coil $L_i$ connected to pick-up coil $L_p$.}
	\label{fig:squid_flux}
\end{center}
\end{figure}

More recently, SQUIDs have been used as low-noise, low-power-dissipation RF and microwave amplifiers \cite{Muck98}. Thanks to their
low dissipation, SQUIDs and SQUID-based devices can be used down to ultracryogenic temperatures where one can take advantage of the
fact that their noise scales with the temperature down to 2 - 300 mK. At lower temperatures, the coupling between the electrons in the
Josephson junction shunt resistors $R_s$ (see Fig. \ref{fig:squid_flux}) and the lattice phonons becomes very weak and the electron gas
undergoes a Joule heating due to the bias current. This heating, the hot-electron effect, causes a deviation from the linear behavior of noise
versus temperature and saturation of the SQUID noise \cite{Vinante}.

Despite this noise saturation, at frequencies below 1 GHz, the SQUID noise is more than an order of magnitude lower than that for a 
state-of-the-art cooled semiconductor-amplifier.

Two-stage commercial dc SQUIDs operating at 100 mK have shown in the audio frequency range an energy resolution referred to the input coil
($e = k_{\rm{B}} T_n/\omega_0$, where $\omega_0/2\pi$ is the operation frequency) of about 30$\hbar$ that is a factor 30 above the quantum limit.

\begin{figure}[h!]
\begin{center}
	\includegraphics[width=0.5\linewidth]{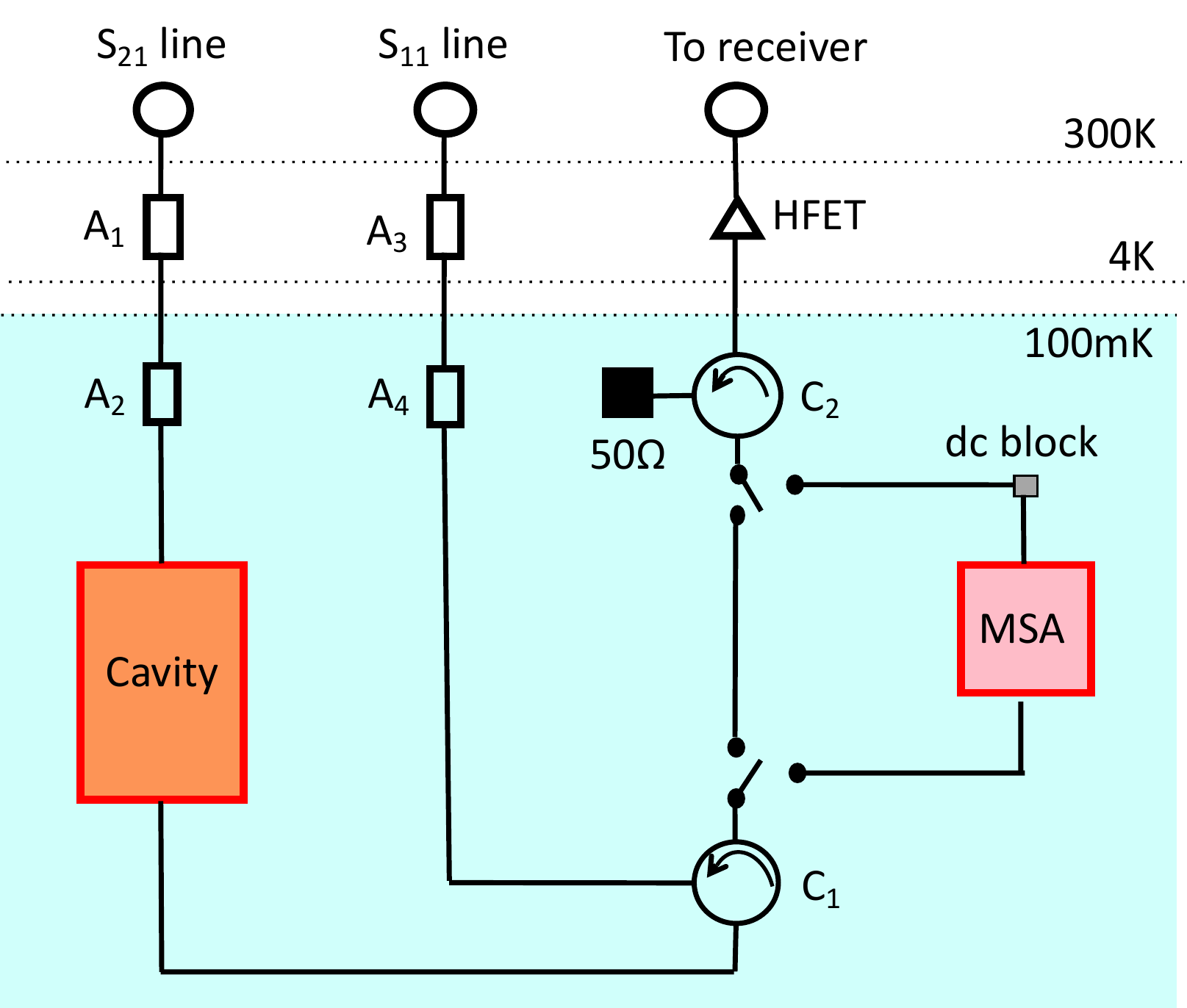}
	\caption{ADMX scheme \cite{ADMX-2}.}
	\label{fig:squid_admx}
\end{center}
\end{figure}

Since dc SQUID readout electronics have been realized with open-loop bandwidth of 300 MHz and flux-locked loop bandwidth up to 130 MHz
\cite{Drung}, the KLASH cavity readout scheme could be configured, at least for the frequency scan 65 -120 MHz, as shown in Fig. \ref{fig:squid_flux}
where the $L_p$ coil is inserted in the cavity to pick up the magnetic field. In this case, however, the high dc magnetic field (about 0.6 T) in which
the pick-up is expected to operate, would turn any micro-vibration of the apparatus into a signal out of the SQUID dynamic range. For this reason,
though the operation frequency band of KLASH is considerably lower, the reference detection scheme remains that of the ADMX experiment
(see Fig. \ref{fig:squid_admx}).

\begin{figure}[ht!]
\begin{center}
	\includegraphics[width=0.5\linewidth]{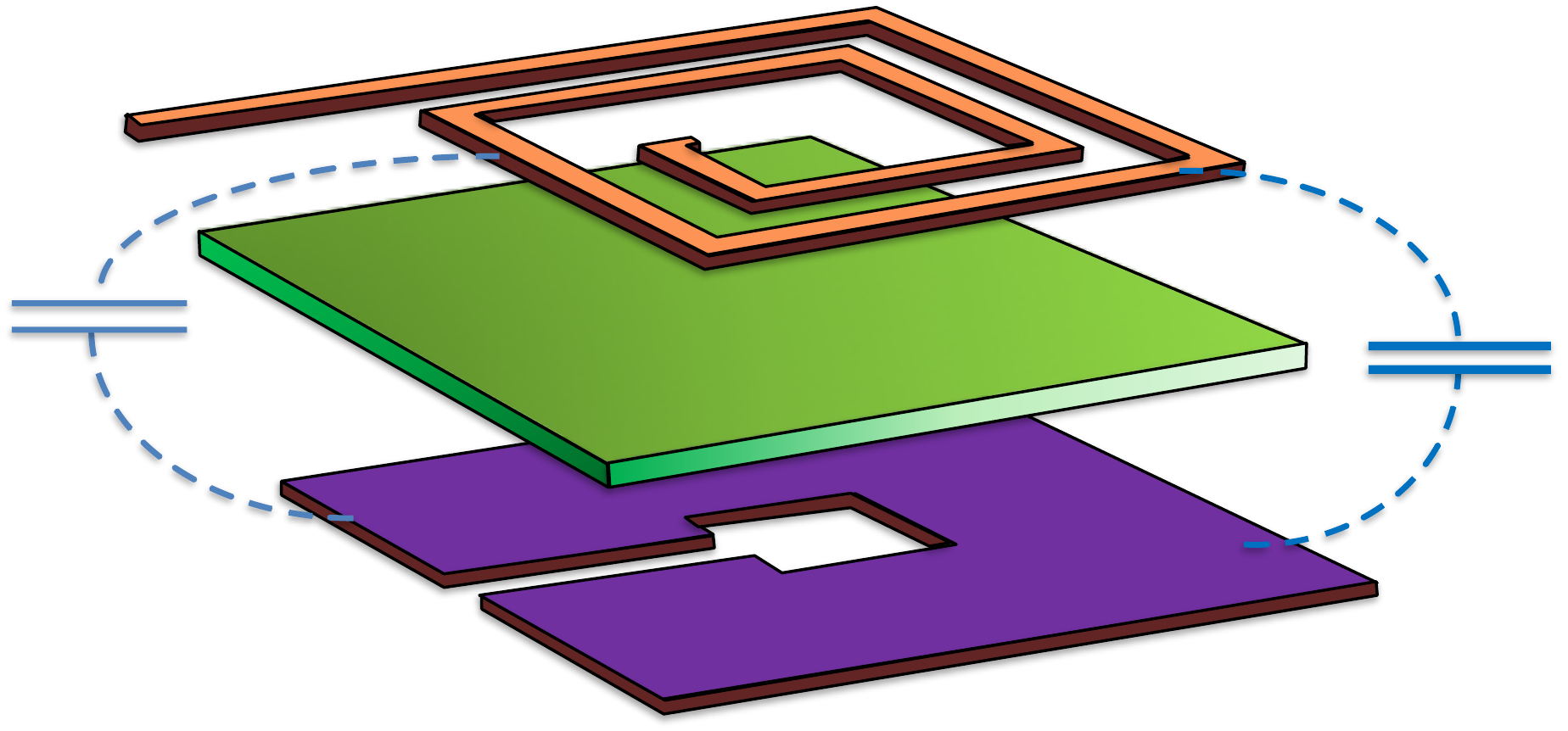}
	\caption{Parasitic capacitance formed by planar input coil on top of SQUID washer \cite{Muck03}.}
	\label{fig:squid_cap}
\end{center}
\end{figure}

\begin{figure}[ht!]
\begin{center}
	\includegraphics[width=0.6\linewidth]{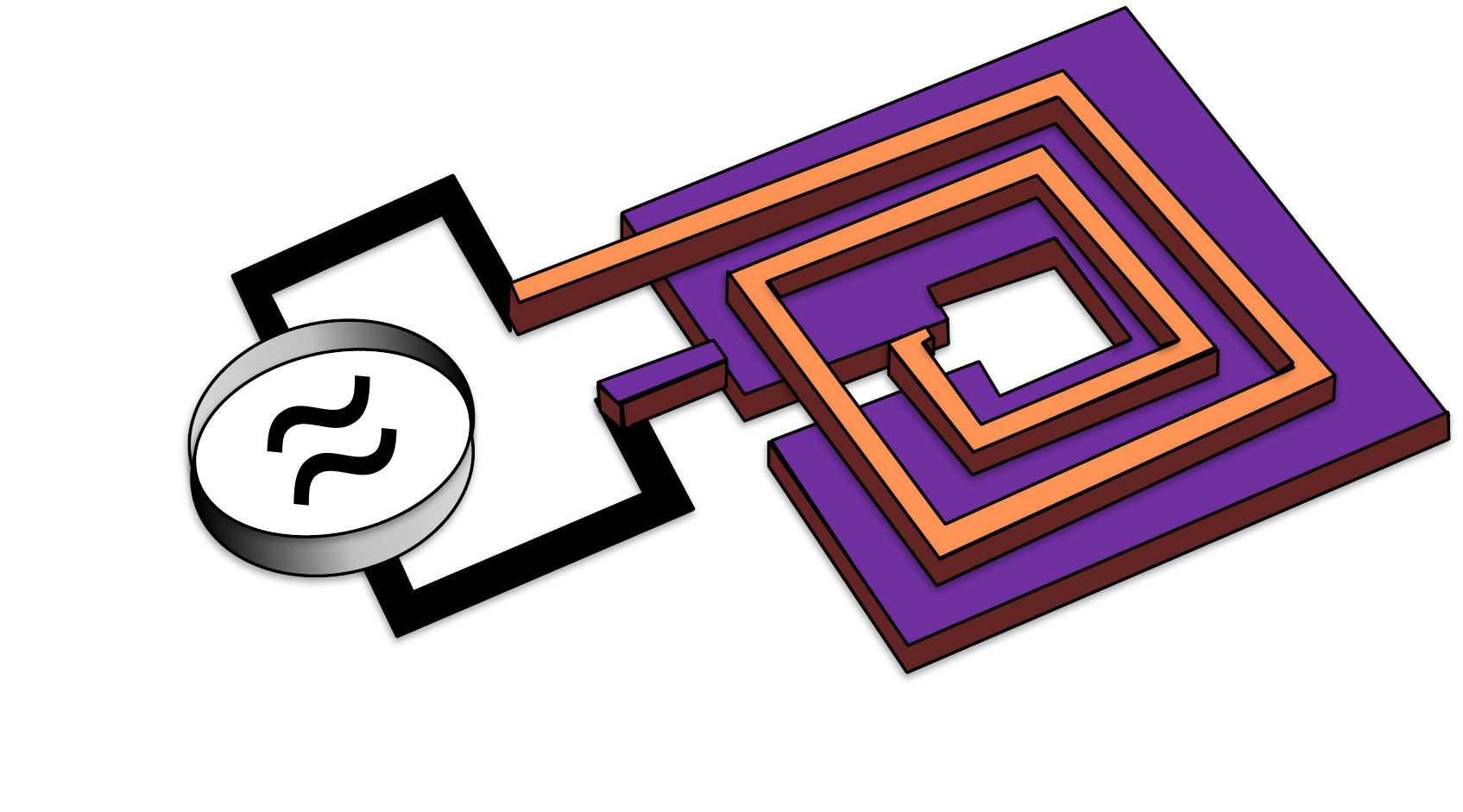}
	\caption{Input coil on top of SQUID operated as a microstrip resonator by coupling input signal to one end of coil and washer, 
	leaving the other end of the coil unconnected \cite{Muck03}.}
	\label{fig:squid_micro}
\end{center}
\end{figure}

\begin{figure}[hb!]
\begin{center}
	\includegraphics[width=0.4\linewidth]{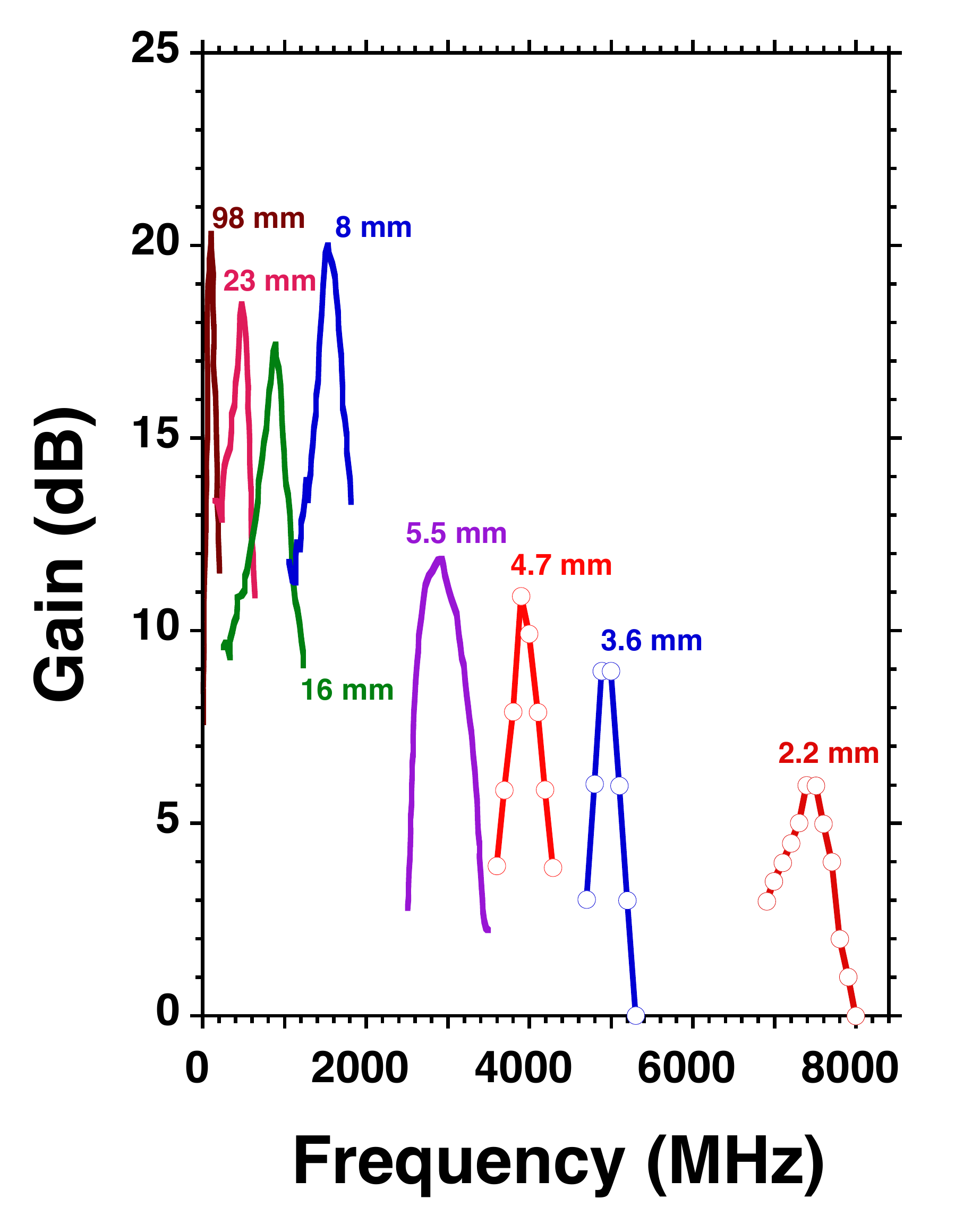}
	\caption{Gain versus frequency for eight microstrip SQUID amplifiers. The length of the input coil (microstrip resonator) 
	is indicated above the corresponding gain curves.}
	\label{fig:squid_gain}
\end{center}
\end{figure}

In the ADMX experiment, which has recently operated in the frequency range 645-680 MHz, a cryogenic heterojunction field-effect transistor (HFET)
amplifier with a noise temperature of 2~K was initially used to amplify the signal picked up by a critically coupled antenna inserted in the cavity.
Subsequently, a Microstrip SQUID Amplifier (MSA) \cite{Muck10} has replaced as front-end amplifier the HFET which is used as second stage
amplifier (see Fig. \ref{fig:squid_admx}). The MSA is an effective solution to the problem of the gain drop of the standard dc SQUID amplifier at
frequencies higher than about 100 MHz. The gain drop can be explained as follows. Input coil and washer-type SQUID, separated by a thin insulating
film, form a capacitor in parallel to the input coil inductance (see Fig. \ref{fig:squid_cap}).
Because of this parasitic capacitance, the input circuit is purely inductive only at frequencies below the self-resonant frequency of the tuned circuit
formed by the coil and the parasitic capacitance. At higher frequencies the signal is shorted by the parasitic capacitance. This deleterious effect
can be avoided by making a virtue of the parasitic capacitance and by operating the input coil of the SQUID as a transmission line resonator
(microstrip) (see Fig. \ref{fig:squid_micro}).
The signal to be amplified is applied between one end of the coil and the washer, while the other end of the coil is left open. Provided that the source
impedance is greater than the characteristic impedance of the microstrip, there is a peak in the gain when the input coil accommodates approximately
(but not exactly) one-half wavelength of the input signal. The frequency of operation of the amplifier is determined by the microstrip length, and thus by
the number of turns of the input coil. The quality factor of the microstrip resonator, and thus the bandwidth of the amplifier, can be set by choosing an
appropriate microstrip impedance. Gains over 20 dB, noise temperatures well below the bath temperature and operation frequencies from 100 MHz
to 8 GHz have been achieved (see Fig. \ref{fig:squid_gain} \cite{Muck03}) .

\begin{figure}[h!]
\begin{center}
	\includegraphics[width=0.5\linewidth]{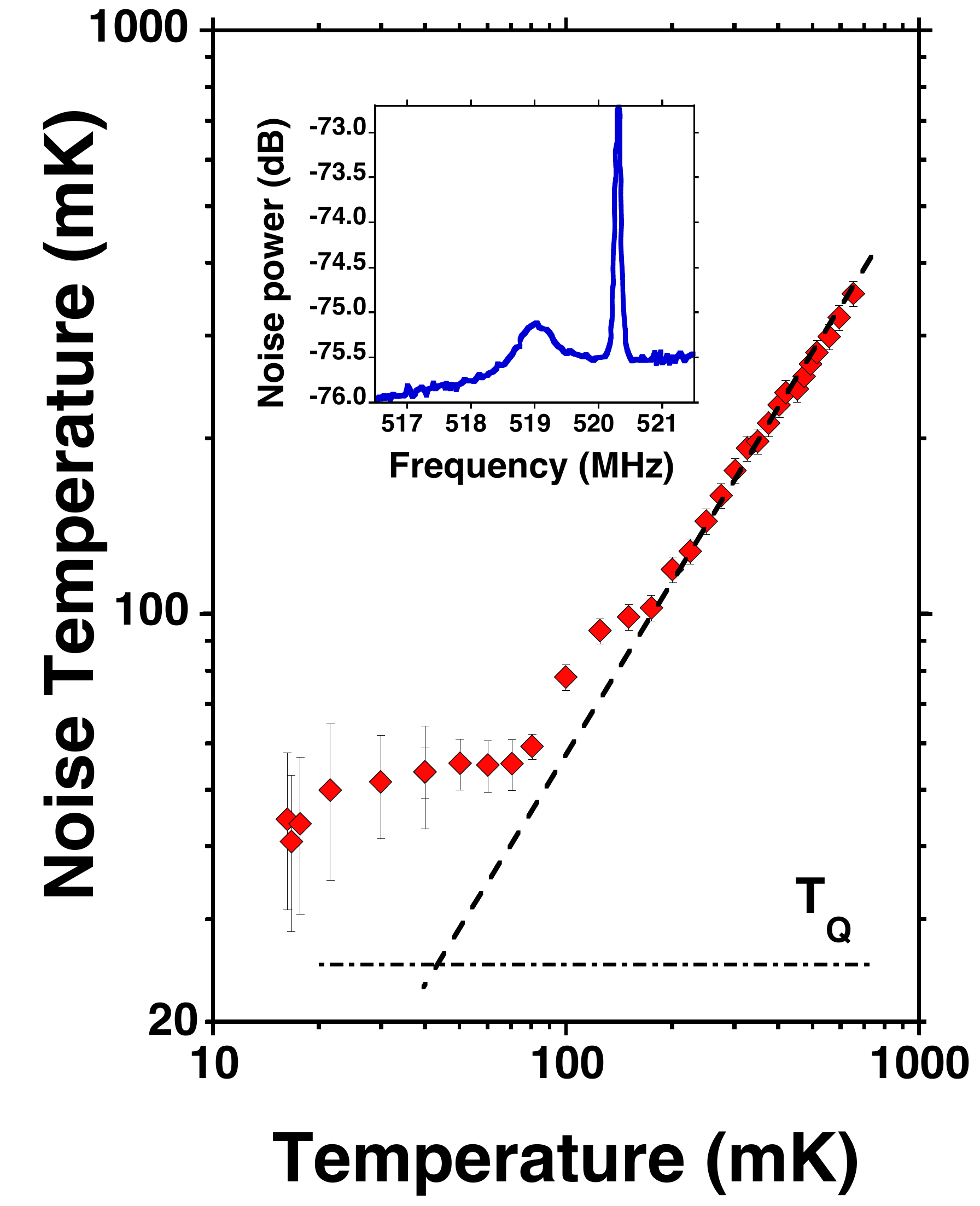}
	\caption{Noise temperature of input microstrip SQUID at 519 MHz vs temperature measured with a resonant source. The dashed 
	line through the data corresponds to $T_N \propto T$, and the dot-dashed line indicates $T_Q = h f/k_\mathrm{B} \simeq 25$ mK. 
	Inset is noise peak produced by LC-tuned circuit at 20 mK. The upward trend of the baseline reflects the fact that the peak in the 
	amplifier gain is at a higher frequency. The peak at 520.4 MHz is a calibrating signal \cite{Muck99}.}
	\label{fig:squid_noistem}
\end{center}
\end{figure}

In the classical limit, the noise temperature $T_n$ of an MSA is expected to scale as $f T$, where $f$ is the operation frequency and $T$ the bath
temperature \cite{Muck01} and measurements show that MSA can approach the quantum-limited performance. For example, a noise temperature
$T_n = 47$ mK has been achieved at 20 mK and 520 MHz \cite{Muck99} (see Fig. \ref{fig:squid_noistem}) where the standard quantum limit
$T_q = h f/k_{\rm{B}}$ is about 25 mK. In the same series of measurements a $T_n = 170$ mK has been obtained at 300 mK. By scaling with the
frequency, noise temperatures of 16 mK and 82 mK are respectively expected at 50 MHz and 250 MHz, the extremes of the frequency range of
interest for KLASH.

\begin{figure}[h!]
\begin{center}
\vspace*{0.5cm}
	\includegraphics[width=0.5\linewidth]{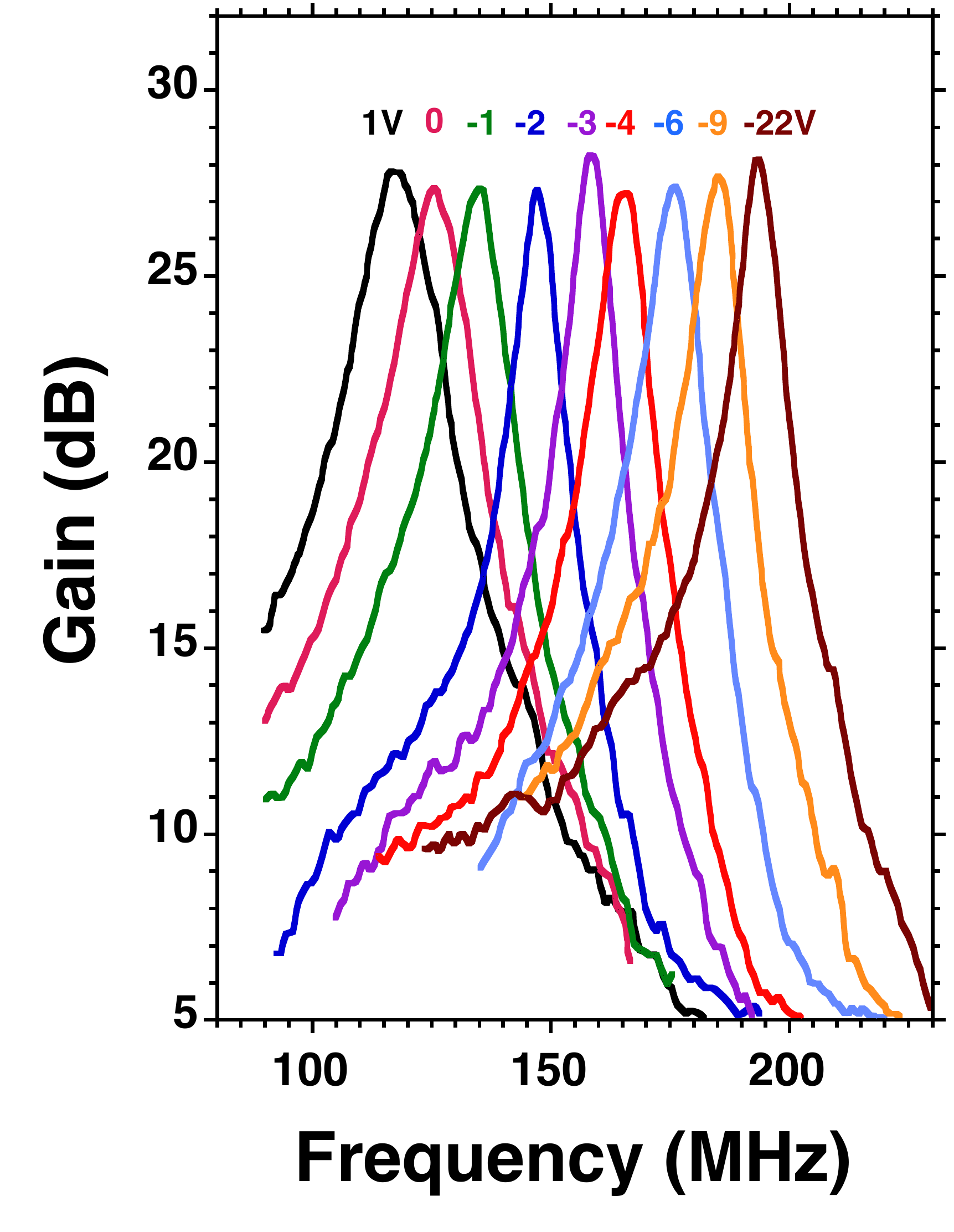}
	\caption{Gain vs frequency for a MSA for nine values of bias voltage applied to the varactor diode (\cite{Muck99})}.
	\label{fig:squid_gain2}
\end{center}
\end{figure}

As mentioned above the bandwidth of the MSA can be set by choosing an appropriate microstrip impedance and it should be possible to cover each of
the three bands planned for KLASH (65-120 MHz, 120-150 MHz, 150-250 MHz) by changing the MSA together with the cavity. In case it is not possible
to cover completely a frequency band one can tune the frequency of the SMA by connecting a varactor diode between the otherwise open end of the
input coil and the washer \cite{Muck99}. The capacitance of the diode can be varied by changing the value of the reverse bias voltage with the effect of increasing
or decreasing the effective length of the microstrip and lowering or raising the peak frequency (see Fig. \ref{fig:squid_gain2}).

For the reasons outlined, we believe that the best choice for the KLASH front-end amplification is the MSA also in view of a measurement of the thermal noise of the 
cavity with a non-optimized antenna. Of course, a series of preliminary tests of different SMA down to 300 mK are necessary in order to identify and characterize those 
that best fit the experiment requirements. ADMX used a circulator in their first experiments with the HEMT preamplifier to avoid back-action noise from the preamp 
entering the cavity and they kept it in the amplification chain when they switched to the MSA. Considering the lower frequencies of KLASH, the circulator should not be 
required \cite{Schupp}. Anyway, preliminary tests will be conducted with high $Q$ LC circuits that simulate the cavity load in order to evaluate the effect of the MSA 
back-action noise and address it with a specific circulator.

\subsection{B Field Shielding} % Di Gioacchino

DC-SQUID amplifiers are very sensitive devices that must operate in a highly shielded volume. Several effects spoil the amplifier sensitivity:
\begin{itemize}
\item[-] Large magnetic fields degrade device performance. The field intensity must be below the device intrinsic-noise level (1mG);
\item[-] Pick up noise from input circuit could induce variation on the induced field (0.01G) and generate spurious signals in the amplifier;
\item[-] Induced noise limits the dynamic range;
\end{itemize}
The MSA must be well thermally and mechanically anchored to provide good thermal stability and isothermality and to avoid spurious signals 
from antenna vibrations inside the cavity volume. Instead of the dipole antenna, a pick-up coil with wire-wound gradiometer \cite{Rigby} geometry 
could be investigated to reduce the effect of DC field variations. Finally, tight RF shielding must be ensured \cite{Claycomb}.

To shield the KLASH magnetic field, $H_\mathrm{DC} = 0.6$ T, 6 to 7 orders of magnitude screening is needed (from 6000 Gauss to milli Gauss). 
Although SQUIDs can operate in larger static field (0.1 G) we show that this conservative screening can be obtained using different approaches:
\begin{itemize}
\item[A.] A passive scheme \cite{Rigby,Claycomb};
\item[B.] A quasi-active system provide by a persistent mode superconducting coil \cite{Bruba,Alken};
\item[C.] An active procedure with a negative feedback electronic circuit system \cite{Okaza,Hilge}.
\end{itemize}
\vspace*{0.5cm}

\noindent
\textbf{A. Passive Magnetic Shield}
%\label{sec:passive}

\begin{figure}[!hb]
  \begin{center}
    \includegraphics[width=0.09\linewidth]{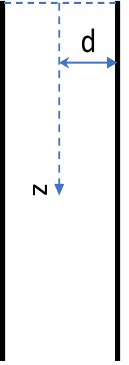}
    \caption{Semi-infinite tube of radius $d$. The $z$-axis originates on the tube entrance and is oriented downward \cite{Claycomb}.}
    \label{fig:SemiTube}
  \end{center}
\end{figure}

%\subsubsubsection{Theoretical estimate}
The passive magnetic-shielding approach is realized with long tubes of high-permeability or superconducting material \cite{BXXu}. Magnetic-field lines are diverted 
around a volume of interest reducing the local magnetic-field in the region within the shielded volume. The analytical expression of the magnetic-field shielded by a ‘semi-infinite’ tube, 
both respect to axial and transverse (radial) directions, is given in \cite{Claycomb} for SC tubes (see Fig. \ref{fig:SemiTube})
\begin{equation}
\label{eq:Hax_sc}
H_\mathrm{axial} = H_{DC}\exp \left(-3.832\,\frac{z}{d}\right)   \qquad\qquad  H_\mathrm{transverse} = H_{DC}\exp \left(-1.84\,\frac{z}{d}\right),
\end{equation}
and for $\mu$-metal tubes
\begin{equation}
\label{eq:Hax_mu}
H_\mathrm{axial} = H_{DC}\exp \left(-2.405\,\frac{z}{d}\right)  \qquad\qquad H_{{DC}-\mathrm{transverse}} = \exp \left(-3.834,\frac{z}{d}\right).
\end{equation}
SC and $\mu$-metal tubes are more effective in dumping axial and transverse fields, respectively.

A combination of the two types of materials serves to effectively screen both transverse and axial field components \cite{Rigby}.
In addition, an external layer of $\mu$-metal screens the inner volume from Earth magnetic-field, avoiding flux trapping in SC shields during cool down and before turning on the KLOE magnet.

\begin{figure}[!ht]
  \begin{center}
    \includegraphics[width=0.8\linewidth]{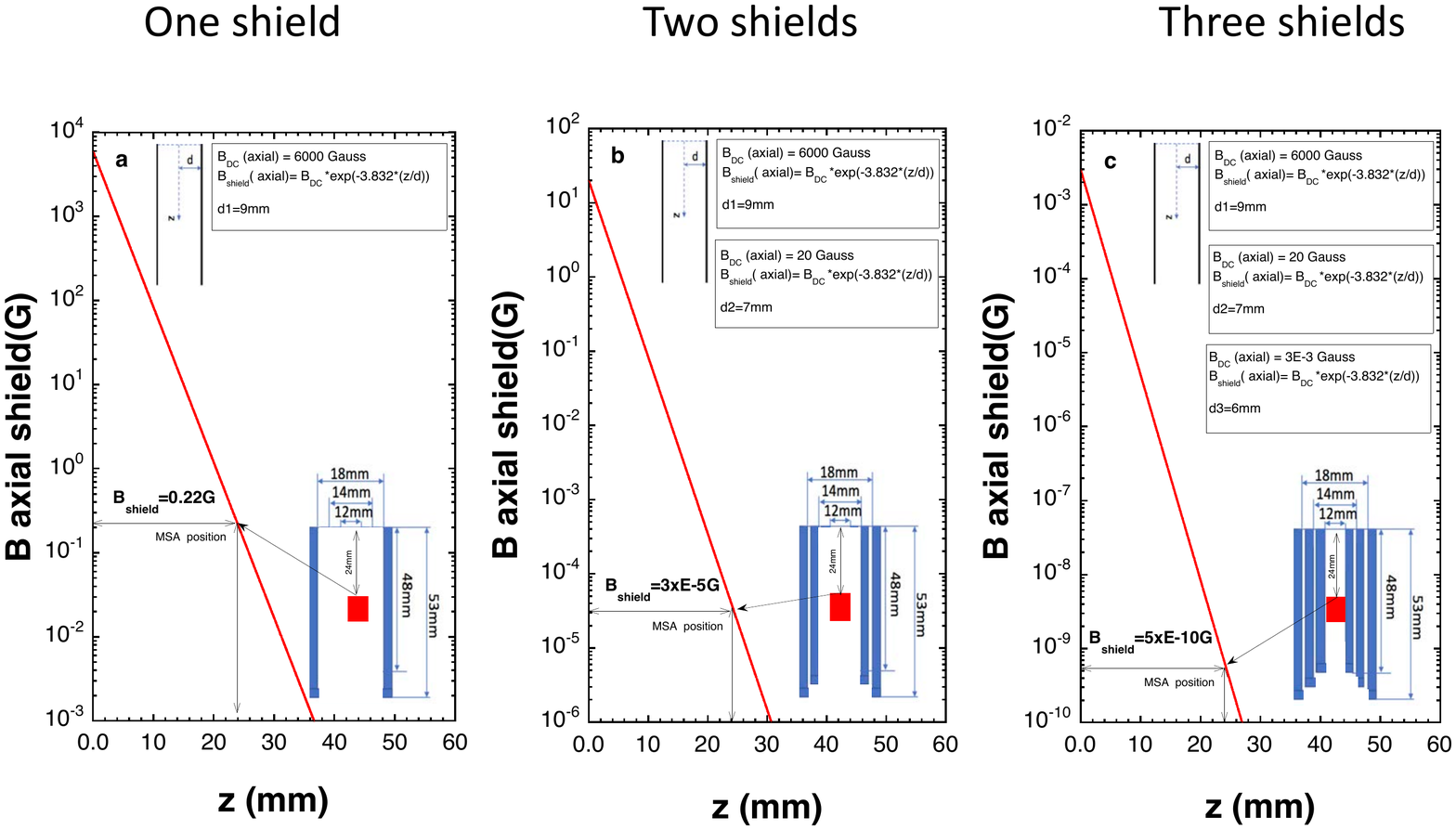}
    \caption{Attenuated magnetic field along the axis coordinate $z$ of a superconducting semi-infinite tube (see the first in Eq. (\ref{eq:Hax_sc}))
    for :1 shield (Left), 2 shields (center), and 3 shields (Right). The case of an external field $H_{DC} = 0.6$ T is assumed. Geometrical setup for 
    the three configurations are shown in Tab. \ref{tab:SCcyl}.}
    \label{fig:Idealsc}
  \end{center}
\end{figure}

We made an estimate on the magnetic shielding using the equation described above for the case of a superconducting cylinder. The MSA device is a square of size approximately 
$10\times10\mbox{mm}^2$. We considered three cylindrical screens of diameters 18, 14, 12\,mm and heights 53, 50, 48\,mm with the MSA device placed at 24\,mm from the top of the cylinders. 
The approximation of a semi-infinite tube is motivated by the small ratio between the MSA dimension and the cylinder real-length (about 5\,cm) and the small demagnetization factor 
D = 0.079 \cite{Goldfarb}.  These are reasonable dimensions and are easily integrated into the KLASH apparatus. In our analysis, the 3 concentric superconducting screens are gradually 
introduced and the results are shown in Fig. \ref{fig:Idealsc}. In Tab. \ref{tab:SCcyl} we show the residual magnetic field $B_{MSA}$ calculated at the MSA position $z_{MSA}$ for three different 
shield geometries. The  final result indicates that there are still 7 orders of  magnitude of margin on the milligauss maximal operative limit of the MSA and well within the 10 $\mu$G limit indicated 
by the ADMX Collaboration, making this approach robust. We tested these conditions measuring the attenuation of the magnetic field with three Nb$_3$Sn cylinders as discussed later.

\begin{table}[h!]
  \begin{center}
  \caption{Residual magnetic field $B_{MSA}$ calculated at the MSA position $z_{MSA}$ for three different shield geometries.}
  \vspace*{0.1cm}
    \begin{tabular}{c|c|c|c}
      & Single Shield & Two Shields & Three Shields \\\hline
      Cylinder diameter& 18\,mm & 18\,mm, 14\,mm & 12\,mm, 14\,mm, 18\, mm \\
			Cylinder height& 53\,mm &50\,mm&48\,mm\\
			z$_{MSA}$ & 24\,mm&24\,mm&24\,mm\\
			B$_{MSA}$ & 0.22\,G & $3\times10^{-5}$ G & $5\times10^{-10}$ G \\
			\hline\hline
    \end{tabular}
  \label{tab:SCcyl}
  \end{center}
\end{table}

\begin{figure}[!ht]
  \begin{center}
    \includegraphics[width=0.4\linewidth]{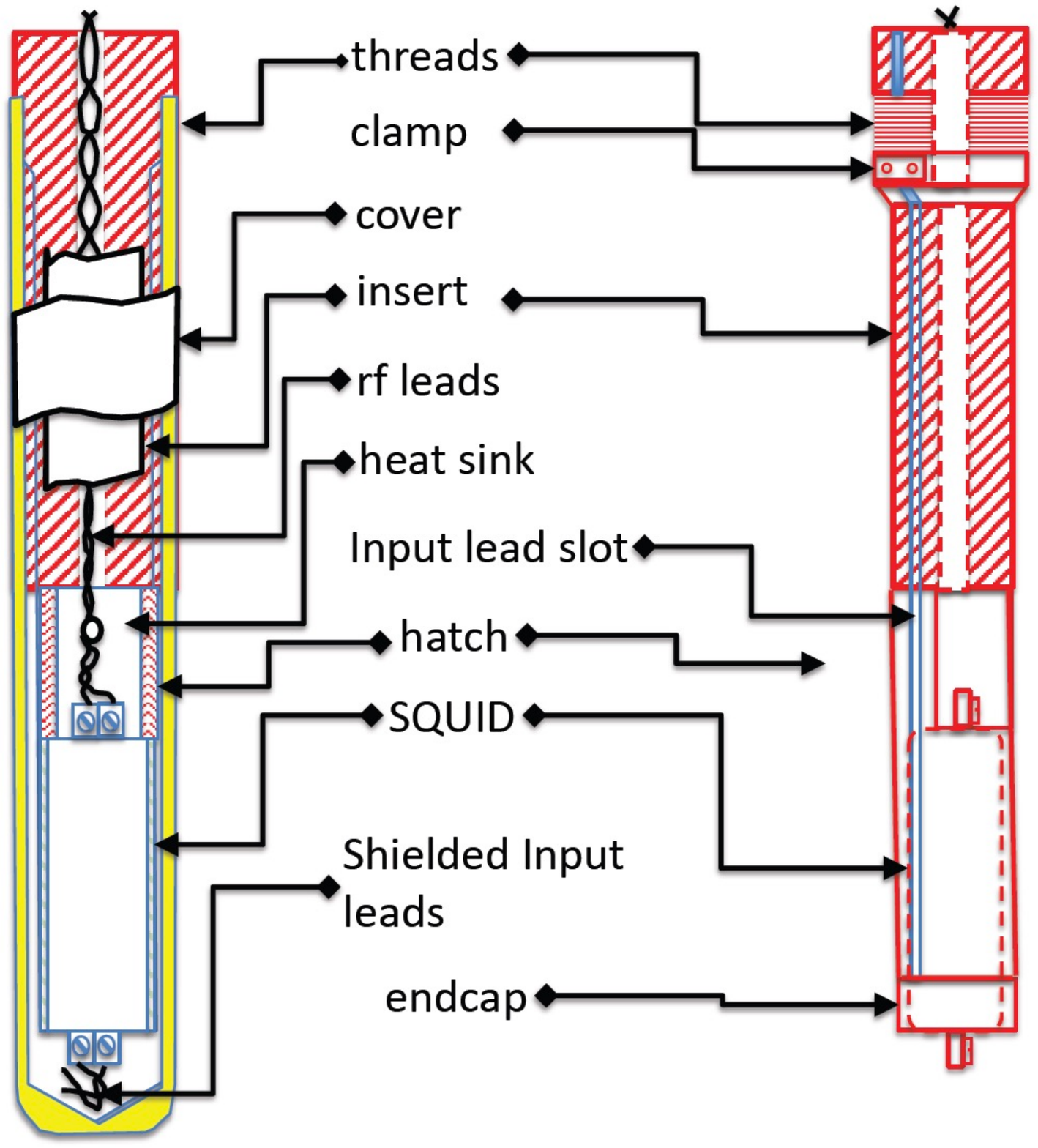}
    \caption{Main components of the SQUID holder \cite{Rigby}.}
    \label{fig:squidholder}
  \end{center}
\end{figure}

%\subsubsubsection{Example of SQUID Holder}
As an example, the result of the measurement of the shielding factor $S$, where $S= B_{\mathrm{appl}}/B_{\mathrm{meas}}$, for a shield composed of two Nb superconductors, with critical field 
$H_{c} \approx 0.3$ T, covering a SQUID \cite{Rigby} is shown in Table \ref{tab:esempioschermo}. The DC field was $H_\mathrm{DC} = 300$ G. Several orientations were considered. The setup is 
shown in Figure \ref{fig:squidholder}.
\begin{table}[h!]
  \begin{center}
  \caption{Measured shielding-factor with magnet and input coil oriented
  transversally and axially \cite{Rigby}.}
  \vspace*{0.1cm}
    \begin{tabular}{c|c|c}
      Magnetic Field Orientation & Input Loop Orientation & Shielding Factor ($S$)\\\hline
      Axial & Transverse & $2\times10^{11}$ \\
      Axial & Axial & $1\times10^{12}$ \\
      Transverse & Transverse & $5\times10^{9}$ \\
      Transverse & Axial & $2\times10^{12}$ \\
      \hline\hline
    \end{tabular}
  \label{tab:esempioschermo}
  \end{center}
\end{table}

\begin{figure}[!ht]
  \begin{center}
    \includegraphics[width=0.4\linewidth]{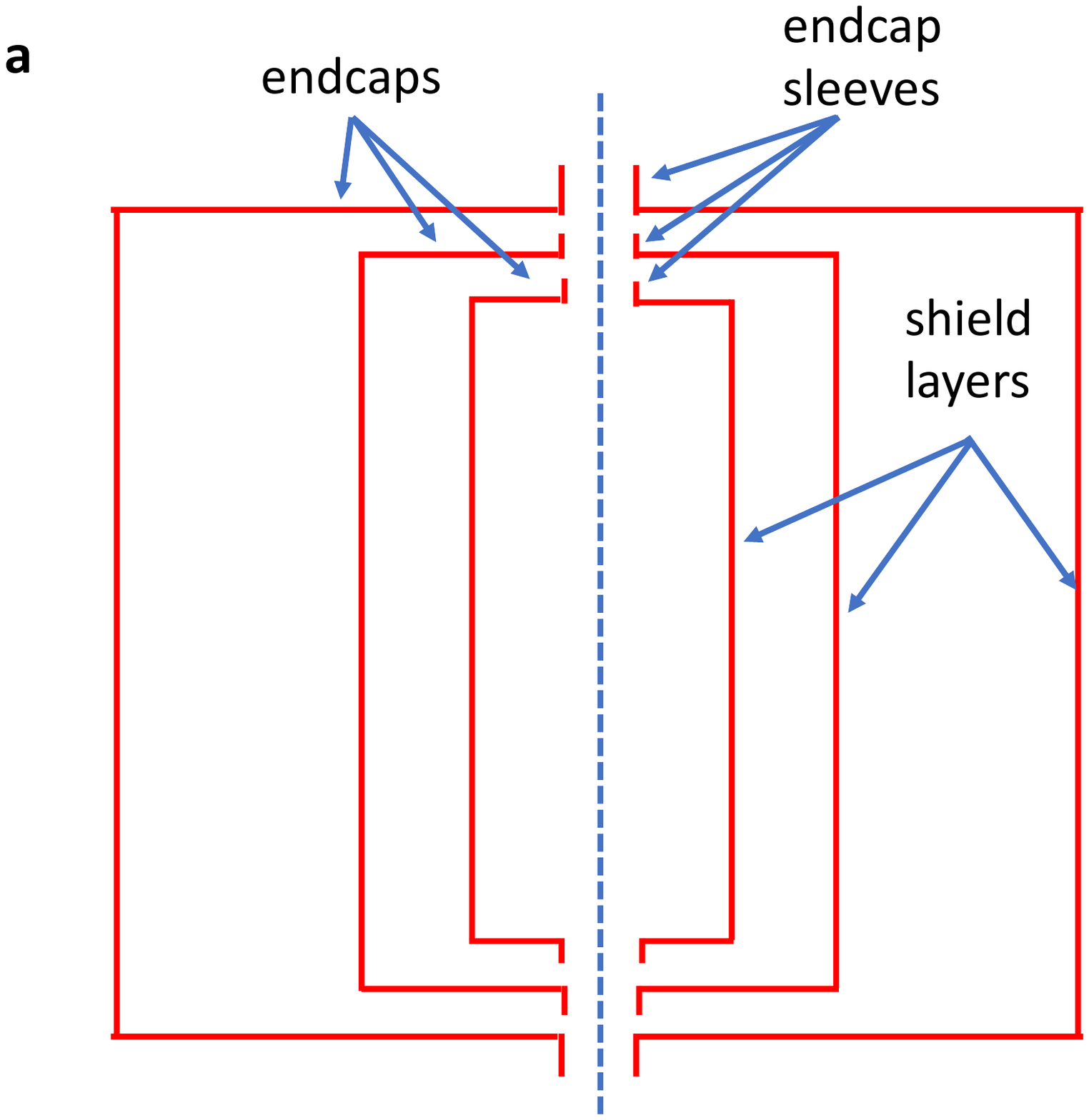}
    \includegraphics[width=0.4\linewidth]{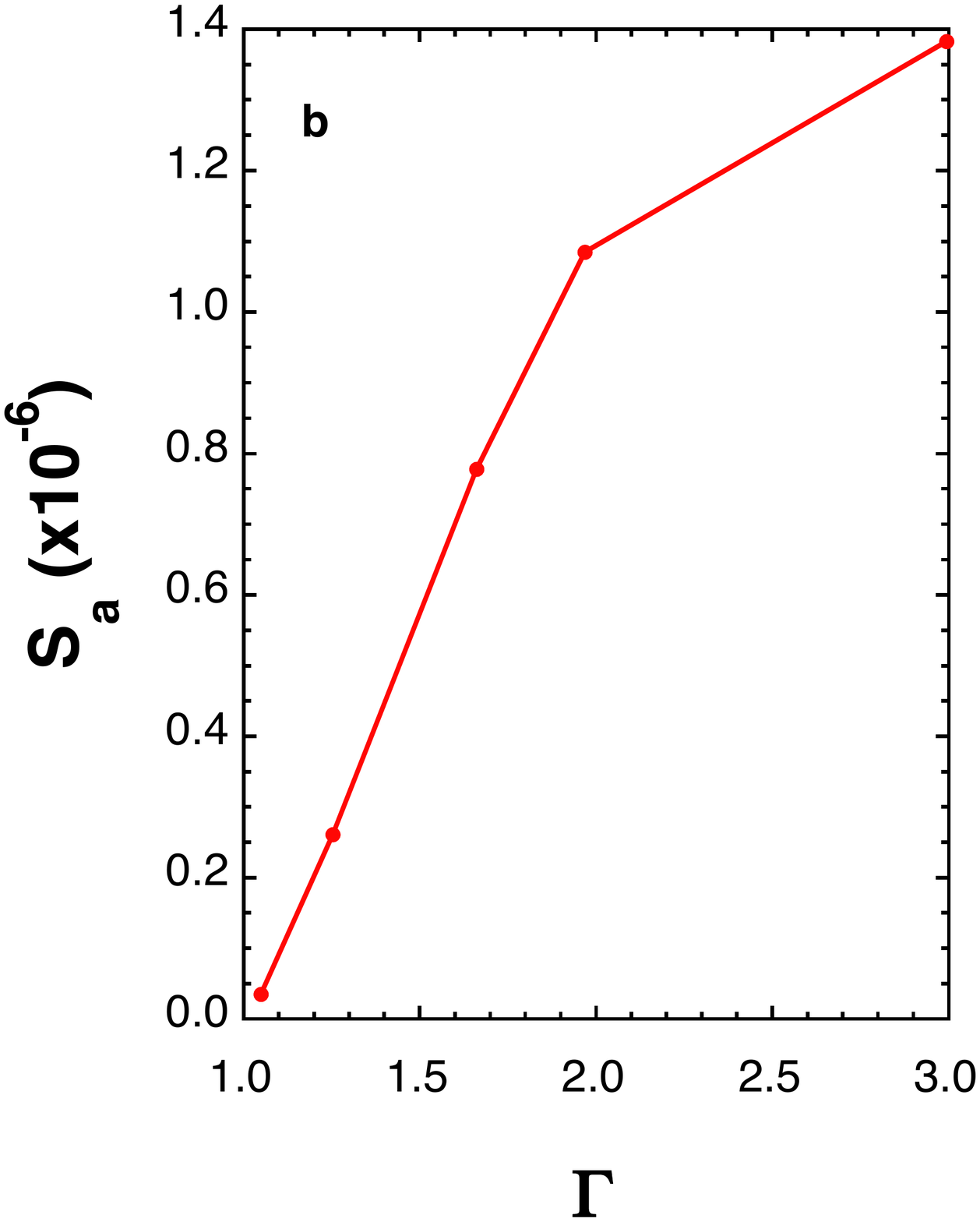}
    \includegraphics[width=0.4\linewidth]{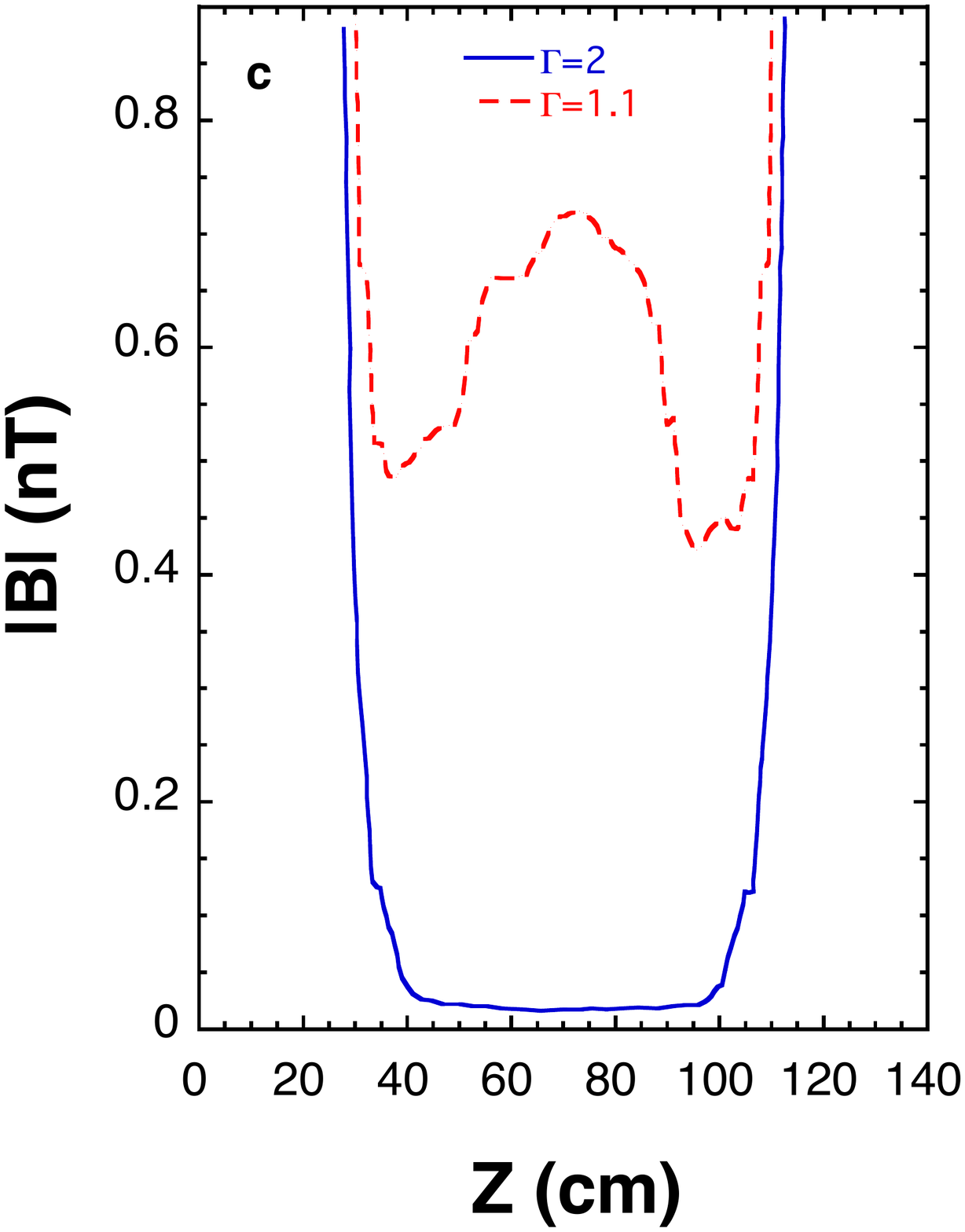}
    \includegraphics[width=0.4\linewidth]{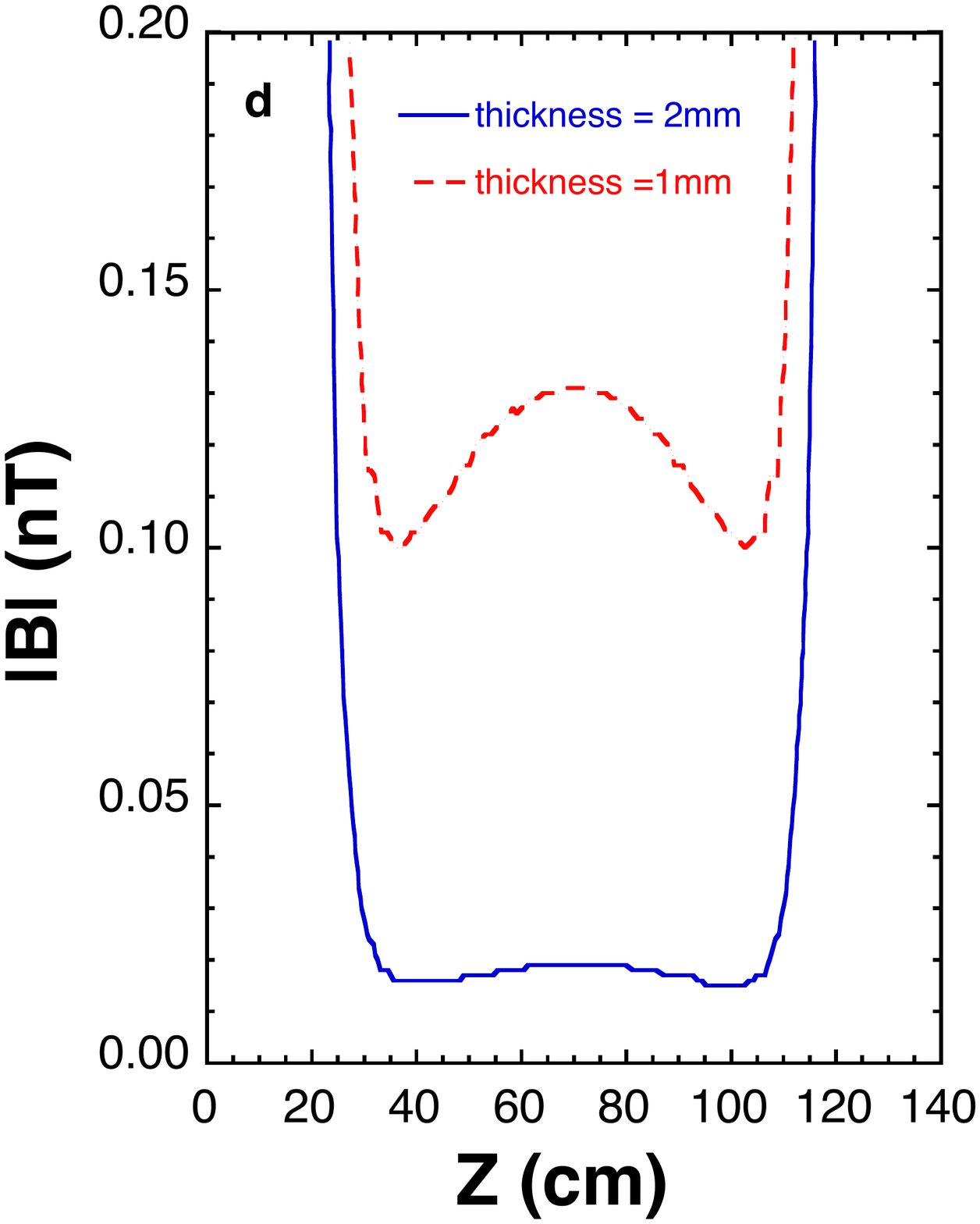}
    \caption{a) View of the shield design; b) Axial shielding-factor S as a function of the radial-shield spacing ratio;
    c) Residual magnetic-field intensity inside along the shield axis for $\Gamma = 1.1$ and $\Gamma = 2$; d) Same as (c) but for different shield thickness \cite{Sumn}.}
    \label{fig:3layers}
  \end{center}
\end{figure}

%\subsubsubsection{Three-layer cylindrical magnetic shields}
The optimal ratio of shield length to shield radius, $L/r$, is 5.5 for a single shield \cite{Sumn} and 2 for three-layer screen \cite{Burt}. The optimal ratio for radial-shield spacing, $\Gamma = r_{i + 1}/r_i$, 
is 2. A layout of a three-layer respecting such proportions is sketched in panel (a) of Fig. \ref{fig:3layers}.

\begin{figure}[htbp]
  \begin{center}
    \includegraphics[width=0.8\linewidth]{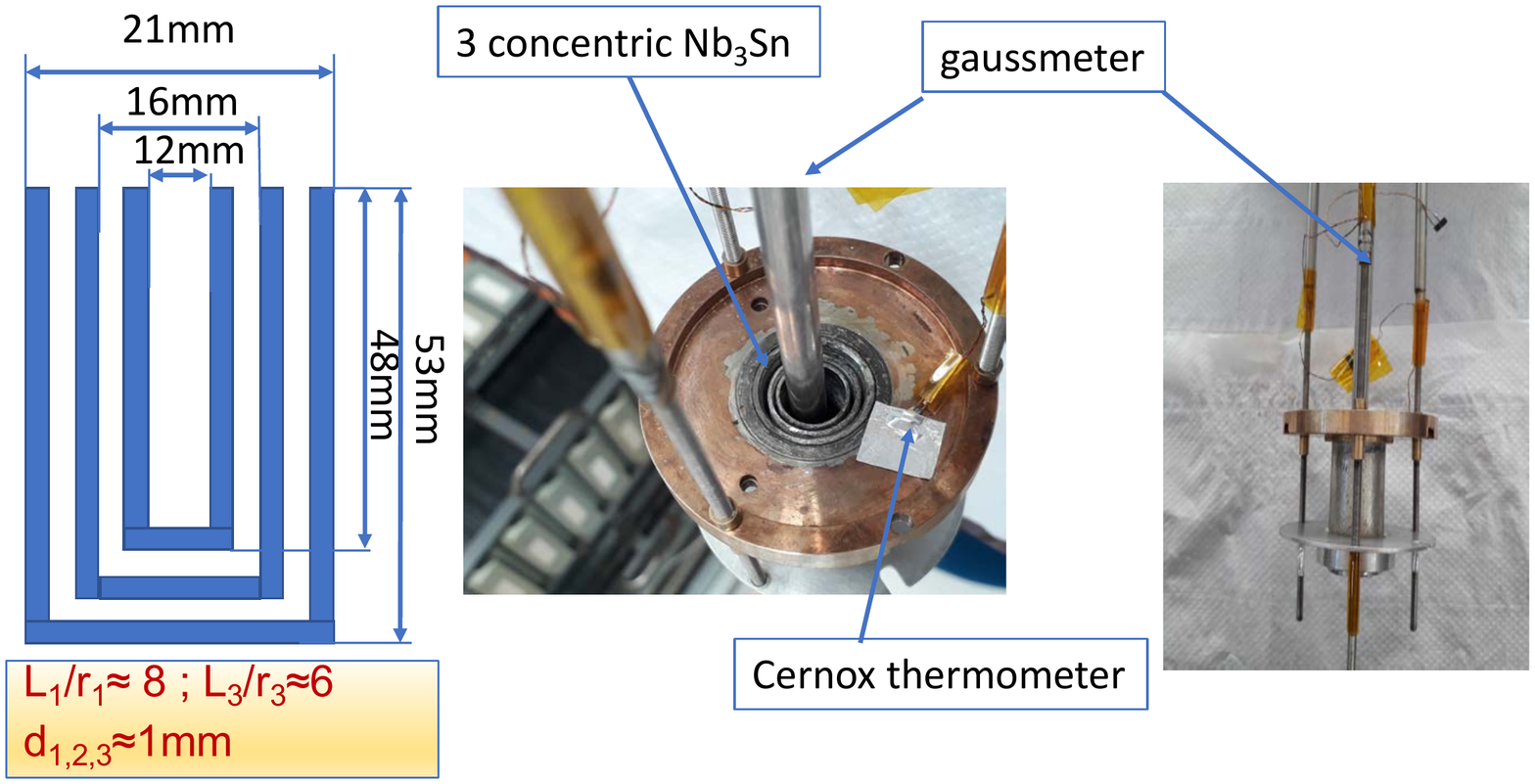}
    \caption{a) Sketch of the Nb3Sn screens, b) photo of the gaussmeter inserted
    in the set of cylinders, c) bottom of the insert to be placed at the centre of the superconducting
    DC magnet inside the cryostat}
    \label{fig:nb3sn}
  \end{center}
\end{figure}

\begin{figure}[htbp]
  \begin{center}
    \includegraphics[width=0.8\linewidth]{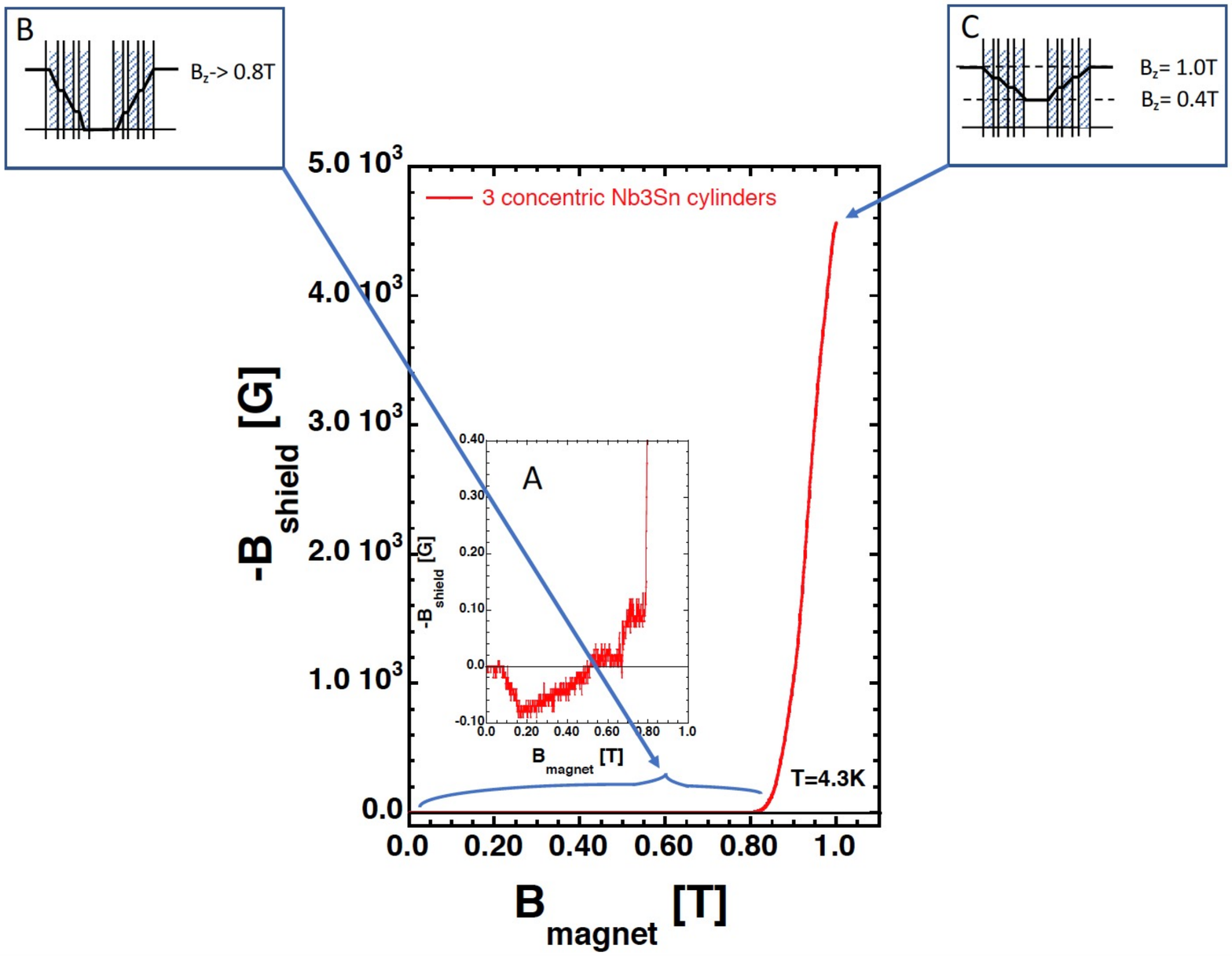}
    \caption{Measured magnetic field in the internal volume of Nb$_3$Sn concentric-cylinders
    versus the applied magnetic field (insert A). Expanded view of the measured vs applied field in the 0-1 T (insert B).
    Trapped magnetic field profile according to the Bean model of flux-pinning \cite{Bean}.}
    \label{fig:bean}
  \end{center}
\end{figure}

\begin{figure}[!ht]
  \begin{center}
    \includegraphics[width=0.5\linewidth]{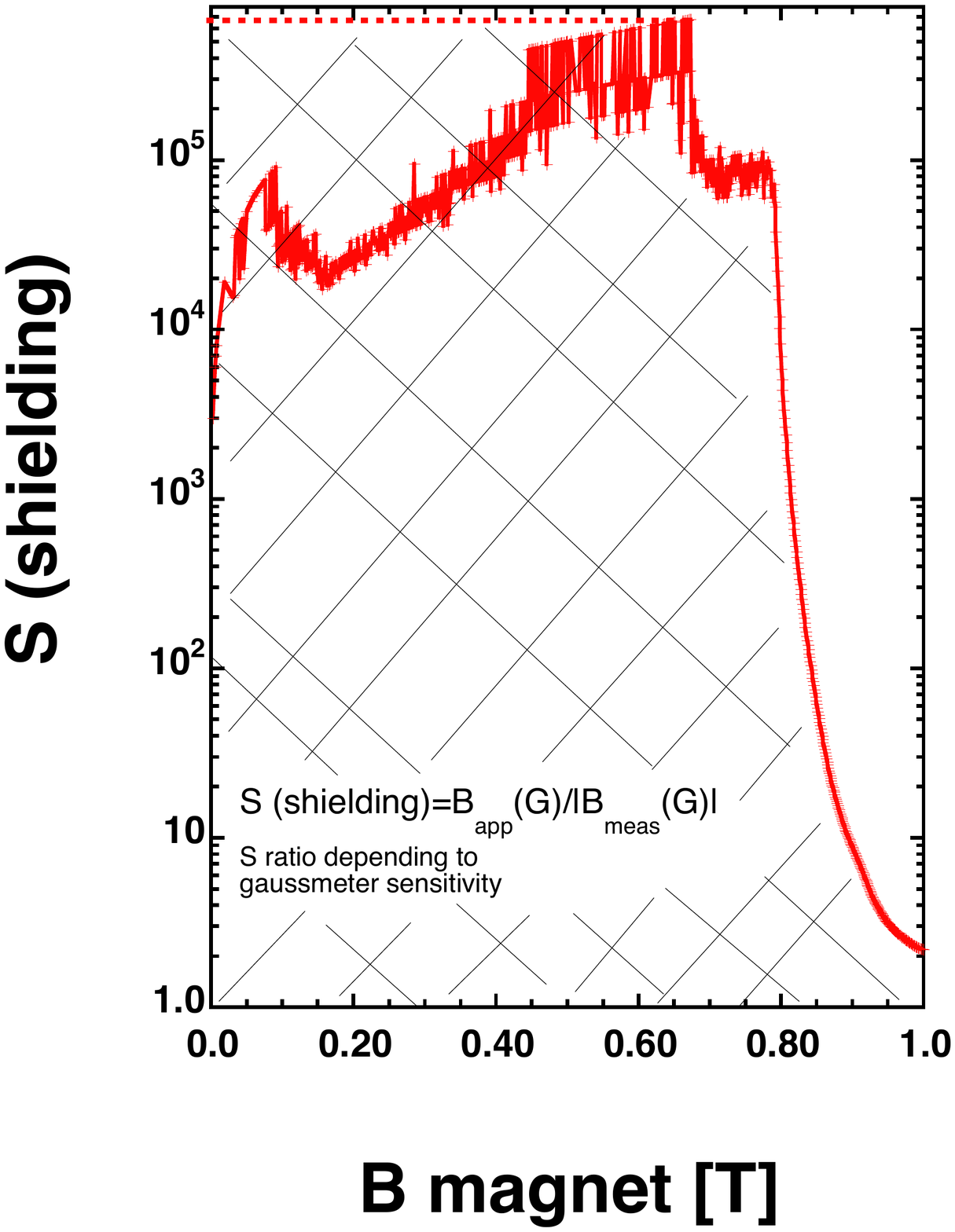}
    \caption{The shielding factor $S$ versus the applied  magnetic field}
    \label{fig:Sfactor}
  \end{center}
\end{figure}

%\subsubsubsection{An experimental test in DC magnetic field with three Nb$_3$Sn superconducting cylindrical shielding}
We performed a measurement with three Nb$_3$Sn cylindrical tubes. Nb$_3$Sn has critical field $H_{c2}$ equal to 23 T suitable for screening the 0.6 T field of the KLOE magnet. Due to the low value 
of the lower critical field, $H_{c1}$ (Nb$_3$Sn) = 35 mT, the field penetrates into the superconducting Nb$_3$Sn bulk where a strong pinning must be present to block it. The measurement is done with a 
magnetic probe, a \texttt{Lake Shore} gaussmeter (model 425), with the Hall sensor placed into the concentric cylinder. The holder is fixed to the bottom of the insert and a Cernox thermometer is attached 
with varnish glue on the top-copper flange of the magnetic screens. The set-up is shown in Fig. \ref{fig:nb3sn}.
We placed the insert in a liquid-He cryostat with flux control at the centre of a NbTi-8T DC-magnet and performed a zero-field-cooling (ZFC) procedure from 300 K to 4.2 K. After the cooling and before 
turning on the magnet, the value of the trapped Earth magnetic-field read by the gaussmeter was set to zero  ("nulling process"). After turning on the magnet power-supply the gaussmeter was measuring 
a field value of 0.02 G $\pm$ 0.01 G. With temperature stable at 4.2 K we cycled the DC magnetic field from 0 T $\to$ 1 T $\to$ 0 T. The temperature was kept stable during this cycle in order to avoid 
lux jumps. In Fig. \ref{fig:bean} we show the field measured as a function of the applied DC-field. Thanks to the strong pinning, the magnetic field is almost completely shielded for values up to 0.8 T. In 
the expanded view shown in panel A of Fig. \ref{fig:bean} variations on the order of 0.1 G are visible. These variations are well described by the strong-pinning Bean-model \cite{Bean}. 
When the field reaches 0.8 T there is a consistent field penetration, up to 0.5T for 1 T of applied field. This situation is still well described by Bean's pinning model as shown in insert C of Fig.\ref{fig:bean}.
In Fig. \ref{fig:Sfactor} we show the measured shielding factor $S$.

This measure indicates that a screen of three concentric cylinders of Nb$_3$Sn could be used to screen efficienctly the KLOE magnetic field. The residual 10$\mu$G will be screened by the Nb case 
of the SQUID amplifier.
\vspace*{0.5cm}

\noindent
\textbf{B. Quasi Active Magnetic Shield}
%\label{sec:quasiactive}

\begin{figure}[htbp]
  \begin{center}
    \includegraphics[totalheight=4.5cm]{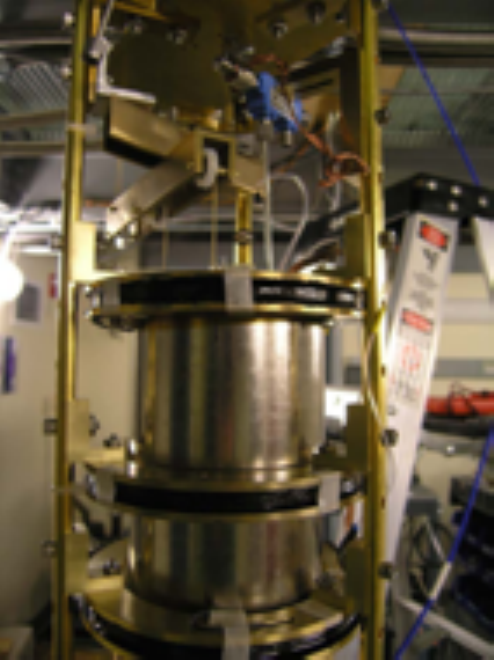}
    \includegraphics[totalheight=4.5cm]{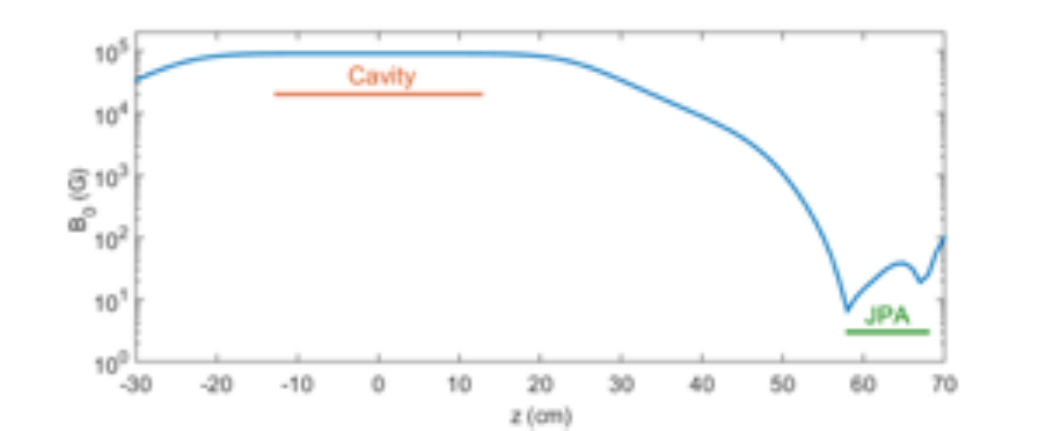}
     \caption{Magnetic shielding of the JPA, including three persistent superconducting coils (a).
     $B_0$ vs $z$ data in HAYSTAC magnet; the axial positions of the cavity and the JPA canister
     are indicated in the absence of the field compensation coil \cite{Bruba,Alken} (b).}
    \label{fig:HAYSTACfield}
  \end{center}
\end{figure}

In this configuration the experimental volume is shielded by a superconducting magnet, for example of NbTi, in a persistent condition. The system must be cooled in a zero field condition (ZFC). Multiple 
magnets can also be arranged. This solution was adopted by the HAYSTAC project \cite{Bruba,Alken}. The set-up is shown in panel (a) of Fig. \ref{fig:HAYSTACfield}; in panel (b) we show the magnetic 
field, without any compensation, present in the region where a Josephson parametric amplifier (JPA) is placed.
\vspace*{0.5cm}

\noindent
\textbf{C. Active Magnetic Shield}
%\label{sec:active}

\begin{figure}[!ht]
  \begin{center}
    \includegraphics[width=0.8\linewidth]{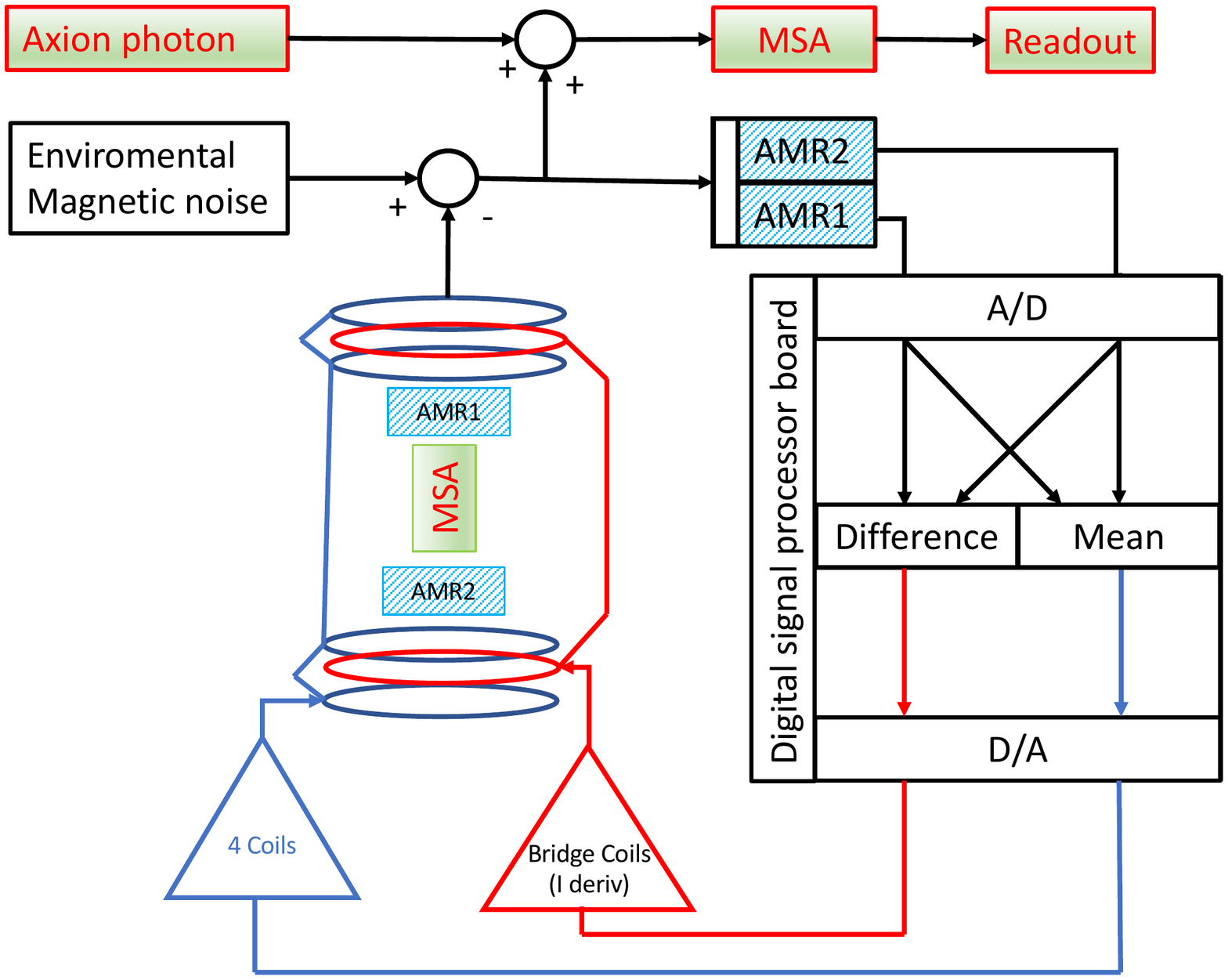}
    \caption{Schematic diagram of the control circuit used for the
    active magnetic field compensation.}
    \label{fig:csys}
  \end{center}
\end{figure}

Active magnetic shielding is obtained by means of canceling coils, like Helmholtz coils, that reduce the external magnetic-field by generating an opposite field \cite{Okaza}.  
A magnetic sensor (see Fig. \ref{fig:csys}) detects the external field, the signal is digitized and a current is generated to cancel the measured field. A simple active shielding 
system can be done using a negative feedback circuit with a magneto-impedance effect sensor to cancel magnetic field noise from dc to higher frequency~\cite{Okaza}.

\section{Secondary Amplification} % Ciambrone
\begin{figure}[h!]
\begin{center}
	\includegraphics[width=0.8\linewidth]{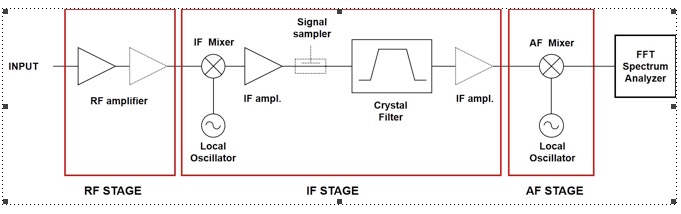}
	\caption{Simplified diagram of readout chain at room
	temperature (the dashed components may not be necessary).}
	\label{fig:chain_room_temp}
\end{center}
\end{figure}

The room temperature electronics does not pose stringent requirements in term of noise due to the high power gain ($>  35$ dB) (foreseen for) of the previous cryogenic stages. 
Commercial low noise components can therefore be used and the simplified readout scheme is shown in Fig. \ref{fig:chain_room_temp}

Three stages can be identified: a radio frequency (RF) amplifier stage, an intermediate frequency (IF) mixer stage and a near audio frequency (AF) mixer stage, the last two 
arranged in a double-heterodyne receiver.

The RF stage is based on low noise RF commercial amplifiers and devices with bandwidth of 500 MHz are adequate. The power gain, required to match the dynamic range 
of the following stages, strongly depends on the effective power of the signal and on the amplification that can be achieved by the cryogenic stages. A power gain between 30 
and 60 dB may be required and two amplification stages may be needed. An equivalent noise temperature of the RF amplifier(s) up to 150 K is also acceptable. In fact with a 
power gain of the previous stages higher than 35~dB, the contribution of the first RF amplifier to the system noise temperature is lower than 0.05~K and, therefore, negligible. 
Commercial Amplifier like \textsf{aptmp3-00200005-1627-D9-lN-2} by \texttt{AmpliTech, Inc.} or \textsf{CITLF3} by \texttt{Cosmic Microwave Technology, Inc.} are suitable for
such purpose but an R\&D phase is anyhow required.

The RF amplifier is placed just outside the cryostat and the necessity to put this stage in a RF shielded cabinet to minimize external noise pick up
should be carefully evaluated. The output of the RF amplifier feeds via a coaxial cable the IF stage of the double heterodyne receiver.
The first component of such stage is an image rejection mixer that shift the RF signal power from the tuneable resonant frequencies of the cavity to
a fixed intermediate frequency of 10.7 MHz while rejecting image noise power. The image rejection is necessary to avoid mixing off-band RF
power into the IF bandwidth and the value of the intermediate frequency is chosen to be compatible with commercial devices.  A mixer, like
\textsf{IRM10-1000} by \texttt{Sirius Microwave}, with an insertion loss of 7 dB and an image rejection better than 25 dB could be used for such purpose.

An IF amplifier to compensate the insertion loss of the mixer and a weakly coupled signal sampler to monitor the internal receiver power level can
be introduced at the mixer output if needed.

A high selectivity crystal bandpass filter is then introduced to suppress noise power outside the final signal bandwidth. The filter is required to prevent
any out-band noise components from being aliased into the signal bandwidth at the output of the following AF mixer stage and its bandwidth (probably
around 30 kHz) must be chosen to be compatible with the quality factor of the resonant cavity. The global receiver transfer function is strongly dependent
by such bandpass filter. The use of a thermalized enclosure box to avoid temperature fluctuation of the filter response should be evaluated.
Finally, a double-balanced mixer down converts the output of the bandpass filter, eventually amplified, to near audio frequency band centred around
30 kHz.

To reach high precision in the down conversion, a frequency synthesizer with low phase noise and good frequency stability is used to generate the local
frequency (LO) of the two mixers. The \textsf{N5171} by \texttt{Keysight}, with a phase noise of -119 dBc/Hz, a frequency accuracy of $\pm\,4 \times 10^{-8}$
and a long-term stability less than $\pm\,1$ ppm/year, is a good candidate.  If higher frequency accuracy and stability are needed an external primary
frequency standard could be used as LO reference clock.

The power spectrum of the AF mixer output is then acquired using a commercial FFT spectrum analyser. A resolution  bandwidth of hundreds hertz
should be reasonable to detect the excess power of the axion signal.

%\include{biblio_sigamp}

%{squid-hand} John Clarke and Alex I. Braginski, "The SQUID Handbook: Fundamentals and Technology of SQUIDs and SQUID Systems, I", (2004) DOI:10.1002/3527603646
%{squid-ref3} M. Mueck, M-O André, J Clarke, J Gail and C Heiden,"Radio-frequency amplifier based on a niobium dc superconducting quantum interference device with microstrip input coupling", Applied Physics Letters 72, 2885 (1998)
%{squid-ref4} M Mueck and R McDemott,"Radio-frequency amplifiers based on dc SQUIDs" Supercond. Sci. Technol. 23 (2010) 093001
%{squid-ref5} M Mueck J Clarke and C Welzel,"Superconducting quantum interference device amplifiers at gigahertz frequencies",Appl. Phys. Lett. 82, 3266 (2003).
%{squid-ref6}  Michael Mück, J. B. Kycia, and John Clarke,"Superconducting quantum interference device as a near-quantum-limited amplifier at 0.5 GHz", Appl. Phys. Lett. 78, 967 (2001).
%{squid-ref7} M Mueck M-O André, J Clarke, J Gail and C Heiden,"Microstrip superconducting quantum interference device radio-frequency amplifier: Tuning and cascading," Ammplied Physics Letters 75, 3545 (1999).
 
 \chapter{Analysis}\label{cha:ana}

\section{Introduction}
In haloscopes, the axion is detected as an excess in measured power spectrum 
in a resonant cavity, in presence of a magnetic field. The axion mass is 
unknown, then the data acquisition procedure is done as a sequence of tuning 
of the resonant frequency in order to span a wider range of possible axion masses. 
The power from axion-to-photon conversion on resonance is given by Eq. \eqref{eq:power}, 
reported here for convenience:
$$
%	\label{eq:power}
	P_{a} = \left( g_{\gamma}^2\frac{\alpha^2}{\pi^2}\frac{\hbar^3 c^3\rho_a}{\Lambda^4} \right) \times
	\left( \frac{\beta}{1+\beta} \omega_c \frac{1}{\mu_0} B_0^2 V C_{mnl} Q_L \right),
%P=(\frac{\alpha}{\pi}\frac{g_{\gamma}}{f_a})^2 V B_0^2 \rho_a C \frac{1}{m_a}min(Q_L, Q_a)
$$
%where $V$ is the volume of the cavity, $B_0$ is the magnetic field strength, 
%$Q_L$ is the cavity quality factor for the interesting propagating mode, $Q_a$ 
%is the axions quality factor, 
%$\rho_a$ is the local dark matter density, $C$ is a mode-dependent form factor, 
%and all the other quantities are defined in the Physics Reach section.
The goal of the analisys is to measure all the experimental parameters 
in order to define the power spectrum that best optimizes the SNR on 
the full frequency range.

From a different perspective, if an axion of mass $m_a$, corresponding to a 
photon frequency $\nu_a$, exists then it should appear as an excess on top 
of the noise in the final spectrum. The power produced by an axion-to-photon 
conversion is in any case very small: of the order of $10^{-22}$ W (in KLASH).
The analysis procedure must be designed to extract the evidence of such 
a small signal with respect a most prominent noise due to the thermal noise 
and the noise introduced by the amplification chain. A big number of power 
spectra must be collected for each value of the resonant frequency. All 
these spectra are combined opportunely before the subtraction of the 
background contribution, obtained from the data, with the help of the modeling 
of the acquisition chain. Simultaneously to the power signal, other ancillary 
data are collected in order to study the stability of the whole system and to 
characterize all the components of the noise, as explained in the following. 
As already discussed above, KLASH is designed to cover the frequency range 
between 60 and 250 MHz, by using two cavities and exploiting the position of 
three metallic rods to modify the resonant frequency for the $TM_{010}$ mode. 
An additional dielectric bar is foreseen for the fine-tuning, mainly in case 
of mode crossing. In this chapter we will briefly describe the analysis procedure, 
based on the experience of \cite{Brubaker} and \cite{ADMX}.
\\
Schematically, the data acquisition procedure is the following:
\begin{itemize}
\item The bars are moved to a new position
\item The antenna critical coupling is measured and optimized
\item The auxiliary parameters are measured (Q of the cavity, resonance frequency, temperatures, pressures,etc.)
\item A given numbers of power spectra are acquired
\end{itemize}
Some of the auxiliary parameters are remeasured to check the stability 
during the data acquisition,
then the procedure restart with the change to a different value of the resonance frequency. 
The integration time, for each value of the resonance frequency, depends on the accuracy needed to measure 
(or to put an upper limit on) the gamma coupling constant. Reversing the Dicke equation (see Eq. \eqref{eq:snr}) 
we have that, for a given SNR, the integration time is:
$$
%\label{ }
\tau = (SNR)^2\Delta\nu_a\,\left(\frac{k_B T_{sys}}{P_a^2}\right)^2 
$$
%where SNR is the signal to noise ratio, $k_B$ is the Boltzmann constant, $P_a$ the power of the signal, $\nu_a$ is the intrinsic 
%width of the galactic axions halo and $T_{sys}$ is the noise of the system, including both the physical temperature 
%of the cavity and the noise temperature of the microwave receiver. 
In the next sessions, we will briefly discuss how to combine the acquired power spectra in order to obtain a “global spectrum” 
to extract the signal.  
Equally important, as stated before, is the measurement of the cavity parameters. Here we will discuss briefly 
the procedure to measure some of them:
\\
 
{\it Q measurement}: the unloaded quality factor for the TM010 mode is very important to define the power of the axions 
conversion signal and the relative contribution with respect to the background. The quality factor depends on the cavity geometry, 
so it must be measured at each step of bars tuning. In order to ensure the stability of Q, influenced by mechanical rods stability 
and by the electrical contact between rods and cavity walls, the measurement is performed before and after each step of the tuning. 
An rf sweep is generated with a VNA (Vector Network Analyser) around the resonance frequency and injected through a very weakly 
coupled calibration port inside the cavity volume. The Q value is obtained from the study of the transmission response. 
To allow this measurement the reading port (signal port) is temporarily disconnected from the acquisition chain. 
\\

{\it Critical coupling measurement}: in order to optimize the scanning rate, it is very important to have a critical coupling between the 
$TM_{010}$ mode and the acquisition chain. The critical coupling is achieved when the incident power in the
 cavity, through the reading port, is completely absorbed without reflections. An attenuated signal, through a directional coupler, 
is injected in the readout port and then acquired. The mechanical insertion of the dipole is adjusted in order to minimize the 
reflected power. Considering that the critical coupling optimization procedure is quite slow, it is reasonable to think that 
this operation is not done at every measurement step, but after a certain time, to be determined based on the experimental 
conditions and stability.
\\

{\it Cryogenic amplifier noise and gain measurement}: the study of the noise in the cryogenic amplifier is very important because ultimately 
is the dominant component with respect to all the other sources of noise. The temperature noise can be measured slowly rising the physical 
temperature and by measuring the reflected power (in critical coupling regime) with the acquisition chain. This measurement is not compatible 
with the standard data acquisition and can be done only periodically. On the other hand, the stability of the noise can be checked each time 
the critical coupling procedure (see the previous point) is performed, looking at the stability of the minimum of the resonance curve. The gain 
of the amplifiers can be periodically checked injecting a small known signal in the input port of the amplification chain. 
\\

{\it Room temperature electronics}: this measurement is less critical with respect to what discussed about the cryogenic amplifier. 
In any case, it is important to check periodically the functioning of all the hot electronics section 
(amplifiers, filters, mixers) since external noise contribution could make unusable the collected data. 
To study the external noise one possibility is to disconnect the cryogenic amplifier from the acquisition chain and to connect 
the chain on an appropriate termination resistor. Excess on the acquired power spectrum is due exclusively to external 
perturbations. The rate at which this kind of test can be performed depends on the possibility to make noise measurement during 
rod movement, on the presence of external noise on the apparatus (that in any case must be reduced) and on the analysis 
procedure. A rescan procedure is in any case foreseen to avoid not persistent external noise.
\\

{\it Local oscillator stability}: an essential part of the analysis procedure is the combination of power spectra eventually 
collected at a very different time. Both power spectra collected in subsequent bars positions and power spectra referring to the 
same resonance frequency but collected at a very different time must be combined in order to obtain the final global spectrum. 
In any case, the stability on long period of the reference clock, used in the heterodyne acquisition chain, must be guaranteed. 
A clock lock system GPS based must be compared periodically with independent source reference. 

{\it Other parameters}: the working parameters of the apparatus must be constantly monitored. Cavity temperature, 
pressures, magnetic field, temperatures of the main electronics components and others are added to the 
power spectra acquired in order to define a set of coherent raw data. 

\section{Spectra Combination and Analysis}
The main task in the analysis is to combine a big number of power spectra opportunely reweighted, in order to obtain a
 single spectrum sensitive to axions signal up to the desired precision level. As described in other places in this document, 
the acquisition is done through a double heterodyne mechanism is such a way to bring the high-frequency signal from the cavity 
to a much lower frequency easier to analyze with commercial electronics. For each value of the frequency, the width of the 
spectrum is defined by the components in the acquisition chain, in particular, the image rejection and the filters. Final width 
of 50 kHz in the final af (audio frequency) spectrum is reasonable to allow the use of a conventional high-speed FFT. This spectrum 
is subdivided in 500 bins of 100 Hz each, to build the power histogram for each acquisition. By considering that the proposed 
acquisition time (see Physics Reach), that is 5 minutes for the big cavity and 10 minutes for the fins mode and the small cavity, 
and that a typical acquisition time for a commercial FFT is in the order of 10 ms, in each step of measurement we collect 
order of 30000 spectra (60000 for the small cavity and the fins). These spectra must be averaged in order to obtain a single 
spectrum for the considered frequency (“base spectrum”). The base spectra are the raw objects considered in the following 
analysis. We can imagine  several steps for the analysis:
\begin{itemize}
\item definition of bad samples and problematic bins, 
\item normalization of the spectra and subtraction of the baseline,
\item combination of spectra centered on different frequencies,
\item candidates definition and rescan,
\item extraction of the confidence limit.
\end{itemize}
In the following, we will outline the general scheme of the procedure.
\\

\subsection{Cut on raw data and problematic bin}
The specific reasons to cut some of the acquired spectra from the final data sample will 
be better defined during the data taking. Some of these reasons could be: too big variation of the resonance 
frequency before and after the spectrum measurement (Q during the acquisition cannot be considered stable), 
too high temperature fluctuations, not constant temperature in the hot electronics, anomalous gain in amplifiers. Most of these reasons 
are connected to failure in the cooling or vacuum systems. Some of them can be related to the mechanical movement of the bars. 
The experiment  will have a slow control system to detect and fix long-stand problems, but we cannot exclude that 
data can be affected by transient problems.
Even in good experimental conditions, some of the bins in an acquired spectrum could be affected by local and temporary problems, 
like electromagnetic interference. To identify problematic bins, we follow two strategies: either we compare the contains of the 
same frequency bin in two spectra acquired in two adjacent acquisition (for two different value of the central resonance frequency) 
or, if the perturbation remains also in subsequent base spectra, repeat the acquisition after a certain time. In the first case, 
we can identify in-time effects and exclude from the analysis the bins involved in the phenomenon. In the second case, one can 
imagine subdividing the integration time into different parts (for instance, in case of the big cavity, we can perform two sweeps 
in the full range with 2.5 minutes of integration per step, instead of 5 minutes). In such a way, the same bin will be acquired 
after a quite long time (order of months) and only very stable external perturbation will give a contribution to 
investigate further. 
\\ 

\subsection{Normalization and baseline subtraction}
Starting from the base spectra it is important to define the “analysis band”, in which the background baseline is calculated. 
Typically cuts on peripheral bins are used to exclude regions of the spectrum influenced by the sharpness of the filters used in 
the acquisition chain to clean-up the noise far from the central resonance value. The shape of each spectrum is influenced by the 
cavity noise and the cryogenic amplifier noise spectrum and the net gain of the receiver. To deduce this shape, two different 
procedures can be followed. In the first case, it is possible to build a model of the cavity based on lumped elements and then 
perform a fit to the measured spectrum. A non-perfect modelization of the device can introduce dilution of the signal, mainly 
in the case in which the signal of the axions spans over several bins. In the case of virialized axions, the intrinsic width of 
the cosmic halo signal is of about 100 Hz and then the signal should appear in 7 or 8 bins. The situation could be slightly better 
in case of non-thermalized axions, for which the power should be relaxed in a single bin. In any case, an alternative solution to 
the fit of the spectrum is the use of a digital Savitzky-Golay (SG)filter. This filter is a polynomial generalization of a moving 
average defined by only two parameters, W and d. The idea is that for each value, $x_0$, of the spectrum a least-squares d-degree 
polynomial fit in a W wide window is performed: the output of the filter is the value of the polynomial in 
$x_0$. Can be demonstrated \cite{Savitzky} that this procedure is equivalent to a filter with a flat passband and 
a mediocre stop-band attenuation. This implies that the shape of the baseline obtained through this filter is not influenced 
by “fast” signals (like the excess due to axions conversions) while the structure of the noise is preserved. The optimal value 
for W and d must be determined from the shape of the base spectrum. Once the baseline is defined, the spectra can be normalized. 
The contains of each bin after normalization is obtained from a Gaussian distribution with average 1 and sigma 
dependent on the bin. To combine bins from different spectra appropriate weights must be assigned. 
The weights depend on many factors:
\begin{itemize} 
\item instability of magnetic field, amplifiers gain, etc. in subsequent acquisitions,
\item differences between the central bin (corresponding to the resonance frequency) and the bins we want to combine (this is due 
to the fact that, in general, the spectra are not flat),
\item different thermal and electronic noise,
\item different number of spectra used to define the base spectrum (in principle this number should be constant, 
but dead time in the acquisition chain and system failure can reduce the number of spectra stored for each frequency step).
\end{itemize}
By applying this procedure iteratively to all the spectra acquired at the various frequencies we have a single spectrum in 
the full range of investigation. Actually, in KLASH, we will have three of these spectra corresponding to the three different 
configurations of the resonant cavity.

\subsection{Candidates and Rescan}
From the global spectrum, it is possible to extract the presence of axions candidates. As already said before, the search 
for axions conversion can be done either in a single bin, in case of non-thermalized axions, or looking to excess in the 
released power in multiple bins, in case of axions with an energy spread generated by the relative motion of the earth with 
respect to the axion halo. In both case the search procedure is very similar:
\begin{itemize}
\item a threshold is defined in order to isolate candidates in the global spectrum (either in single or multiple bins),
\item for all the bins passing the first threshold, a “rescan” is done to acquire more statistics for specific frequencies,
\item the events surviving to the application of a new threshold after the rescan will be examined individually.
\end{itemize}
There is a small difference between the multiple bins search with respect to the single bin search. Essentially there are 
two effects: first the Maxwellian velocity distribution of the axions means that not all the power will be released on a 
pre-defined number of bins, second a simple a priori group of multiple bins in larger bins not centered on all the possible value 
of the frequency scanned, can have the effect of signal splitting. 

In both searches, the main problem is the definition of 
the threshold both in the first phase and in the rescan. After the base spectra merging, for each bin the spectrum we have three 
relevant quantities: the power value $\delta_i$, the fluctuation $\sigma_i$ extracted from the 
Gaussian distribution corresponding to the 
bin, and the SNR $R_i$ . 
In the absence of correlations between the bins in the power distribution, the definition of the threshold depends 
only on these three quantities and on the desired confidence level:
\begin{equation}
%\label{ }
\Theta = R_T - \Phi^{-1}(c_1),
\end{equation}
where $R_T$ is the average of the SNR in the various bins while the function $\Phi$ is 
the cumulative distribution function of the standard normal distribution. 
The threshold in the rescan data depends on the new statistics acquired and, then, on the time of the rescan data acquisition. 
This duration depends on the level of precision needed to excluded fluctuations at a level higher with respect to the other bins. 
It is possible to derive a formula from the Dicke equation \cite{Brubaker} 
to consider both the statistics acquired in the normal scan and in the 
rescan phase. From this new level of SNR obtained it is possible to define the new threshold to apply to the rescan candidates. 
Any survivors to the rescan phase must be examined one-by-one, checking experimental effects (for instance, by disconnecting the 
acquisition chain from the cavity to understand if there are effects on the hot electronics due to external noise and terminating 
with a resistor the two main sources of rf in the cavity, the weakly coupled port and the readout port, or by improving the 
electromagnetic isolation, or by checking with dedicated search the sources of electromagnetic interferences in the experimental 
area).

\subsection{Limits}
We have no a priori knowledge of which bin corresponds to axions mass. On the other hand, in the presence of noise, 
the only difference between the true signal and a power fluctuation is that the axions signal is persistent across different 
spectra and scans at the same frequency. The only things we can do it to evaluate the ratio between the expected signal and noise 
level and the sigma of the Gaussian distribution in each bin and to require to be below a given threshold: the 
probability to have a signal above this value can be considered as the confidence level of the measurement. Since in the analysis 
procedure for each bin different values of $\sigma_i$ are obtained, due to experimental effects and different statistics collected 
(for instance due to the rescan procedure) the exclusion limit will depend on the position of the bin considered. 

\section{Computing}
Although the information collected in every single acquisition is relatively small and simple, the huge quantity of spectra 
needed to search for a fluctuation greater than the spectral baseline at the desired confidence level requires the definition of 
a strategy for the storage and data processing. In Tab. \ref{tab:tuning} in Sec. \ref{ch1:sec_sum-halo}, the total number of steps needed to 
complete the scan in the range accessible by KLASH is presented. The total of the steps is around 230000. Each base spectrum is 
obtained from the average of 30000 or 60000 spectra, for a total of the order of $10^{10}$ single spectra produced. In addition, 
ancillary information must be stored in the raw data needed by the analysis. It is difficult to establish a priori the correct 
number of this information (temperatures, pressures, resonance frequency, amplifiers gain, etc.) but it is something in the order 
of 10 words of 32 bits that means the order of 1 Gb of storage. The dimension of the single spectrum depends on the digitizer. 
If we want to have a spectrum 50 kHz wide with 100 Hz wide signal, a factor of 10 in the digitization in order to avoid structures
 due to axions conversion is probably needed. By assuming 1 kHz and an ADC of 16 bits, we have that a single spectrum is 80 kb. This means that the 
total space needed for the storage is 1100 TB, a quantity not easy to be managed. Moreover, we are assuming a commercial FFT, while 
other solution can be adopted. One possibility is to use a dedicated fast digitizer with the possibility to have more information 
on the spectrum and apply FFT and filters in software. Obviously, this solution while increasing further the quantity of data to
 be acquired. A solution to this problem is to design a system to have an online pre-processing during data acquisition, exploiting
 the RAM of the PC as a temporary buffer. The easier things to do in this pre-processing is to perform the average of the single 
spectra to obtain the base spectrum, but, eventually, also other steps of the proposed analysis can be moved in the online part 
(for instance the algorithm for the identification of the problematic bins). Probably the time needed to move the bars between 
two steps in the procedure is enough to accommodate the online processing. If this time is not sufficient we can think to 
parallelize the processing of the step N-1 while we start to acquire the data for the step N. Additional more complex processing 
(as for instance the FFT in software) would benefit on the graphics processing units (GPU) processing in order to allow a fast 
processing in real-time without back pressure on the digitizer.

%
%\begin{thebibliography}{}
%%\cite{Brubaker:2017rna}
%\bibitem{Brubaker:2017rna}
%  B.~M.~Brubaker, L.~Zhong, S.~K.~Lamoreaux, K.~W.~Lehnert and K.~A.~van Bibber,
%  %``HAYSTAC axion search analysis procedure,''
%  Phys.\ Rev.\ D {\bf 96} (2017) no.12,  123008
%  doi:10.1103/PhysRevD.96.123008
%  [arXiv:1706.08388 [astro-ph.IM]].
%  %%CITATION = doi:10.1103/PhysRevD.96.123008;%%
%  %19 citations counted in INSPIRE as of 04 Jul 2019
%
%%\cite{Asztalos:2001tf}
%\bibitem{Asztalos:2001tf}
%  S.~J.~Asztalos {\it et al.} [ADMX Collaboration],
%  %``Large scale microwave cavity search for dark matter axions,''
%  Phys.\ Rev.\ D {\bf 64} (2001) 092003.
%  doi:10.1103/PhysRevD.64.092003
%  %%CITATION = doi:10.1103/PhysRevD.64.092003;%%
%  %141 citations counted in INSPIRE as of 04 Jul 2019
%
%%\cite{Savitzky}
%\bibitem{Savitzky}
%  A.~Savitzky and M.~J.~E.~Golay,
%  Anal.\ Chem.\ D {\bf 36} (1964) 1627.
%\end{thebibliography}
%

%
%
\chapterimage{./Pictures/figs_ALP.pdf} %partimage can be used to assign a figure to a part cover page, not used in the final version
 \part{Bibliography}

\end{document}